\documentclass{elsart}

\usepackage{graphicx}

\usepackage{amssymb}

\usepackage{url}

\newcommand\noff{n_{\rm off}}     
\newcommand\non{n_{\rm on}}       
\newcommand\ntot{n_{\rm tot}}     
\newcommand\toffoverton{\tau}     
\newcommand\musignal{\mu_{\rm s}} 
\newcommand\mubkgnd{\mu_{\rm b}}  
\newcommand\muon{\mu_{\rm on}}    
\newcommand\muoff{\mu_{\rm off}}  %
\newcommand\mutot{\mu_{\rm tot}}  
\newcommand\estb{\hat\mu_{\rm b}} 
\newcommand\mles{\tilde{\mu}_{\rm s}}
\newcommand\mleb{\tilde{\mu}_{\rm b}}
\newcommand\mlemleb{\tilde{\tilde{\mu}}_{\rm b}}
\newcommand\sigmab{\sigma_{\rm b}}

\newcommand\binparam{\rho}        
\newcommand\ratmean{\lambda}      

\newcommand\pbi{p_{\rm Bi}}
\newcommand\zbi{Z_{\rm Bi}}

\newcommand\pn{p_{\rm N}}
\newcommand\zn{Z_{\rm N}}
\newcommand\zp{Z_{\rm P}}

\newcommand\pgamma{p_\Gamma}
\newcommand\zgamma{Z_\Gamma}

\newcommand\zpl{Z_{\rm PL}}

\newcommand\zsb{Z_{\rm sb}}
\newcommand\zssb{Z_{\rm ssb}}
\newcommand\zzr{Z_{\rm ZR}}

\newcommand\vfive{V_{\rm nn}}
\newcommand\zfive{Z_{\rm nn}}
\newcommand\znn{Z_{\rm nn}}

\newcommand\vfivep{V_{\rm bo}}
\newcommand\zfivep{Z_{\rm bo}}
\newcommand\zbo{Z_{\rm bo}}

\newcommand\vnine{V_{\rm BiN}}
\newcommand\znine{Z_{\rm BiN}}
\newcommand\zbin{Z_{\rm BiN}}

\newcommand\lhood{{\cal L}}
\newcommand\lhoodp{{\cal L}_{\rm P}}
\newcommand\lhoodg{{\cal L}_{\rm G}}
\newcommand\lratio{\Lambda}
\newcommand\llr{-2\ln{\lratio\left(\musignal\right)}}
\newcommand\llrzero{-2\ln{\lratio\left(\musignal=0\right)}}
\newcommand\confinta{100(1-2\alpha)\%}

\newcommand\zclaim{Z_{\rm claim}}
\newcommand\ztrue{Z_{\rm true}}
\newcommand\zdiff{\ztrue-\zclaim}

\newcommand\nbi{{\rm NBi}}
\newcommand\bi{{\rm Bi}}
\newcommand\nfail{N_{\rm on}}
\newcommand\nsuc{N_{\rm off}}

\newcommand\epsdir{.}

\begin{document}

\begin{frontmatter}



\title{Evaluation of three methods for calculating statistical
significance when incorporating a systematic uncertainty into a test
of the background-only hypothesis for a Poisson process}


\author[label1]{Robert D. Cousins},
\ead{cousins@physics.ucla.edu}
\author[label2]{James T. Linnemann},
\ead{linnemann@pa.msu.edu}
\author[label1]{Jordan Tucker}
\ead{tucker@physics.ucla.edu}
\address[label1]{Dept.\ of Physics and Astronomy, University of 
California, Los Angeles, California 90095, USA}
\address[label2]{Dept.\ of Physics and Astronomy,
Michigan State University, East Lansing, Michigan 48840, USA}

\begin{abstract}
Hypothesis tests for the presence of new sources of Poisson counts
amidst background processes are frequently performed in high energy
physics (HEP), gamma ray astronomy (GRA), and other branches of
science.  While there are conceptual issues already when the mean rate
of background is precisely known, the issues are even more difficult
when the mean background rate has non-negligible uncertainty.  After
describing a variety of methods to be found in the HEP and GRA
literature, we consider in detail three classes of algorithms and
evaluate them over a wide range of parameter space, by the criterion
of how close the ensemble-average Type I error rate (rejection of the
background-only hypothesis when it is true) compares with the nominal
significance level given by the algorithm.  We recommend wider use of
an algorithm firmly grounded in frequentist tests of the ratio of
Poisson means, although for very low counts the over-coverage can be
severe due to the effect of discreteness.  We extend the studies of
Cranmer, who found that a popular Bayesian-frequentist hybrid can
undercover severely when taken to high $Z$ values.  We also examine
the profile likelihood method, which has long been used in GRA and
HEP; it provides an excellent approximation in much of the parameter
space, as previously studied by Rolke and collaborators.
\end{abstract}

\end{frontmatter}

\section{Introduction}
\label{intro}
The incorporation of systematic uncertainties into hypothesis tests
(and by implication into confidence intervals and limits) remains a
murky area of data analysis in spite of much study in the professional
statistics community and in high energy physics, in gamma ray
astronomy, and in other branches of science \cite{cousinsoxford}.
Exact methods using the frequentist definition of probability
typically do not exist, while purely Bayesian methods, as commonly
used in high energy physics, invoke uniform priors which make the
resulting probability statements hard to interpret if not completely
arbitrary.

The foundational issues already arise in startlingly simple prototype
problems such as the one that we examine in this paper: $\non$ events
are observed from the Poisson process with mean $\musignal+\mubkgnd$,
where $\musignal$ is the unknown parameter of interest (the mean
number of signal events), while $\mubkgnd$ is the mean number of
background events (mimicking signal events), measured to have a value
$\estb$ with some uncertainty from subsidiary observations.  One
wishes to test the hypothesis $H_0$ that $\musignal=0$, i.e., that the
observed number of events is statistically consistent with being all
background.  In this paper, we focus on the {\em significance level}
$\alpha$ of the hypothesis test, also known as the {\em size} of the
test, and in particular consider the very small values of $\alpha$
corresponding to a statistical significance of up to five standard
deviations.  In the formal theory of Neyman-Pearson hypothesis
testing, $\alpha$ is specified in advance; once data are obtained, the
{\em $p$-value} is the smallest value of $\alpha$ for which $H_0$
would be rejected.  In a real application, the {\em power} of the
test, which depends on the alternative hypothesis, should be
considered as well, but we do not explore that complementary aspect of
the test here \cite{Kendall}.  Also, we do not address the complex
issue of the utility of $p$-values, which is discussed by Berger and
others (e.g., Refs.~\cite{Berger87,Sellke}); we merely remind the
reader that at best, a $p$-value conveys the probability under $H_0$
of obtaining a value of the test statistic at least as extreme as that
observed, and that it should not be interpreted as the probability
that $H_0$ is true.  Having said that, given the ubiquity of
$p$-values in the literature, we study in detail the efficacy of three
methods for calculating $p$-values in the presence of systematic
uncertainties.

Frequently the $p$-value is communicated by specifying the
corresponding number of standard deviations in a one-tailed test of a
Gaussian (normal) variate; i.e., one communicates a $Z$-value (often
called $S$ in HEP) given by
\begin{equation}
\label{zdef}
Z = \Phi^{-1}(1-p) = -\Phi^{-1}(p)
\end{equation}
where
\begin{equation}
\label{Phidef}
\Phi(Z) = \frac{1}{\sqrt{2\pi}} \int_{-\infty}^Z \,\exp(-t^2/2)\,dt
\ =\ \frac{1 + {\rm erf}(Z/\sqrt{2})}{2},
\end{equation}
so that
\begin{equation}
\label{eqn:z}
Z = \sqrt{2}\, {\rm erf}^{-1}(1-2p).
\end{equation}
Thus, for example, $Z=3$ corresponds to a $p$-value of 
$1.35 \times 10^{-3}$.
This relation can be approximated to better than 1\% for 
$Z > 1.6$ as 
\begin{equation}
\label{eqn:zapprox}
Z \approx \sqrt{u - \ln \ u}, 
\end{equation}
where $u = - 2 \ln (p \sqrt{2 \pi} )$. (See Appendix
\ref{app:zapprox}.)  This form fortuitously is much more accurate than
directly inverting the full asymptotic expansion to second or third
order.  Asymptotically, $Z$ goes as $\sqrt{- \ln\ p}$ at large $Z$.

If the uncertainty on $\estb$ vanishes (so that $\estb=\mubkgnd$),
some controversy exists as to the best way to proceed, but at least in
that case there seems to be some clarity about the different methods,
their performance, and their merits and demerits.  In contrast, if the
uncertainty on $\estb$ is non-negligible, then the nature of the
subsidiary measurement of $\mubkgnd$ becomes crucial, and the
interpretation of results of various recipes (algorithms for computing
the $p$-value) becomes much more difficult.  We take a pragmatic point
of view that the performance of a recipe is of more interest than the
foundational solidity of the recipe, and evaluate this performance by
the frequentist criterion of how well the nominal significance level
of a test corresponds to the true frequency of Type I errors
(rejecting $H_0$ when it is true).

As in Ref.~\cite{Linnemann}, we consider two variations of this
prototype problem (described in Sec.~\ref{twoproblems}), which differ
in the specification of the subsidiary measurement of $\mubkgnd$.  In
the first case, it is a (typically small-integer) Poisson measurement
in a signal-free control region, and in the second case it is a
Gaussian (normal) measurement with known rms deviation.
Section~\ref{binom} describes the little-used fact
\cite{Linnemann,zhang} that the standard frequentist solution to the
ratio-of-Poisson-means problem can be directly applied to the first
prototype problem at hand, which makes evaluation of $Z$ easy with
modern software tools.  In Sec.~\ref{bayes}, we outline the
frequentist-Bayesian hybrid which is commonly used in HEP, noting its
lack of foundational solidity and ambiguity due to choice of the
Bayesian prior.  We note the remarkable mathematical connection
between one choice of prior and the frequentist solution of
Sec.~\ref{binom}.  In Sec.~\ref{profilelikelihood}, we explore the
profile likelihood method (well-known in HEP as the MINUIT MINOS
method \cite{minuit} and in gamma ray astronomy (GRA) as popularized
by Li and Ma \cite{li}), which gives approximate results based on
likelihood ratios.  In Sec.~\ref{othermethods}, we briefly describe
other methods, and in Sec.~\ref{comparison} we compare some results
obtained with all the methods.

In the remaining sections we focus on the three main methods
introduced in Secs.~\ref{binom}-\ref{profilelikelihood}, and study in
detail the relations among the computed $Z$ values and the
Type I error rates, as one spans the space of true values of the
parameters.  We conclude in Sec.~\ref{conclusion} that the little-used
frequentist solution should have much broader use, and we even
advocate its prudent use in the second prototype problem, in which it
applies only via a rough correspondence.  As found in
Refs.~\cite{rolkelopez,rolkeconrad}, (which advocate some
modifications) the profile likelihood method provides remarkably good
results over a wide range of parameters.  Given the richness of
results even for these simple prototype problems, there remains much
work to be done, beyond the scope of this paper, in exploring
performance of other recipes and further generalizations to more
complicated problems \cite{cousinsoxford,cranmerslac,cranmer,banff}.

Appendix~\ref{app:notation} contains a summary of our notation.
Appendix~\ref{app:zapprox} has a derivation of Eqn.~\ref{eqn:zapprox},
followed in Appendix~\ref{app:proof} by a proof of the ``remarkable
connection'' mentioned above.  Calculational details of the various
$Z$-values are in Appendix~\ref{app:ztrue}, and some implementation
examples are in Appendix~\ref{app:root}.

\section{\boldmath
Two prototype problems differing in the measurement of $\mubkgnd$}
\label{twoproblems}

\subsection{The on/off problem}
\label{onoff}
In the first prototype problem, which we refer to as the ``on/off''
problem, the subsidiary measurement of $\mubkgnd$ consists of the
observation of $\noff$ events in a control region where no signal
events are expected.  In HEP, the control region is commonly referred
to as a ``sideband'' since it is typically a sample of events which is
near the signal region in some measured parameter, i.e., in a band of
that parameter alongside but disjoint from the parameter values where
the signal might exist.

This HEP prototype problem has an exact analog in gamma ray astronomy
(GRA), upon which we base our notational subscripts ``on'' and
``off''.  The observation of $\non$ photons when a telescope is
pointing at a potential source (``on-source'') includes both
background and the source, while the observation of $\noff$ photons
with the telescope pointing at a source-free direction nearby
(``off-source'') is the subsidiary measurement.  In both the HEP and
GRA examples, we let the parameter $\toffoverton$ denote the ratio of
the expected means of $\noff$ and $\non$ under $H_0$, i.e., when
$\muon = \mubkgnd$:
\begin{equation}
\label{toffoverton}
\toffoverton \equiv \muoff/\mubkgnd.
\end{equation}
In GRA, $\toffoverton$ in the simplest case is the ratio of observing
time off/on source (subject to corrections in more complicated cases),
while in HEP the calculation of $\toffoverton$ might involve
background shapes, efficiencies, etc., determined by Monte Carlo
simulation.  In the prototype problems studied in detail in this paper
we assume that $\toffoverton$ itself is known exactly or with
negligible uncertainty.  Thus, since the point estimate of $\muoff$ is
$\noff$, the point estimate of $\mubkgnd$ is
\begin{equation}
\label{bnofftau}
\estb = \noff / \toffoverton.
\end{equation}

\subsection{The Gaussian-mean background problem}
\label{gaussianmean}
In a second prototype problem, which we refer to as the
``Gaussian-mean background'' problem, the subsidiary measurement of
$\mubkgnd$ is assumed to be drawn from a Gaussian (normal) probability
density function (pdf) with rms deviation $\sigmab$.  We emphasize
that while the measurement of the background {\it mean} has a Gaussian
pdf, the number of background counts obeys Poisson statistics
according to the fixed but unknown true background mean as described
above.  In this paper, we consider two cases, one in which $\sigmab$
is known absolutely, and one in which $\sigmab$ is known to be a
fraction $f$ of $\mubkgnd$, and therefore the experimenter estimates
$\sigmab$ by $f\estb$ in analyzing the data from an experiment.

\subsection{Correspondence between the two problems}
These two problems have an approximate correspondence since a {\it
rough} estimate of the uncertainty in estimating $\muoff$ by $\noff$
is $\sqrt{\noff}$, so that a rough estimate of the uncertainty on
$\estb$ in the first problem is $\sqrt{\noff}/\toffoverton$. Thus, the
correspondence is
\begin{equation}
\label{sigmab_corr}
\sigmab = \sqrt{\noff}/\toffoverton, 
\end{equation}
which when combined with
Eqn.~\ref{bnofftau} yields
\begin{equation}
\label{eqn:corr}
\toffoverton = \estb/\sigmab^2.
\end{equation}
We emphasize that {\em in using this rough correspondence in
equations, one takes both conceptual and numerical liberties.}
Nonetheless, it is useful to study the pragmatic consequences of
transferring recipes between the two prototype problems based on the
correspondence in Eqns.~\ref{sigmab_corr} and \ref{eqn:corr}, while of
course keeping in mind the lack of firm foundation.

\section{Frequentist solution to the on/off problem}
\label{binom}

The on/off problem above maps exactly onto one of the classic problems
in statistics, namely that of constructing hypothesis tests for the
ratio of Poisson means (solved by Przyborowski and Wilenski
\cite{Przy}). Each of $\non$ and $\noff$ is a sample from a Poisson
probability with unknown means $\muon$ and $\muoff$; the
background-only hypothesis $H_0$ is therefore that the ratio of
Poisson means $\ratmean=\muoff/\muon$ is equal to the corresponding
ratio with background only, $\toffoverton$.

The joint probability of observing $\non$ and $\noff$ is the product
of Poisson probabilities for $\non$ and $\noff$, and can be rewritten
as the product of a single Poisson probability with mean $\mutot =
\muon+\muoff$ for the total number of events $\ntot$, and the binomial
probability that this total is divided as observed if the binomial
parameter $\binparam$ is $\binparam = \muon/\mutot = 1/(1+\ratmean)$:
\begin{eqnarray}
P(\non,\noff) & = & 
 \frac{\e^{-\muon} \muon^{\non}}{\non !} \times
 \frac{\e^{-\muoff} \muoff^{\noff}}{\noff !} \nonumber \\ & = &
 \frac{\e^{-(\muon+\muoff)}\,
(\muon+\muoff)^{\ntot}}{\ntot !}  \times \ \\
&& \frac{\ntot!}{\non!(\ntot-\non)!}\, 
\binparam^{\non}\,(1-\binparam)^{(\ntot-\non)}.
\label{eqn-jointProb}
\end{eqnarray}
That is, rewriting in terms of observables $(\non,\ntot)$ and
parameters $(\ratmean,\mutot)$:
\begin{eqnarray}
P(\non,\noff; \muon, \muoff) &=& 
P(\ntot; \muon+\muoff)\ P(\non| \ntot; \binparam)\\
            &=& P(\ntot; \mutot)\ P(\non| \ntot;1/(1+\ratmean) ),
\end{eqnarray}
where on the right-hand side the probabilities $P$ are Poisson and
binomial, respectively.  In this form, all the information about the
ratio of Poisson means $\ratmean$ (and hence about $H_0$) is in the
{\em conditional} binomial probability for the observed ``successes''
$\non$, {\em given} the observed total number of events $\ntot =
\non+\noff$.  In the words of Reid~\cite{Reid95}, ``\dots it is
intuitively obvious that there is no information on the ratio of rates
from the total count\dots''.  The same result was obtained in the HEP
community by James and Roos \cite{JamesRoos} and in the GRA community
by Gehrels \cite{gehrels}.  Therefore one simply uses $\non$ and
$\ntot$ to look up a standard hypothesis test result for the binomial
parameter $\binparam$, and rewrites it in terms of $\toffoverton$ and
hence $H_0$.  To be more explicit, in the notation thus far $H_0$ can
be variously expressed as:
$\musignal=0$; 
$\muon=\mubkgnd$; 
$\muoff/\muon = \toffoverton$;
$\ratmean = \toffoverton$; or as most relevant here,
$\binparam = 1/(1+\toffoverton)$.
In the last form, the standard frequentist binomial parameter test can
be used; this dates back to the first construction of confidence
intervals for a binomial parameter by Clopper and Pearson in 1934
\cite{Kendall,Clopper}.

The $p$-value for the test of $\binparam = 1/(1+\toffoverton)$, and
hence of $H_0$, is then the one-tailed probability sum:
\begin{equation}
\label{pbisum}
\pbi = \sum_{j=\non}^{\ntot} P(j|\ntot;\binparam).
\end{equation}
This can be computed from a ratio of
incomplete and complete beta functions (both denoted by $B$ and
distinguished by the number of arguments):
\begin{equation}
\label{betaincomplete}
\pbi = B(\binparam,\non,1+\noff)/B(\non,1+\noff).
\end{equation}
The corresponding $Z$-value, $\zbi$, then follows using
Eqn.~\ref{eqn:z}.  This {\em ratio} in Eqn.~\ref{betaincomplete} is
itself called ``the'' incomplete beta function in {\em Numerical
Recipes} \cite{numrep}, which contains an algorithm for calculating
it.  This algorithm is implemented in the analysis software package
ROOT \cite{root}; examples of the ROOT implementation are in Appendix
\ref{app:root}.  This implementation, however, runs into numerical
troubles for large values of its parameters; for the calculations in
this paper we use a different implementation of the incomplete beta
function due to Majumder and Bhattacharjee~\cite{newbetainc}, which
exhibits good precision over the parameter space studied here.

As reviewed by Cousins \cite{cousinsratio}, the above construction for
tests of the ratio of Poisson means (or equivalently, confidence
intervals for the ratio of Poisson means) is used broadly in science
and engineering.  This use of conditional binomial probabilities in a
problem with discrete observations is discussed in
Ref.~\cite{cousinsratio}, which observes that these need not
correspond to uniformly most powerful unbiased tests, since the
theorem of Lehmann and Scheff\'e assumes continuous observables.
Ref.~\cite{cousinsratio} constructs a set of binomial confidence
intervals which are subsets of the standard ones (and therefore at
least as short in any metric).  However, use of such intervals remains
controversial because of the importance with which conditioning is
regarded in statistical inference \cite{Reid95}, as also discussed in
Ref.~\cite{cousinsratio}.  For the demonstrations in this paper, we
use the standard set, which is more conservative, particularly for
small numbers of counts, due to the discreteness.

Remarkably, while the ratio-of-Poisson-means problem and solution are
widely known, its straightforward application to the central problem
of this paper seems to have escaped both the GRA and HEP communities,
except for the 1990 paper by Zhang and Ramsden \cite{zhang} in GRA and
the recent paper by one of us \cite{Linnemann}, which is the only
paper we know of that cited Zhang and Ramsden.
 
\section{Bayesian-frequentist hybrid recipes for the two problems}
\label{bayes}

Recipes which combine Bayesian-style averaging with frequentist
calculation of tail-integral probabilities may have intuitive appeal
and some adherents in the professional statistics community
\cite{Berger99,box}, but such mixing of paradigms must be viewed with
care: either one is introducing the foreign notion of a pdf of an
unknown true value into a frequentist calculation, or one is
introducing the foreign notion of a tail probability (i.e.,
probability of obtaining data not observed) into a Bayesian
calculation.  Once a hybrid method is used to calculate a $p$-value 
or a $Z$-value significance, then it is by definition attempting a
frequentist claim and is appropriate to evaluate it by those
standards, and in particular to test if the true Type I error rate of
the method is consistent with claimed significance levels: if not,
this is a weakness of the method.

Thus, the properties of such hybrid calculations must be understood,
in the present context by computing the true Type I error rate of a
hypothesis test with significance level corresponding to some chosen
stated $Z$-values.  Cousins and Highland \cite{Cousinshighland}
recommended such a hybrid for the prototype problem of small-count
upper limits in which one wishes to incorporate an uncertainty in the
normalization.  The resulting upper limits as applied in HEP (which
typically take uniform prior for the background mean) appear to be
conservative, i.e., the Type I error rate of the corresponding
hypothesis test is less than implied by the quoted $Z$-value.  The
basic idea has been extended to problems in which the uncertainty is
on the mean background, with studies such as that of Tegenfeldt and
Conrad \cite{Tegenfeldt} indicating continued conservatism in the
results, at least for low $Z$-values.  However, Cranmer has warned
\cite{cranmer} that for $Z=5$, gross over-statement of the
significance can result.  Thus it is important to define the recipe(s)
precisely and study the performance.

For the two prototype problems in Sec.~\ref{twoproblems}, if there is
no uncertainty in $\estb$, then $\estb=\mubkgnd$ and the $p$-value
(denoted by $p_{\rm P}$) can be obtained immediately by computing the
Poisson probability of obtaining $\non$ or greater counts given true
mean $\mubkgnd$:
\begin{equation}
\label{pp}
p_{\rm P} = \sum_{j=\non}^\infty \e^{-\mubkgnd}\mubkgnd^{\,j}/j!
= \Gamma (\non,0,\mubkgnd)/\Gamma(\non),
\end{equation}
here written \cite{Kendall_vol1}
in terms of the lower incomplete $\Gamma$ function,
\begin{equation}
\Gamma(n,0,x) = \int_0^x \ t^{n-1}\,e^{-t}\,dt.
\end{equation}
With uncertainty in $\estb$, then with the Bayesian definition of
probability (degree of belief), one can encapsulate the result of the
background measurement into a pdf $p(\mubkgnd)$, assumed to be
normalized here.  While this is sometimes considered to be a prior
pdf, Refs.~\cite{Linnemann,cranmerslac,Cousinshighland} consider it to
be the posterior pdf of the background measurement, which is the
product of the prior pdf for the background measurement as well as its
likelihood function from the subsidiary measurement.  In any case,
ignoring foundational issues, one can then attempt to introduce this
uncertainty by averaging $p_{\rm P}$ over different values of
$\mubkgnd$, weighted by $p(\mubkgnd)$, so that the hybrid $p$-value so
obtained is
\begin{equation}
\label{eqn:bayesavg}
\int\, p_{\rm P}\, p(\mubkgnd)\, d\mubkgnd.
\end{equation}

While the above approach was viewed by Cousins and Highland as adding
some Bayesian reasoning to a frequentist $p$-value, the same
mathematical result is obtained if one starts from the Bayesian
prior-predictive distribution and adds on a frequentist-style tail
probability calculation to obtain the prior-predictive $p$-value, as
advocated by Box \cite{box}; the different points of view simply
correspond to reversing the order of summing/integrating
\cite{Linnemann,cranmer,demortier}.

\subsection{Hybrid recipe using Gaussian likelihood for 
the Gaussian-mean background problem: $\zn$} 
A common assumption in HEP (even when the underlying statistics of the
measurement of $\mubkgnd$ is Poisson) is that of uniform prior and
Gaussian likelihood so that $p(\mubkgnd)$ is Gaussian.  Then $\pn$
denotes the resulting hybrid $p$-value obtained from
Eqn.~\ref{eqn:bayesavg}, and $\zn$ denotes the $Z$-value derived from
it via Eqn.~\ref{eqn:z}. (The subscript N is for ``normal'', the usage
preferred by statisticians.)  For the results in this paper, we
implemented our own program and checked that it gave the same results
as one of several such programs of which we are aware,
Ref.~\cite{bityukov}, except where renormalization caused a
difference.

In typical programs (including ours), the low tail of the Gaussian is
truncated to avoid negative values of $\mubkgnd$ (and the result
renormalized).  If this truncation is not negligible (so that the
renormalization makes a non-negligible difference), then conceptual as
well as procedural problems arise.  Conceptually, the problem is a
nonzero density for the true background at $\mu_b=0$, despite a
nonzero measurement.  As emphasized in Ref.~\cite{Cousinshighland}, if
truncation makes a material difference, the Gaussian form of the pdf
may not be appropriate, and a form which goes to zero at the origin
(such as log-normal) may be a better model; in the next subsection,
the Gamma function density arises naturally and is well-behaved in
this respect.  As Cranmer et al. have noted \cite{cranmeratlas}, one
must also understand the $Z\sigmab$ contours of the background in
order to claim that $Z$-value.  Thus, a sign that the Gaussian form is
almost certainly inadequate is if one finds $Z$ such that
$Z\sigmab>\mubkgnd$, since in this case the computation assumes that
the high tail of the Gaussian is reliable in a region where the
corresponding low tail is in the non-physical negative region.

Furthermore, for $Z\sigmab>\mubkgnd$ and large enough $\mubkgnd$, the
systematic uncertainty $\sigmab$ is much larger than the statistical
fluctuations in $\non$ (which are of order $\sqrt{\mubkgnd}$).  The
circumstance in which ones observes high $Z$ is then essentially a
measurement $\estb$ which is lower than $\mubkgnd$ by $Z\sigmab$.  But
since $\estb$ is constrained to be non-negative, $\mubkgnd/\sigmab$
becomes an effective upper limit on the observed $Z$, which is only
rarely significantly surpassed by anomalously high statistical
fluctuations in $\non$.

For both these reasons, $Z\sigmab>\mubkgnd$ leads to unreliable $Z$;
since $\sigmab=f\mubkgnd$, the criterion for unreliable $Z$ is then
roughly \cite{cranmeratlas}
\begin{equation}
\label{zf}
Z>1/f;
\end{equation}
of course statistical fluctuations superimposed on the mean-background
uncertainty complicate the argument, but we take Eqn.~\ref{zf} as a
useful rule of thumb, and care should be taken as $Z$ approaches $1/f
= \mubkgnd/\sigmab$.

\subsection{Hybrid recipe using Poisson likelihood 
for the on/off problem: $\zgamma=\zbi$} If the underlying statistics
of the measurement of $\mubkgnd$ is Poisson, then an alternative
advocated by one of us some years ago \cite{jtlclw}, and which is also
known to the GRA community \cite{alexandreas}, again uses the uniform
prior, but with the likelihood function for $\mubkgnd$ appropriate to
the on/off problem ($\noff$ events observed in a Poisson sample from a
control region with mean that is $\toffoverton$ times that of the
background in the signal region):
\begin{equation}
\label{eqn:lhood}
\lhood(\mubkgnd) = \frac{(\toffoverton \mubkgnd)^{\noff}
                 \e^{-\toffoverton \mubkgnd}}{\noff!}.
\end{equation}
With uniform prior, the posterior pdf $p(\mubkgnd)$ is the same
mathematical expression, which is a Gamma function.  Inserting this
into Eqn.~\ref{eqn:bayesavg} results in a $p$-value denoted by
$\pgamma$ (given explicitly in Eqn.~\ref{eqn:pgexplicit}) with a
corresponding $Z$-value denoted by $\zgamma$.

{\em Remarkably, the values computed for $\zgamma$ are identical to
those computed for the frequentist result $\zbi$} of Sec.~\ref{binom}!
This is quite surprising even if not unprecedented as a mathematical
``coincidence'' of results from Poisson-based Bayesian and frequentist
calculations; one can recall for example that upper limits with
uniform prior (and lower limits with $1/\mu$ prior) are identical to
corresponding frequentist results, due to an identity which connect
integrals of the Poisson probability over $\mu$ with sums over the
observed integers \cite{cousinsajp}.  In the present case, after the
identity was suggested by numerical results in preparation of
Ref.~\cite{Linnemann}, an unpublished proof was worked out \cite{kim}.
Our more recent, shorter proof, is presented in
Appendix~\ref{app:proof}.  The identity of $\zgamma$ and $\zbi$
guarantees good frequentist properties for hybrid Bayesian-derived
$\zgamma$.  Of course there is no such guarantee for hybrid
Bayesian-derived $\zn$.

\section{The profile likelihood method}
\label{profilelikelihood}
The profile likelihood method (based on asymptotic theory and
therefore not exact for finite sample sizes) has long been widely used
for evaluating approximate confidence intervals and regions in HEP,
notably using the method called MINOS in the CERN Program Library
package MINUIT \cite{minuit,jamescpc}.  (Further discussion, with some
modifications, is in Refs.~\cite{rolkelopez,rolkeconrad}.)  In GRA the
application to the on/off problem by Li and Ma \cite{li} is widely
cited.  Using the correspondence between confidence intervals and
significance tests discussed in Ref.~\cite{Kendall}, the test of the
hypothesis $H_0$ that $\musignal=0$ at significance level $\alpha$
corresponds to a test if $\musignal=0$ is contained in the $\confinta$
C.L. central confidence interval for $\musignal$.  Thus the
profile-likelihood-derived $p$-value for an obtained data set is
obtained by first finding the smallest C.L. for which $\musignal=0$ is
included in the profile-likelihood-derived approximate central
confidence interval, and then $p = (1-{\rm C.L.})/2$.  To obtain the
approximate confidence interval, one begins with the likelihood
function; for the on/off problem, this is
\begin{equation}
\label{eqn:LP}
\lhoodp = \frac{\left( \musignal + \mubkgnd \right)^{\non}}{\non !}
             e^{-\left( \musignal + \mubkgnd \right)}
            \frac{\left(\toffoverton \mubkgnd\right)^{\noff}}{\noff !}
             e^{-\toffoverton \mubkgnd},
\end{equation}
while for the Gaussian-mean background problem with either absolute or
relative $\sigmab$, it is
\begin{equation}
\label{eqn:LG}
\lhoodg = \frac{\left( \musignal + \mubkgnd \right)^{\non}}{\non !}
             e^{-\left( \musignal + \mubkgnd \right)}
             \frac{1}{\sqrt{2\pi\sigmab^2}}
             \exp{\left(-\frac{\left(\estb - \mubkgnd\right)^2}
                              {2\sigmab^2}
                  \right)},
\end{equation}
where as discussed below we have explored the effect of truncating the
Gaussian pdf in $\estb$ and renormalizing prior to forming $\lhoodg$.

Using either $\lhoodp$ or $\lhoodg$, one obtains the log-likelihood
ratio
\begin{equation}
\label{eqn:lratio}
\lratio(\musignal) = 
\frac{\lhood\left( \musignal, \mlemleb \left(\musignal\right) \right)}
            {\lhood\left( \mles, \mleb \right)},
\end{equation}
where $\mles$ and $\mleb$ are the maximum-likelihood estimates of
$\musignal$ and $\mubkgnd$, respectively, obtained by minimizing the
appropriate likelihood function with respect to both $\musignal$ and
$\mubkgnd$, and $\mlemleb\left(\musignal\right)$ is the result of
minimizing the likelihood function only with respect to $\mubkgnd$,
left as a function of $\musignal$. The log-likelihood ratio in
Eqn.~\ref{eqn:lratio} has one free parameter, so under regularity
conditions and in the limit of large sample counts $\ntot$, Wilks's
asymptotic theorem~\cite{Wilks} says that under the null hypothesis,
$\llr$ is distributed as a chi-square statistic with one degree of
freedom (d.o.f.).  The $\confinta$ confidence interval would therefore
be the set of $\musignal$ for which
\begin{equation}
\label{eqn:lratioCI}
\llr < F_{\chi^2_1}^{-1}(1-2\alpha),
\end{equation}
where $F_{\chi^2_1}^{-1}$ is the inverse cumulative distribution
function for the chi-square with one d.o.f.  The background-only
hypothesis $H_0$ would then be rejected at significance level $\alpha$
if the so-formed $\confinta$ C.L. confidence interval for $\musignal$
does not contain the value $\musignal=0$.

In the present case, as emphasized to us by Cranmer, the regularity
conditions of Wilks's theorem are in fact not satisfied since the null
hypothesis ($\musignal=0$) is on the boundary of allowed $\musignal$.
This affects the lower endpoint of the confidence interval and changes
the confidence level of the full intervals.  However, the asymptotic
Type I errors associated with the upper endpoint and tail appear to be
unaffected, and we thus proceed using the nominal results for
significance claims. As noted above, the $p$-value is then the
smallest value of $\alpha$ for which $H_0$ would be rejected.  As the
chi-square with one d.o.f., is the positive half of a Gaussian under
an appropriate transformation of variables, the $Z$-value
corresponding to the $p$-value for an obtained data set can be
computed directly from the likelihood ratio as
\begin{equation}
\label{eqn:lratioCIequiv}
\zpl =  \sqrt{\llrzero},
\end{equation}
where the likelihood ratio is computed using $\lhoodp$ or $\lhoodg$,
as appropriate for the problem.

For the on/off problem and $\lhoodp$, the explicit result obtained
from Eqn.~\ref{eqn:lratioCIequiv} was given by Li and Ma (their
Eqn.~17) \cite{li}:
\begin{equation}
\label{eqn:lima}
\zpl = \sqrt{2}\  \left(\non \ln \frac{\non (1+\toffoverton)}{\ntot} 
  + \noff 
  \ln\frac{\noff (1+\toffoverton)}{\ntot\toffoverton}\ \right)^{1/2}.
\end{equation}

\section{\boldmath Other Methods for Estimating $Z$}
\label{othermethods}
Other methods for estimating $Z$ found in the literature are typically
of the form of the ratio of the inferred signal size to its rms
deviation, i.e., $Z = s/\sqrt{V}$, where in the on/off problem the
signal $s$ is estimated by $s = \non - \estb = \non -
\noff/\toffoverton$, and where $V$ is an estimate of the variance of
$s$.

One widely used form is 
\begin{equation}
\zsb = \frac{s}{\sqrt{\estb}}
\end{equation}
(sometimes \cite{astrostats} imprecisely called the ``signal to noise
ratio'').  While this ignores the uncertainty in the background
estimate, it is often used for optimizing selection criteria, because
of its simplicity.

Occasionally one also sees
\begin{equation}
\zssb =  \frac{s}{\sqrt{\non}} = \frac{s}{\sqrt{\estb + s}}.
\end{equation}
Aside from recommending $\zpl$, Ref.~\cite{li} mention this in their
Eqn.~11.  Our experience is that this expression typically results
from confusing a test of the null hypothesis ($\musignal=0$) with
estimating $\musignal$ and its 1-$\sigma$ uncertainty once the
existence of a signal has been established.  For example, if $\non=9$
and $\estb = 0.1$ with small uncertainty, then a correct $Z$-value
will be very high, even though a estimate of $\musignal$ will have a
relative uncertainty of roughly $1/\sqrt{\non} = 1/3$. (If there is a
paradox due to the notion that the estimate of $\musignal$ is ``only
3$\sigma$ from zero'', it is resolved by carefully considering
confidence intervals and noting the non-Gaussian behavior.)  In
another extreme, if $\sigmab$ is large, $\zssb$ can badly
over-estimate the significance.

Ref.~\cite{li} also gives as another example method (their Eqn.~5),
\begin{equation}
 \vfive = \non + \noff/\toffoverton^2,
\end{equation}
(subscript nn for no null) which as the authors note treats $\non$ and
$\noff$ as independent, and therefore does not consistently calculate
$V$ under the null hypothesis, $\muoff/\muon = \toffoverton$. In fact
it biases against signals for $\toffoverton > 1$ by overestimating
$V$.  In the limit of large $\toffoverton$, $\vfive \rightarrow \non =
s + \estb$, where $\estb$ has negligible uncertainty.  Then using
$\vfive$ leads to $\zssb$, which as noted above is not appropriate.

Ref.~\cite{Linnemann} has derived a related formula, 
\begin{equation}
\vfivep = \noff(1+\toffoverton)/\toffoverton^2,
\end{equation}
(subscript bo for background-only) by using only the off-source counts
$\noff$ to estimate the mean and variance; while not optimal, it at
least is consistent with the null.  Ref.~\cite{li} also provides
(their Eqn.~9)
\begin{equation}
\vnine = (\non + \noff)/\toffoverton, 
\end{equation}
(subscript BiN for Binomial Normal) which better implements the null
hypothesis.  It is interesting to note that taking a normal
approximation to the binomial test $\zbi$ (that is, comparing the
difference of estimate of binomial parameter from its expected value
$\binparam$ , to the square root of its normal-approximation variance)
yields $(\non/\ntot - \binparam)/\sqrt{\binparam(1-\binparam)/\ntot}$,
which can be shown to be identical to $\znine = s/\sqrt{\vnine}$.

Zhang and Ramsden \cite{zhang} used a variance-stabilizing
transformation to derive an asymptotically normal variable with nearly
constant variance (their Eqn.~23),
\begin{equation}
\zzr = \frac{2}{\sqrt{1 + 1/\toffoverton} }\ 
\left( \sqrt{\non + 3/8} -
\sqrt{(\noff + 3/8)/\toffoverton}\right).
\end{equation}
The $3/8$ speeds convergence to normality from the underlying
discreteness.

One can also calculate a $\zp$ from the Poisson probability $p$-value
in Eqn.~\ref{pp} and substituting $\estb$ for $\mubkgnd$, but such a
$\zp$ ignores the uncertainty in $\estb$.  Occasionally one sees
substitutions of $\mubkgnd \rightarrow \estb + \sigmab$ into
Eqn.~\ref{pp} in an attempt to incorporate the uncertainty in $\estb$.

A different approach, known as the Fraser-Reid method, attempts to
move directly from likelihood to significance by using a 3rd-order
expansion \cite{fraser,fraserreidwu}.  The mathematics is interesting,
combining two first order estimates (which give significance to order
$1/\sqrt{n}$) to yield a $1/\sqrt{n^3}$ result.  Typically, the
first-order estimates are of the form of a normal deviation, $Z_t$
(like $\znine$), and a likelihood ratio like $\zpl$; of these, the
likelihood ratio is usually a better first-order estimate.  The two
are then combined into the third order estimate by a formula such as
\begin{equation}
Z =\zpl + \frac{1}{\zpl} \ln (Z_t/\zpl).
\end{equation}
Generically, $Z_t = \Delta/\sqrt{V}$ is a Student t-like variable,
where $\Delta$ is the difference of the maximum likelihood value of
$\theta$ (the parameter of interest) from its value under the null
hypothesis, and $V$ is a variance estimate derived from the Fisher
Information $\partial ^2 L/\partial ^2 \theta$.  The attraction of the
method is to achieve simple formulas with accurate results.  However,
the mathematics becomes more complex \cite{fraserreidwu} when nuisance
parameters are included, as is needed when the background is
imperfectly known.  In the present paper, we do not apply this method.

\section{Comparison of results for some example data}
\label{comparison}
In this section, we illustrate the various methods using several
interesting test cases from the HEP and GRA literature.  The input
values and published $Z$-value results are shown in Table 1 in
boldface at the top of the table; typically in HEP cases, the values
reported in the papers are $\non$, $\estb$, and $\sigmab$, while in
GRA, the reported values are $\non$, $\noff$, and $\toffoverton$. We
also include a few artificial cases for further illustration. We take
$\zbi=\zgamma$ as a reference standard because of its frequentist
foundation.  None of these published $Z$-values differed materially
from $\zbi$.  In the remainder of the table, results from the various
formulas above are given, and as explained in the caption departures
from $\zbi$ highlighted.  More detailed results for $\zbi$, $\zn$, and
$\zpl$ are in Sec.~\ref{performance}.

There are numerical issues to be faced in evaluation of the more
complex methods.  The Binomial is straightforward in its Beta function
representation.  The Bayes $p$-value methods may involve an infinite
sum, and are touchy and slow for large $n$; Ref.~\cite{alexandreas}
suggests approximating the summation by an integral. The Bayes
$p$-value summation results are also sensitive numerically for large
$n$; integer-based ``exact'' calculations become slow (e.g. in
Mathematica), while floating point algorithms may have convergence
difficulties.  An alternative approach is to leave the $p_{\rm P}$ as
a $\Gamma$ function ratio and trade an integration for the infinite
sum. Doing so in the Bayes Gaussian case is less unstable than
summing, but for large $n$ requires hints on the location of the peak
of the integrand.

The method most used in HEP, $\zn$, produces $Z$'s that are always
larger than those from $\zbi=\zgamma$. This is confirmed in the wider
scan of the parameter space described in Sect.~\ref{performance}, and
can be understood by the fact that the gamma pdf, for the same inputs
as the normal, shifts $\mubkgnd$ to higher values and smears it more
broadly than the normal does, resulting in larger tail probabilities
and thus smaller $Z$ values. Viewing the calculation as averaging
the Poisson $p$-value $p_{\rm P}$ over the posterior for $\mubkgnd$
(Eqn.~\ref{eqn:bayesavg}), the shorter tails of the normal compared to
the gamma place less weight on the larger probabilities (smaller
$p$-values) obtained when the off-source measurement happens to
underestimate the true value of $\estb$. The difference is most
striking for small values of $\toffoverton$, that is, when the
background estimate is performed with less sensitivity than the signal
estimate; in this case, results in $Z$ differing by over 0.5 units can
occur.

The most common method in GRA, $\zpl$, also is always larger than
$\zbi$ in a wide scan of parameter space, but seems less vulnerable to
problems at small $\toffoverton$.  As further evident in
Sec.~\ref{performance}, the relative size of $\zn$ and $\zpl$ varies
with the input parameters.

The variance stabilization method $\zzr$ presented in Ref.~\cite{zhang}
does not appear to be in general use in GRA, but produces results of
similar quality to $\zbi$ and $\zpl$.  These methods agree for
$N>500$, where the normal approximations are good, even out to 3-6
$\sigma$ tails.

The ``not recommended'' methods all produce results off by more than
0.5 for several low-statistics cases. $\znine$, which approximates
$\zbi$, does best; $\zfive$ is indeed biased against real signals
compared to other measures, and its alternative $\zfivep$, while
curing that problem, overestimates significance as the price for its
less efficient use of information compared to $\znine$.

As expected, ignoring the uncertainty in the background estimate leads
to overestimates of the significance.  $\zsb = s/\sqrt{\estb}$ is much
more over-optimistic than an exact Poisson calculation of
Eqn.~\ref{pp}. The implicit Gaussian approximation underestimates the
Poisson tail at large $\non$; there is in addition a smaller bias
towards $\zsb > \zp$ from ignoring the discreteness of the Poisson
sum.  Any method ignoring background uncertainty overestimates
significance, particularly for small $\ntot$, or $\toffoverton < 1$,
where the background uncertainty is most important.  For $s>0$, one
can show that $\zsb > \zssb > \znn$ and that $\zsb > \zbo > \znn$.
(The best that can be said for $\zsb$ is that it is mostly monotonic
in the true significance, so that when used for a speedy optimization
of selection criteria with $\non$ varying by an order of magnitude at
most, it is not too misleading).

One can also show that $\zbo > \zbin$; that $\znn > \zbin$ for
$\toffoverton<1$, i.e., poorly determined background; that
$\zbo>\zssb$ for $\toffoverton>\mubkgnd/s$, i.e., for well-determined
backgrounds; and that $\zzr<\zbo$ unless $\toffoverton$ is very small.
Thus most of the non-recommended methods over-estimate $Z$, except for
$\znn$ and $\zssb$, which are too low for moderate $\toffoverton$, and
too high for small $\toffoverton$.  In general, small $\toffoverton$
(poorly measured backgrounds) gives many methods problems; results are
generally more stable for an adequate control region.

Of the ad-hoc corrections for signal uncertainty, none are reliable;
the ``corrected'' Poisson calculation is less biased than the
uncorrected, but still widely overestimates significance for
$\toffoverton < 1$. The attempt to include background uncertainty with
$s/\sqrt{\estb+\sigmab}$ isn't much better than its ``un-corrected''
version.

To summarize our provisional conclusions from these examples, most bad
approximations overestimate significance (the only exceptions are
$\zfive$ for $\toffoverton > 1$, $\zssb$, and Poisson with $\estb
\rightarrow \estb + \sigmab$). Thus, prudence demands using a formula
with well-understood properties, in order to not overstate the true
significance.  In the next sections, we study the most promising of
these methods in detail.

\begin{table*}[t]
\hspace{-2.5cm}
\begin{tabular}{|l|cccccccccc|} \hline
\hfill Reference: & \cite{Alexe} & \cite{cdftop} & \cite{d0topsearch}
 & \cite{d0topobs} & \cite{Zhangex} & \cite{Zhangex} & \cite{alarge} 
 & \cite{Hegra}  & 
\cite{Whipple} & \cite{Milagro} \\ \hline

$\non$  & {\bf 4} & {\bf 6} & {\bf 9} & {\bf 17} & {\bf 50} & {\bf 67}
 & {\bf 200} & {\bf 523} &   {\bf 498426}  & {\bf 2119449}  \\
$\noff$ & {\bf 5} & 18.78   & 17.83   & 40.11    & {\bf 55} & {\bf 15}
 & {\bf 10} & {\bf 2327}  &  {\bf 493434}  & 
23650096 \\
$\toffoverton$  & {\bf 5.0}  &  14.44  &  4.69   &  10.56  
  &  {\bf 2.0}  &  {\bf 0.5}  &  {\bf 0.1}  &  {\bf 5.99}  & {\bf 1.0}
  &  {\bf 11.21} \\
$\estb$ & 1.0 & {\bf 1.3} & {\bf 3.8} & {\bf 3.8}
 & 27.5 & 30.0 & 100.0 & 388.6 & 493434 & {\bf 2109732}  \\
$s = \non - \estb$ & 3.0 & 4.7 & 5.2 & 13.2 & 22.5 & 37 & 100 
 & {\bf 134} & {\bf 4992} & {\bf 9717} \\
$\sigmab$ & 0.447 & {\bf 0.3} & {\bf 0.9} & {\bf 0.6} & 3.71 & 7.75
 & 31.6 & 8.1  &  702.4  &  433.8 \\
$f= \sigmab/\estb$ & 0.447 & 0.231 & 0.237 & 0.158 & 0.135 & 0.258
 & 0.316 & 0.0207  & 0.00142 & 0.000206 \\
Reported $p$ &  & {\bf 0.003} & {\bf 0.027} & \bf{2E-06} &  &  & 
 &  &  &    \\
Reported $Z$ &  & 2.7 & 1.9 & {\bf 4.6} &     &     &  & {\bf 5.9} 
 & {\bf 5.0} & {\bf 6.4}  \\ \hline
See conclusion:&&&&&&&&&&\\
$\zbi=\zgamma\ \ $Binomial & \bf 1.66 & \bf 2.63 & \bf 1.82 & \bf 4.46
 & \bf 2.93 & \bf 2.89 & \bf 2.20 & \bf 5.93  & \bf 5.01 & \bf 6.40 \\
$\zn$ Bayes Gaussian
 & 1.88 & 2.71 & 1.94 & 4.55 & 3.08 & \it 3.44 & \it 2.90 & \bf 5.93 
 & \bf 5.02 & \bf 6.40 \\
$\zpl$, $\lhoodp$ Profile Lik'hood & 1.95 & 2.82 & 1.99 & 4.57 & 3.02
 & 3.04  & 2.38 & \bf 5.95  & \bf 5.01 & \bf 6.40 \\
$\zpl$, $\lhoodg$ Profile Lik'hood & 2.00 & 2.83 & 2.02 & 4.62 & 3.10 
 & \it 3.45 & \it 2.90 & \bf 5.96  & \bf 5.02 & \bf 6.40 \\
$\zzr$ variance stabilization
 & 1.93 & 2.66 & 1.98 & 4.22 & 3.00 & 3.07 & 2.39 & 5.86  & \bf 5.01
 & \bf 6.40 \\
\hline
Not Recommended: &  & &&&&&&&&\\
$\znine=s/\sqrt{\ntot/\toffoverton} $ & \it 2.24 & \it 3.59 & 2.17
 & \it 5.67 & 3.11 & \bf 2.89 & \bf 2.18 & 6.16  & \bf 5.01
 & \bf 6.41 \\
$\zfive = s/\sqrt{\non + \noff/\toffoverton^2}$ & 1.46 & \it 1.90
 & 1.66 & \it 3.17 & 2.82 & 3.28 & \it 2.89 & \it 5.54  & \bf 5.01
 & \bf 6.40 \\
$\zssb = s/\sqrt{\estb + s} $ &      1.50& {\it 1.92}  &  1.73
 & {\it 3.20}  & 3.18 & {\it 4.52}   &  {\it  7.07}  & 5.88 
 &  {\it  7.07}  &  6.67    \\
$\zfivep = s/\sqrt{ \noff(1+\toffoverton)/\toffoverton^2}$
 & \it 2.74 & \it 3.99 & \it 2.42 & \it 6.47 & \it 3.50 & \it 3.90
 & \it 3.02 & 6.31 & \bf 5.03 & \bf 6.41 \\
\hline
Ignore $\sigmab$:&&&&&&&&&&\\
 $\zp\ \ $Poisson: ignore
$\sigmab$ & 2.08 & 2.84 & 2.14 & 4.87 & \it 3.80 & \it 5.76
 & \it 8.76 & \it 6.44  & \it 7.09 & 6.69 \\
$\zsb\ \  =s/\sqrt{\estb}$ & \it 3.00 & \it 4.12 & \it 2.67 & \it 6.77
 & \it 4.29 & \it 6.76 & \it 10.00 & \it 6.82  & \it 7.11 & 6.69 \\
 \hline
 Unsuccessful ad hockery:&&&&&&&&&&\\

Poisson: $\mubkgnd \rightarrow  \estb +
\sigmab$ & 1.56 & 2.51 & 1.64 & \bf 4.47 & 3.04 & \it 4.24 & \it 5.51
 & 6.01  & \it 6.09 & \bf 6.39 \\
s / $\sqrt{\estb + \sigmab}$
 & \it 2.49 & \it 3.72 & \it 2.40 & \it 6.29 & \it 4.03 & \it 6.02
 & \it 8.72 & \it 6.75  & \it 7.10 & 6.69 \\

\hline
\end{tabular}
\label{results} 
\caption{Test Cases and Significance Results.  In the top section, the
primary input numbers from the papers are in boldface, with derived
numbers (using Eqns.~\ref{toffoverton}-\ref{eqn:corr}) in normal font.
$\zpl$ is shown for both $\lhoodp$ and $\lhoodg$ 
regardless of the primary input numbers.
The test cases are ordered in data counts; \protect\cite{Zhangex},
\protect\cite{alarge}, and \protect\cite{Whipple} have small values of
$\toffoverton$, troublesome for some methods.  Below the top section,
$Z$-values in boldface are nearly equal to the reference $\zbi$, while
$Z$-values in italics differ by more than 0.5.  }
\end{table*}

\section{Application of three recipes to the two problems}
\label{applyingrecipes}
For detailed coverage studies, we examine the three recipes for
$Z$-values in Secs.~\ref{binom}-\ref{profilelikelihood}:
\begin{itemize}
\item $\zbi$ ($=\zgamma$) takes as input $\non$, $\noff$, and
$\toffoverton$.
\item $\zn$ takes as input $\non$, $\estb$, and $\sigmab$.
\item $\zpl$ takes either set of inputs, as appropriate for computing
either $\lhoodp$ or $\lhoodg$ for the problem at hand.
\end{itemize}
It is interesting to explore the performance of each of the first two
recipes not only for the problem for which it was designed, but also
(by using the ``rough correspondences'' of Eqns.~\ref{bnofftau}
through \ref{eqn:corr}) for the other problem.  (One can also imagine
studying the performance of $\zpl$ using the wrong likelihood function
for the problem at hand, e.g. using $\lhoodp$ for the Gaussian-mean
background problem or vice versa; however, we do not pursue those
combinations of methods and problems here.)  Since there are two cases
of the Gaussian-mean background problem, each recipe is then applied
in three situations:
\begin{enumerate}
\item On/off problem: One has $\non$, $\noff$, and $\toffoverton$, so
$\zbi$ and $\zpl$ are computed immediately.  To compute $\zn$, the
inputs are $\non$; $\estb$ from Eqn.~\ref{bnofftau}; and $\sigmab$
from Eqn.~\ref{sigmab_corr}.
\item Gaussian-mean background problem with exactly known $\sigmab$:
One has $\non$, $\estb$, and $\sigmab$, so $\zn$ and $\zpl$ are
computed immediately.  For the remaining inputs required for $\zbi$,
$\toffoverton$ is obtained from Eqn.~\ref{eqn:corr}, and then $\noff$
is obtained from Eqn.~\ref{bnofftau}.
\item Gaussian-mean background problem with exactly known relative
uncertainty $f$: One has $\non$, $\estb$, and $f$, from which
$\sigmab$ is estimated by $f\estb$, and then $\zn$ and $\zpl$ are
computed.  One can then also proceed to compute $\zbi$ as in the
previous case.
\end{enumerate}

We emphasize again that only $\zbi$ applied to the on/off problem is
guaranteed not to undercover based on the formal theory of statistics.
The recipe for $\zn$ mixes frequentist and Bayesian statistics even
for the Gaussian-mean background problem, and when applied to the
on/off problem further approximates the Poisson background as
Gaussian.  Applying $\zbi$ to the Gaussian-mean background problem
does the reverse, by approximating the Gaussian background as
Poisson. As noted, $\zpl$ is an approximation based on an asymptotic
theorem.

\section{Frequentist evaluation of the performance of the various 
recipes}
\label{performance}

In the frequentist evaluation of $p$-values, one considers particular
true values of the background mean $\mubkgnd$ in the signal region and
of another parameter characterizing the experimental setup, namely
$\toffoverton$ for on/off experiments or $f =\sigmab/\mubkgnd$ for the
Gaussian-mean background experiments.  For each fixed pair of such
parameters and each recipe, an ensemble of experimental measurements
is considered appropriate to the relevant problem described above.
For each set of measurements corresponding to an experiment, one
proceeds as follows.  In evaluating the performance of $\zbi$, $\zn$,
and $\zpl$, $Z$ is computed according to a recipe and compared to a
value $\zclaim$ (e.g., $\zclaim=5$).  In the ensemble of experiments,
one calculates the fraction of those experiments which obtain
$Z\ge\zclaim$ according to the recipe; this is the true Type I error
rate for that recipe and a significance level corresponding to that
value of $\zclaim$.  One can then substitute this true Type I error
rate for $p$ in Eqn.~\ref{eqn:z} in order to obtain the $Z$-value that
we call $\ztrue$.

We note that each recipe implicitly chooses its own ordering of the
points in the $(\non,\noff)$ space (or equivalently in the
$(\non,\estb)$ space): contours of equal $Z$ in each space will be
different for each recipe.  If two recipes both faithfully provide the
significance levels, then as Neyman and Pearson pointed out, to
distinguish between them one must compare their power for rejecting
relevant alternative hypotheses.  As noted in the Introduction, in
this paper we do not pursue such considerations of power.

A recipe is ``conservative'' and we say that it ``overcovers''
(borrowing language from confidence intervals) with respect to a
particular problem and a particular $\zclaim$ if the true ensemble
Type I error rate is smaller than implied (so that $\ztrue>\zclaim$).
We say that it ``undercovers'' if the Type I error rate is higher (so
that $\ztrue<\zclaim$).  While neither departure from the correct Type
I error rate is desirable, undercoverage is generally considered to be
more of a flaw than overcoverage.  Of the combinations of problems and
recipes under consideration here, {\it only the application of $\zbi$
to the Poisson on/off problem is guaranteed by construction to have
$\ztrue\ge\zclaim$}, i.e., not to have undercoverage.

For purposes of illustration, we have selected three values of
$\zclaim$ (1.28, 3, and 5), corresponding via Eqn.~\ref{eqn:z} to
$p$-values of 0.1, $1.35 \times 10^{-3}$,and $2.87 \times 10^{-7}$,
respectively.  In order to calculate the Type I error rate, one needs
the probability of obtaining $Z\ge\zclaim$.  Although we compute this
probability directly, we mention the alternate method of Monte Carlo
simulation, which we use as a crosscheck for our results.  For
example, for the on/off problem, given $\mubkgnd$, $\toffoverton$, and
$\zclaim$, one samples $\non$ and $\noff$ from the appropriate
distributions and counts the number of times the recipe yields a value
of $Z>\zclaim$.  While this method remains useful as a cross-check,
for more efficient evaluation of $\ztrue$, we calculate discrete
probabilities directly from the Poisson formula and sum them, and
evaluate tail integrals of normal probabilities using the error
function erf, using a binary search to find how much of the tail
yields results with $Z\ge\zclaim$.  Details for each case are
described in Appendix~\ref{app:ztrue}.

\begin{figure}[htbp]
\centering
\includegraphics*[width=2.7in]{\epsdir/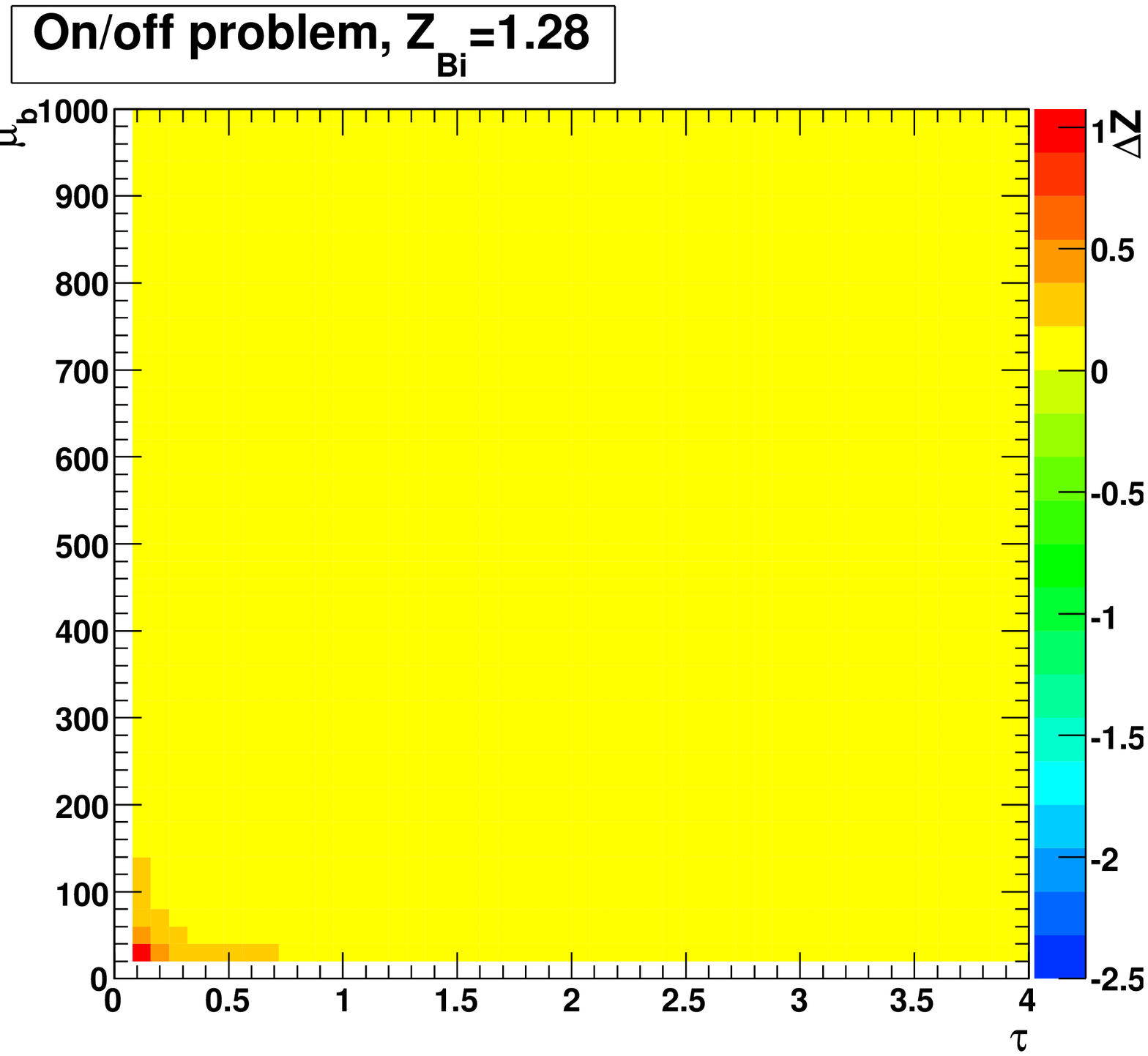}
\includegraphics*[width=2.7in]{\epsdir/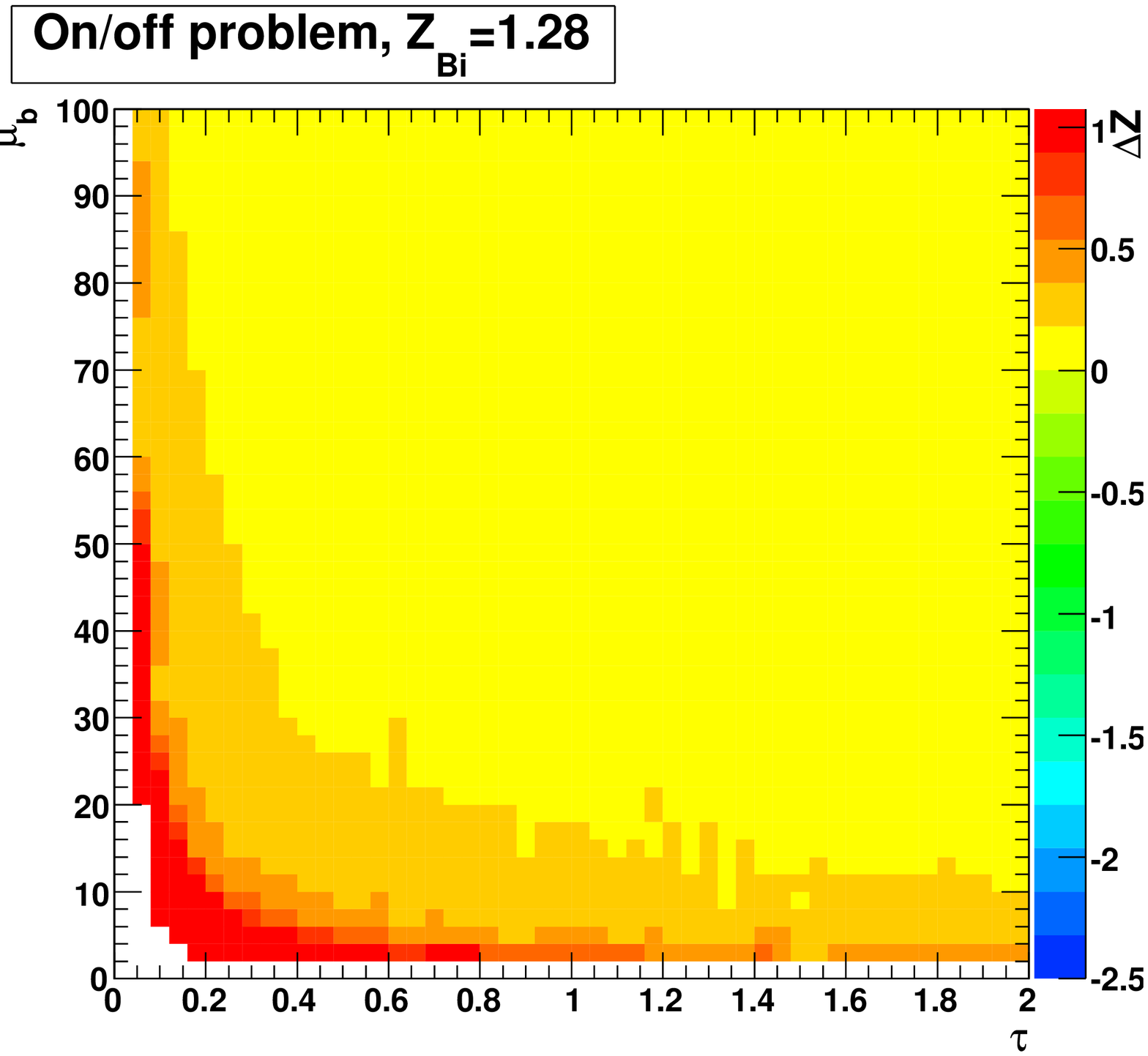}
\caption{ For the on/off problem analyzed using the $\zbi$ recipe, for
each fixed value of $\toffoverton$ and $\mubkgnd$, the plot indicates
the calculated $\zdiff$ for the ensemble of experiments quoting a
$\zclaim\ge 1.28$, i.e., a $p$-value of 0.1 or smaller. }
\label{zbi_lp_1.28}
\end{figure}

\begin{figure}[htbp]
\centering
\includegraphics*[width=2.7in]{\epsdir/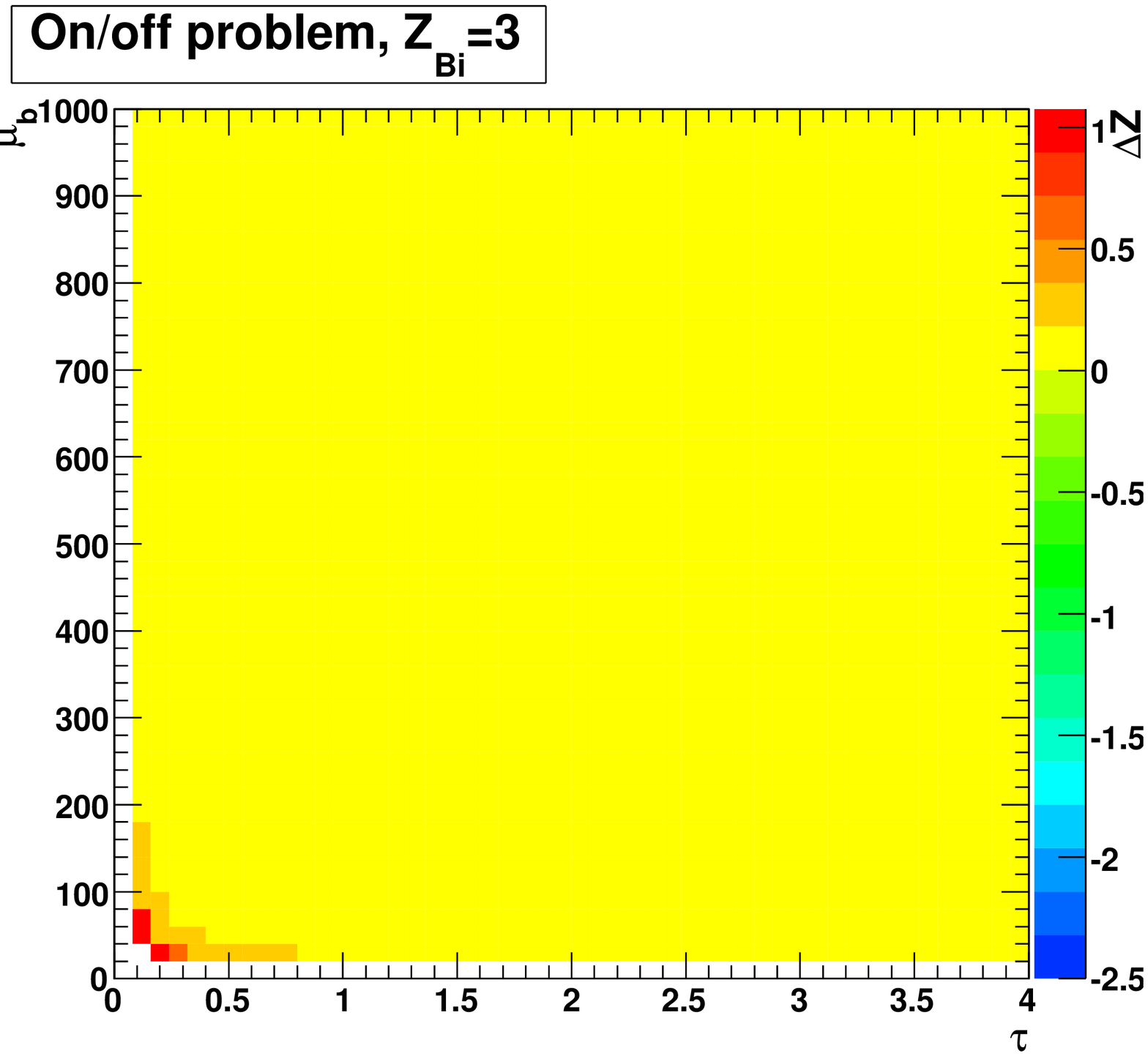}
\includegraphics*[width=2.7in]{\epsdir/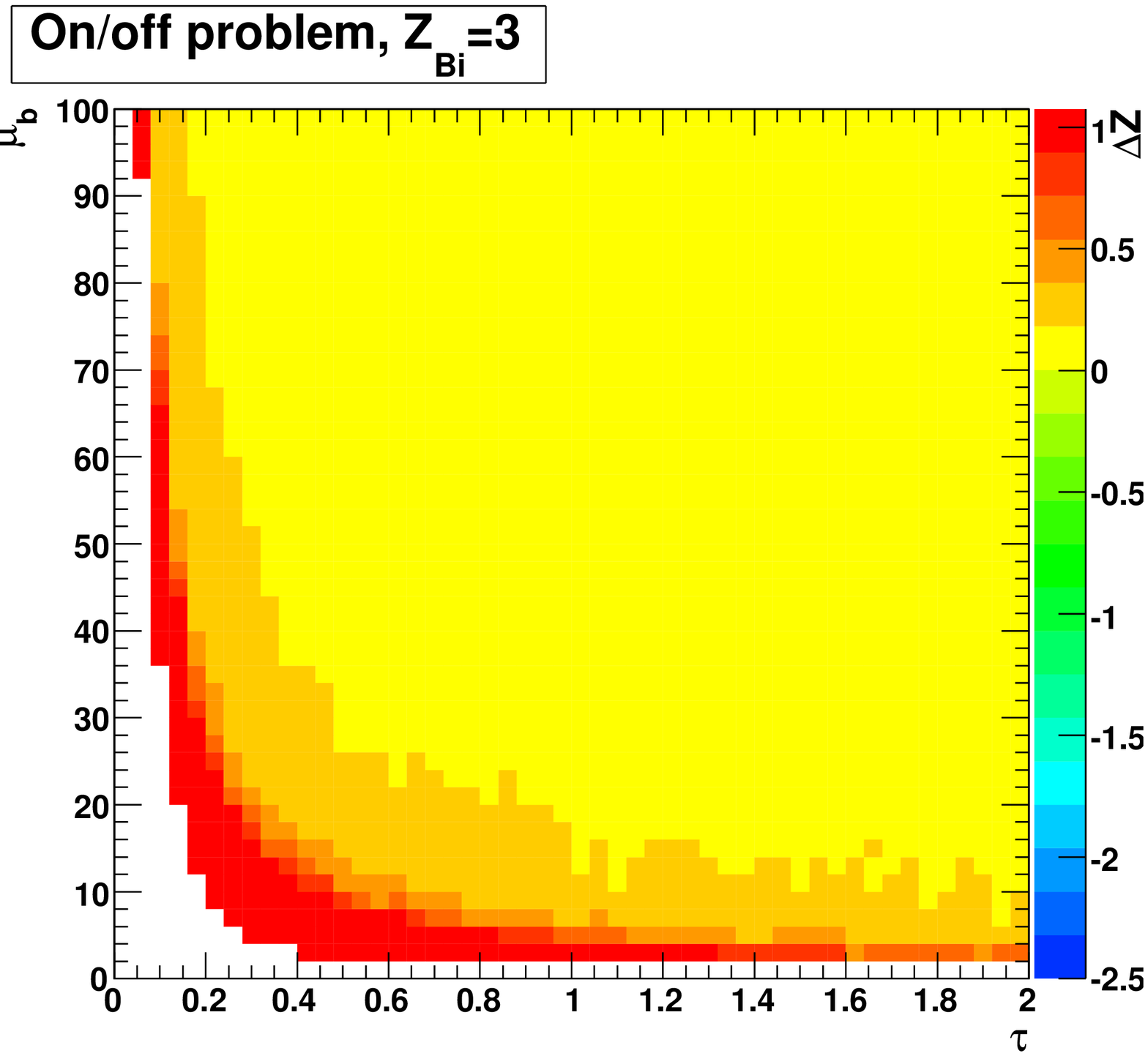}
\caption{ For the on/off problem analyzed using the $\zbi$ recipe, for
each fixed value of $\toffoverton$ and $\mubkgnd$, the plot indicates
the calculated $\zdiff$ for the ensemble of experiments quoting
$\zclaim \ge 3$, i.e., a $p$-value of $1.35 \times 10^{-3}$ or
smaller.  }
\label{zbi_lp_3}
\end{figure}

\begin{figure}[htbp]
\centering
\includegraphics*[width=2.7in]{\epsdir/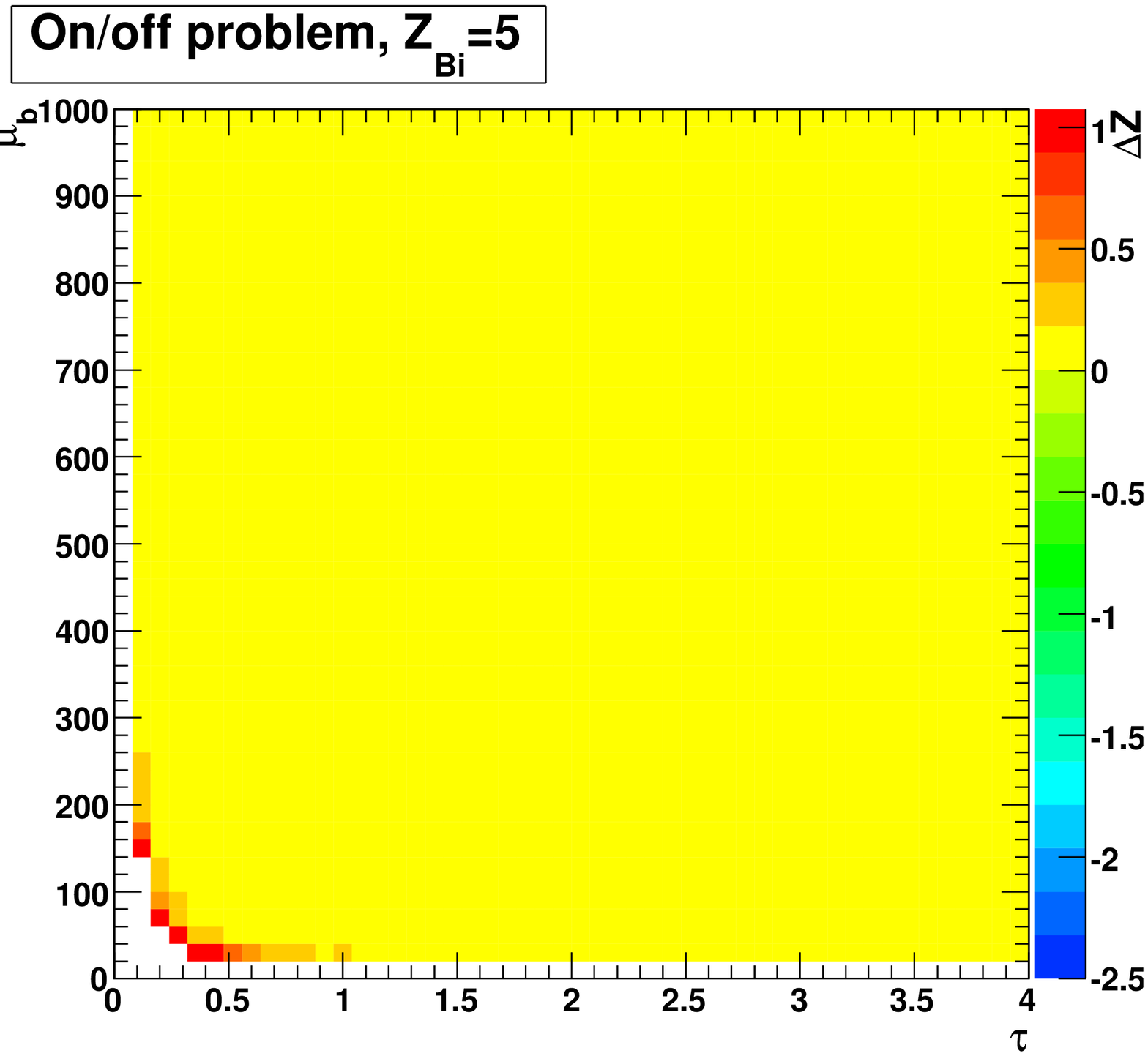}
\includegraphics*[width=2.7in]{\epsdir/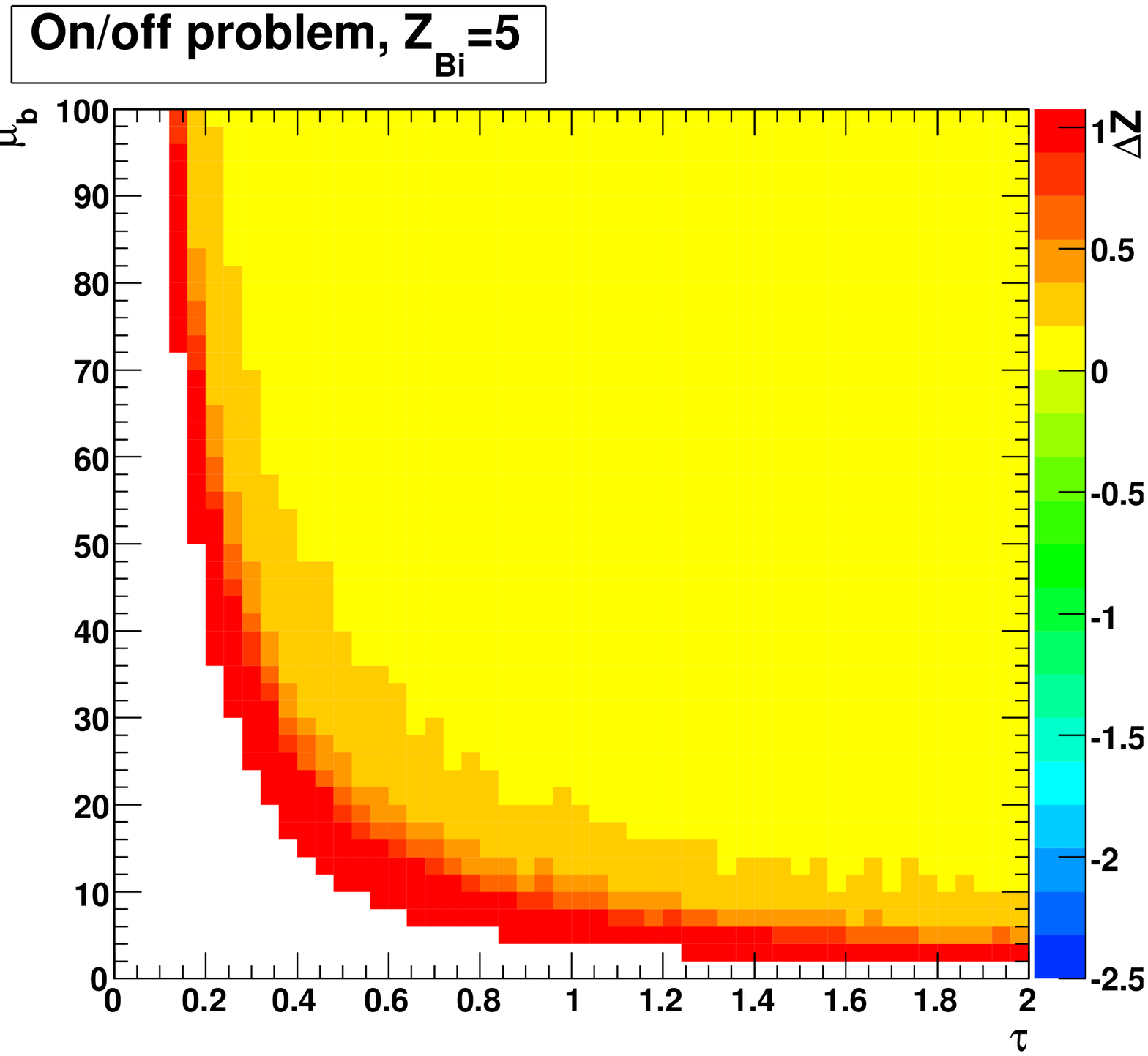}
\caption{ For the on/off problem analyzed using the $\zbi$ recipe, for
each fixed value of $\toffoverton$ and $\mubkgnd$, the plot indicates
the calculated $\zdiff$ for the ensemble of experiments quoting
$\zclaim \ge 5$, i.e., a $p$-value of $2.87 \times 10^{-7}$ or
smaller.  }
\label{zbi_lp_5}
\end{figure}

For the on/off experiments analyzed using the $\zbi$ recipe, the
results are displayed in Figs.~\ref{zbi_lp_1.28} through
\ref{zbi_lp_5}.  Each plot corresponds to a particular value of
$\zclaim$, and for each point $(\toffoverton,\mubkgnd)$ chosen on a
fine grid of 50 by 50 points $\ztrue-\zclaim$ is indicated.  As with
all these figures, the right plot is a zoomed-in version of the left.
The value indicated in each pixel is calculated using the
$(\toffoverton,\mubkgnd)$ of its lower left corner.  As expected from
the construction, $\ztrue\ge\zclaim$ everywhere; the overcoverage is
significant for small values of counts, where the discreteness is most
relevant, as seen in the lower left corner of the zoomed-in version of
each figure.  This overcoverage could be reduced by using the
non-standard intervals for the ratio of Poisson means in
Ref.~\cite{cousinsratio}, but we do not pursue that option in this
paper.

At the limit of numerical precision in our implementation, it turns
out that the result errs in the conservative direction, but of course
extreme caution should be used to avoid quoting a result badly
affected in this way.  The highest calculated value of $\ztrue$ is
nearly 7.6 (corresponding to a $p$-value of $\sim10^{-14}$) due to the
machine limit of our implementation of the calculation of $\ztrue$
from the $p$-value; this can be alleviated by using approximation in
Eqn.~\ref{eqn:zapprox}, but we do not pursue that option in this
paper, and leave blank those regions in the plot where the associated
$p$-value is less than $\sim10^{-14}$.

\begin{figure}[htbp]
\centering
\includegraphics*[width=2.7in]{\epsdir/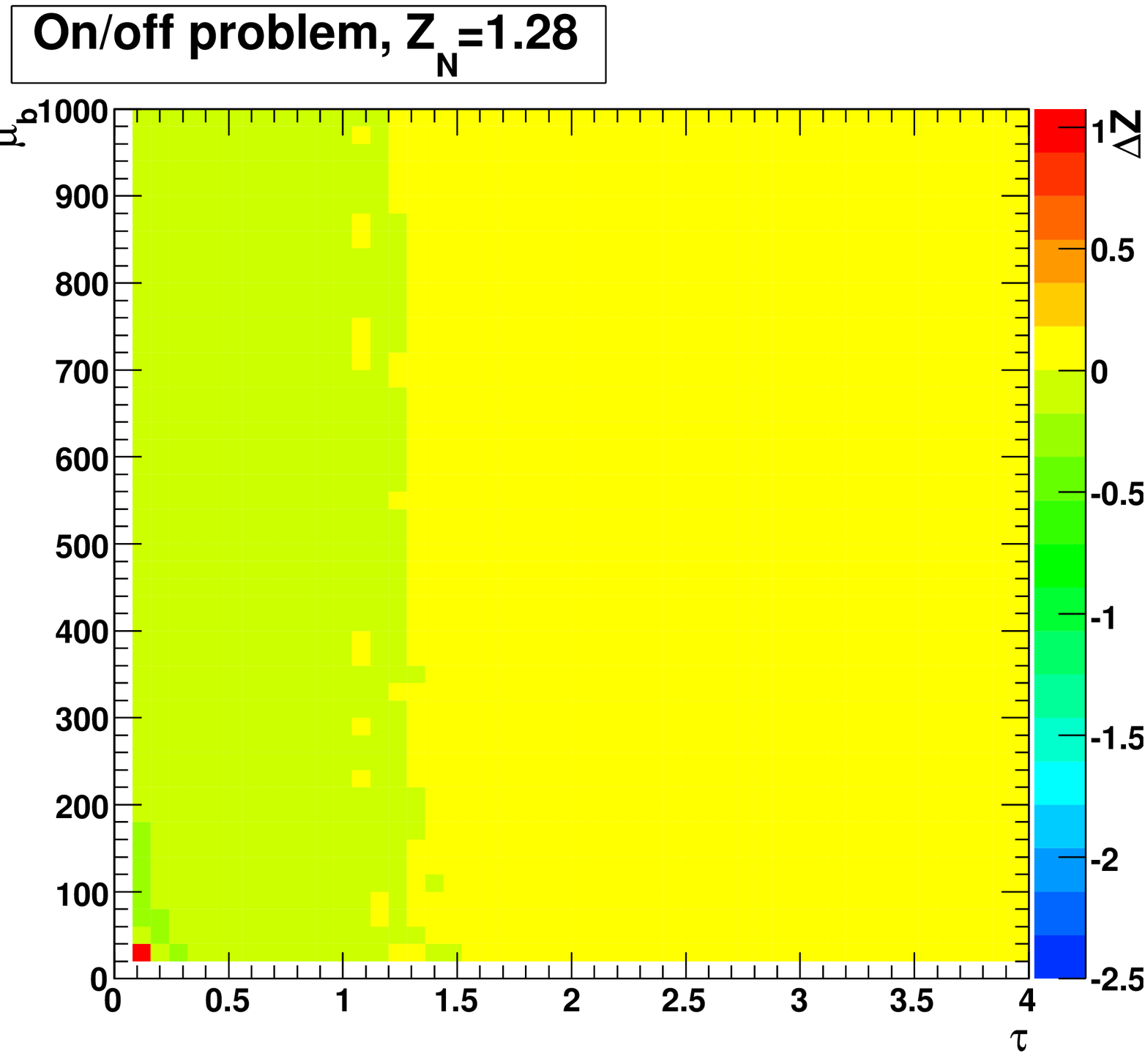}
\includegraphics*[width=2.7in]{\epsdir/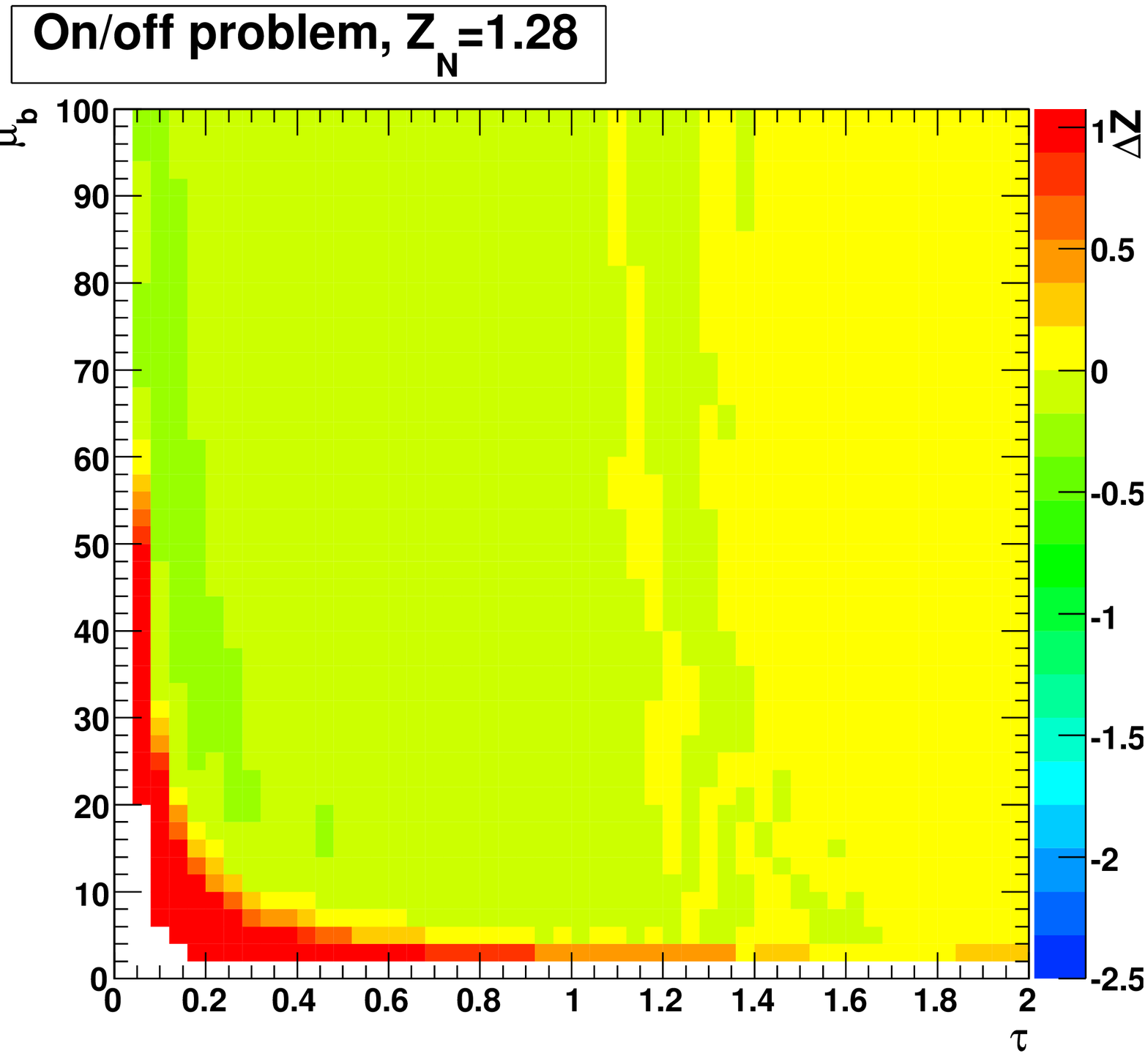}
\caption{ For the on/off problem analyzed using the $\zn$ recipe, for
each fixed value of $\toffoverton$ and $\mubkgnd$, the plot indicates
the calculated $\zdiff$ for the ensemble of experiments quoting
$\zclaim \ge 1.28$, i.e., a $p$-value of $0.1$ or smaller.  }
\label{zn_lp_1.28}
\end{figure}

\begin{figure}[htbp]
\centering
\includegraphics*[width=2.7in]{\epsdir/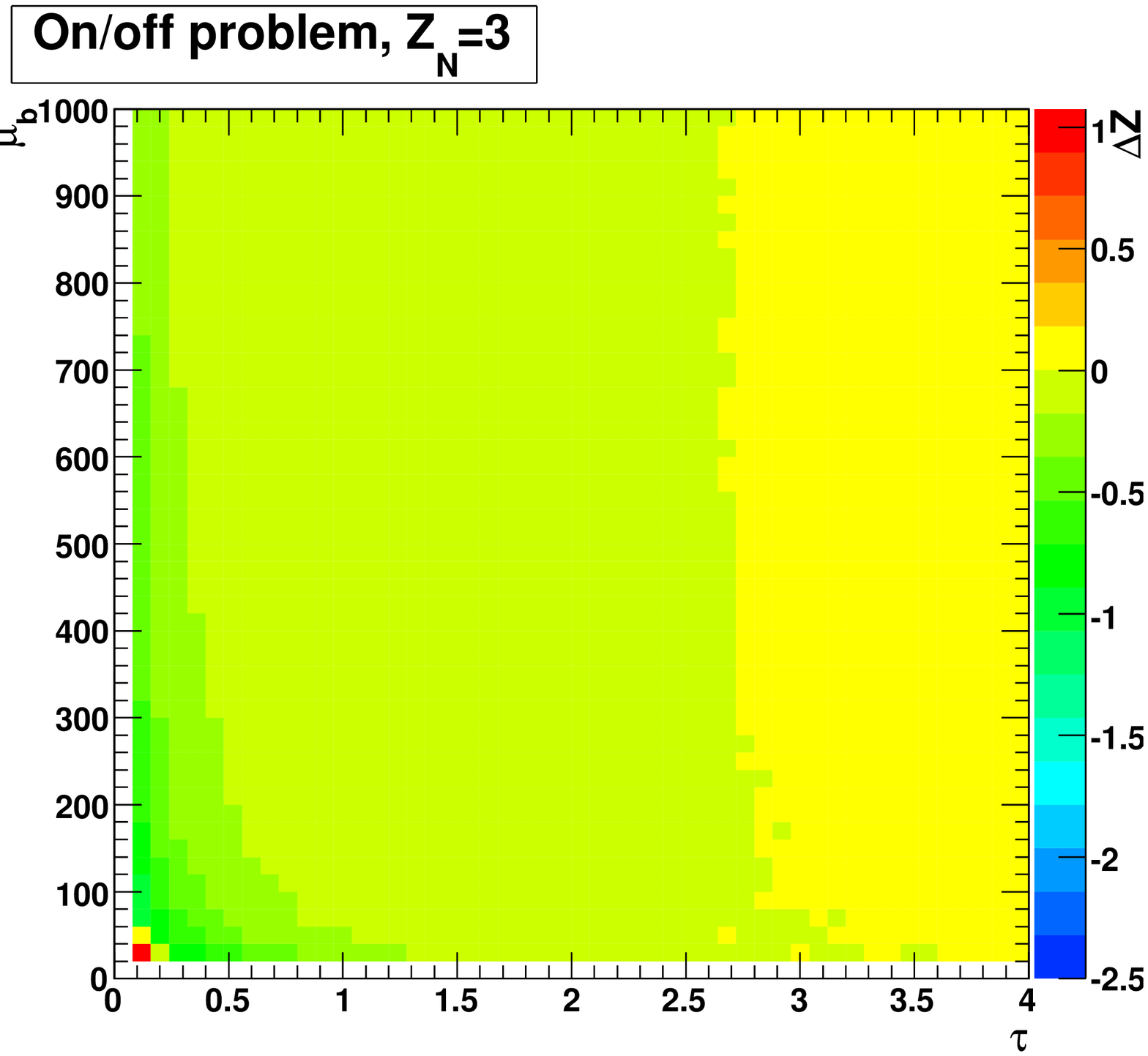}
\includegraphics*[width=2.7in]{\epsdir/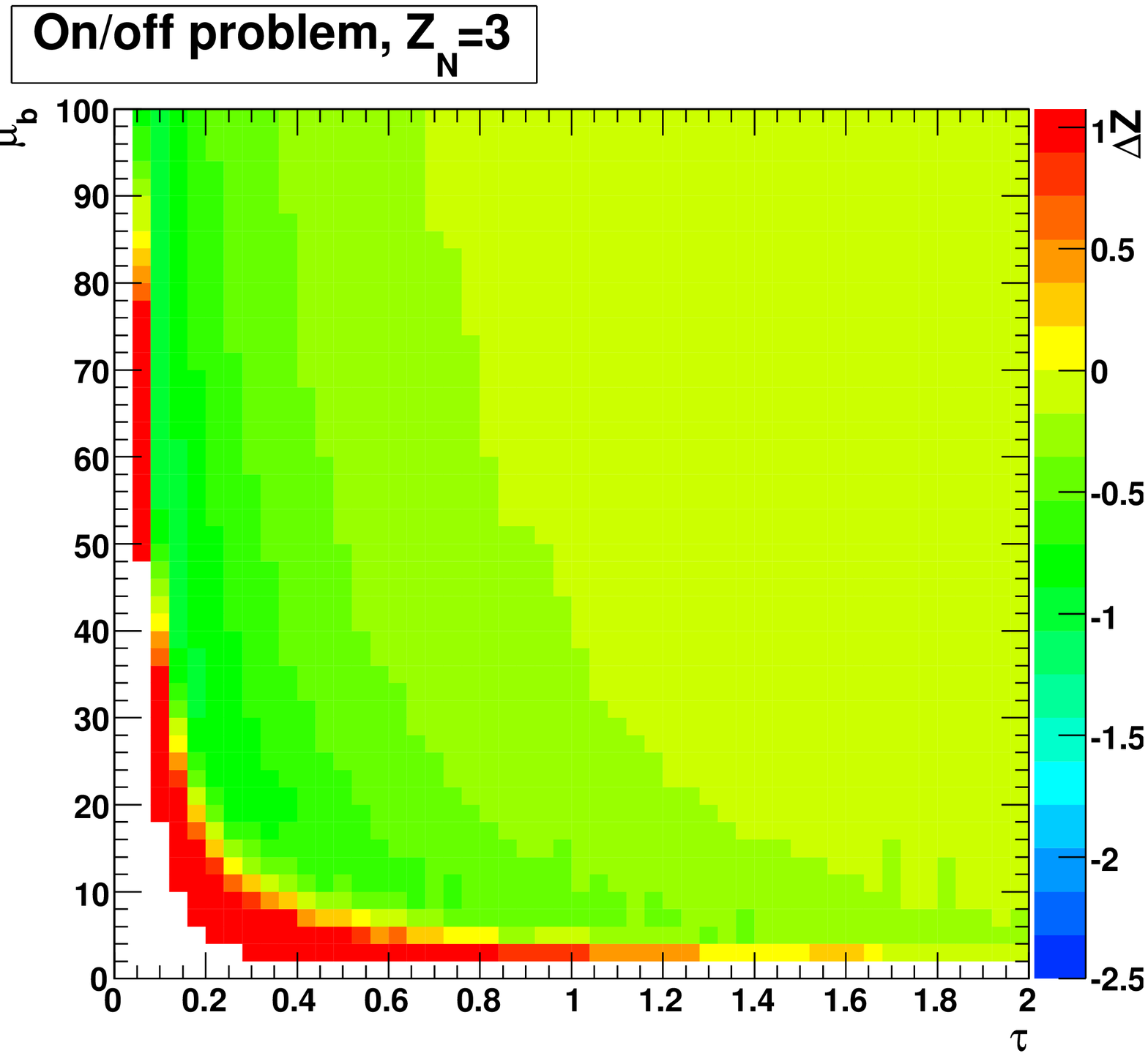}
\caption{ For the on/off problem analyzed using the $\zn$ recipe, for
each fixed value of $\toffoverton$ and $\mubkgnd$, the plot indicates
the calculated $\zdiff$ for the ensemble of experiments quoting
$\zclaim \ge 3$, i.e., a $p$-value of $1.35 \times 10^{-3}$ or
smaller.  }
\label{zn_lp_3}
\end{figure}

\begin{figure}[htbp]
\centering
\includegraphics*[width=2.7in]{\epsdir/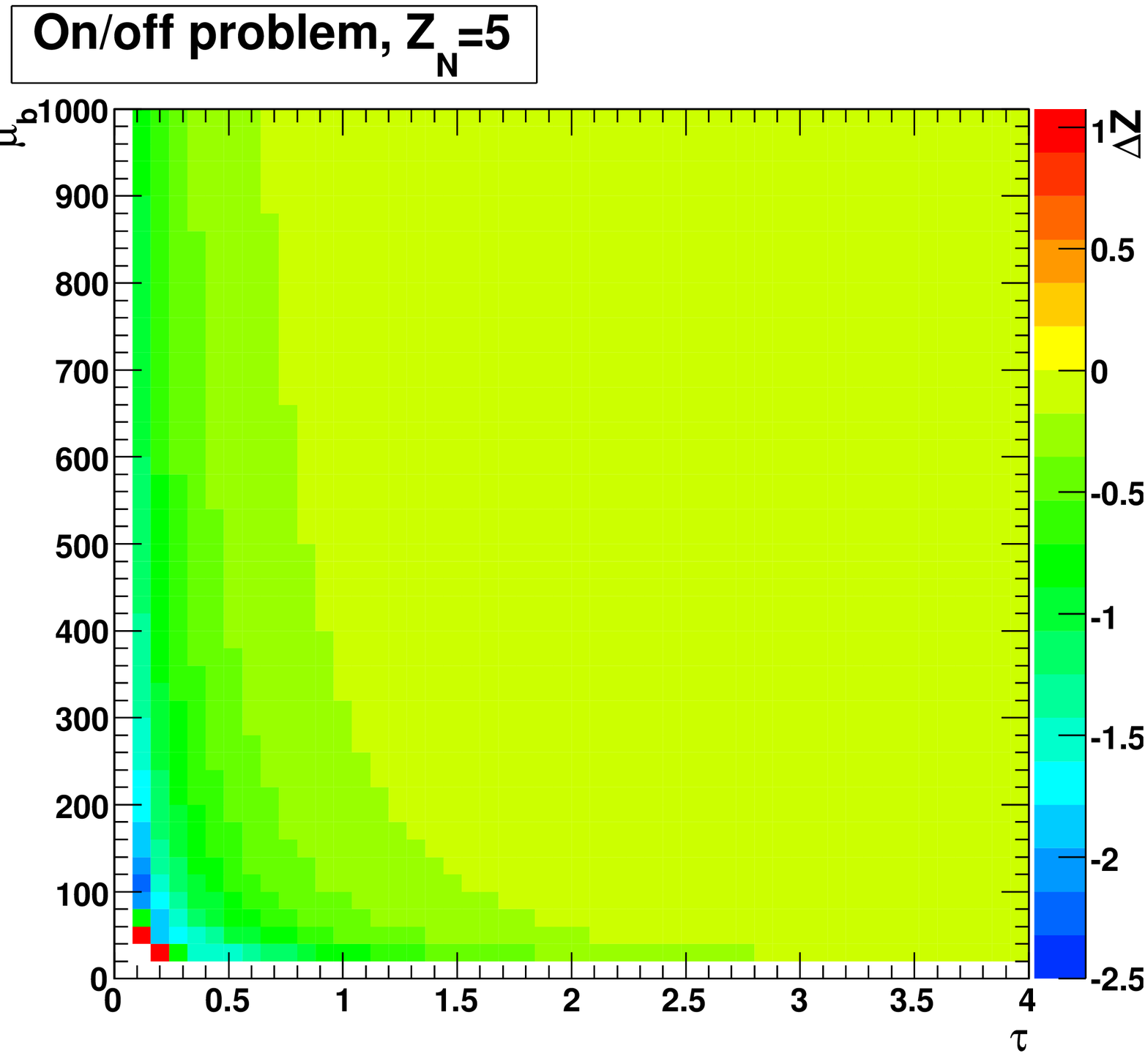}
\includegraphics*[width=2.7in]{\epsdir/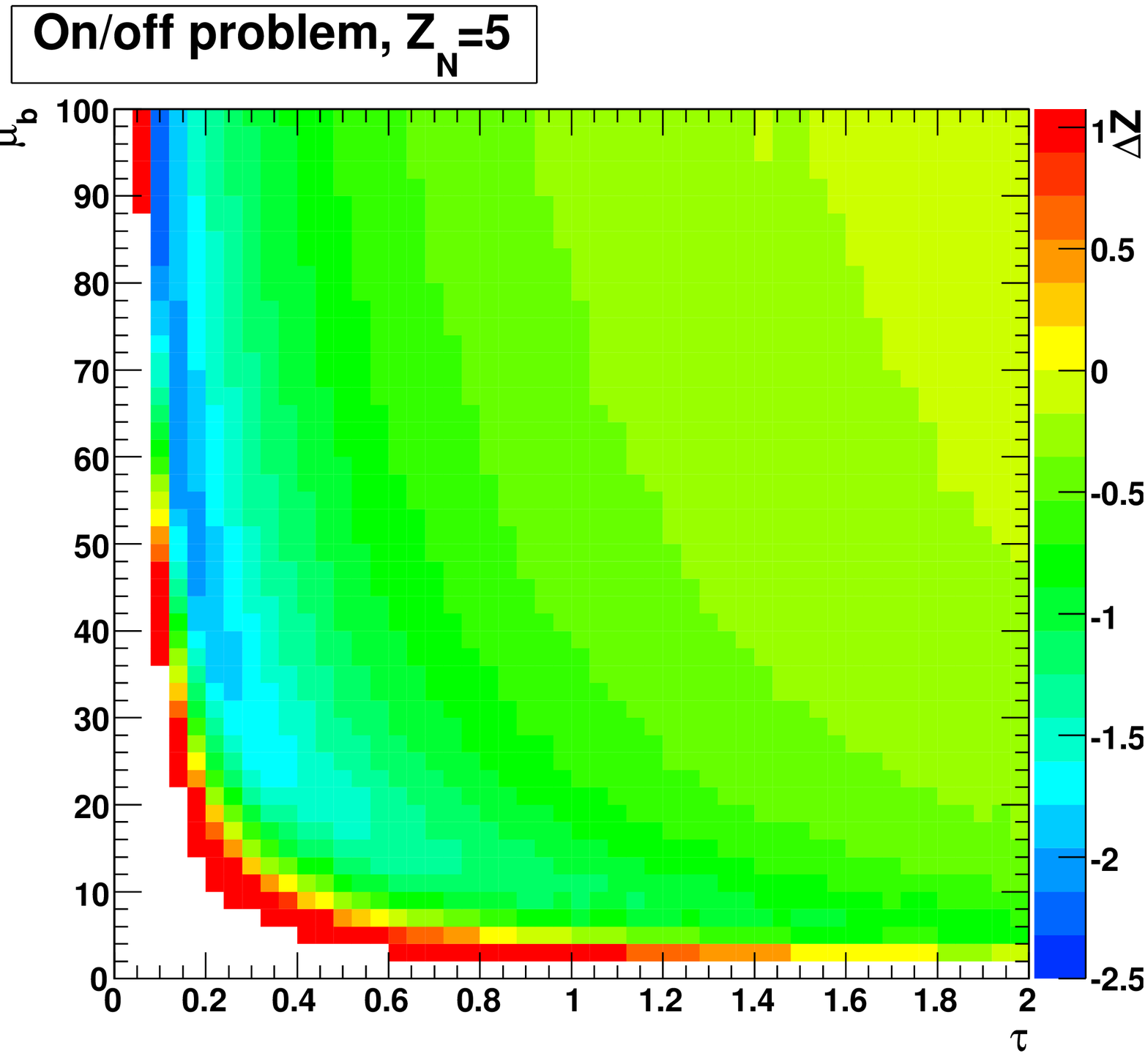}
\caption{ For the on/off problem analyzed using the $\zn$ recipe, for
each fixed value of $\toffoverton$ and $\mubkgnd$, the plot indicates
the calculated $\zdiff$ for the ensemble of experiments quoting
$\zclaim \ge 5$, i.e., a $p$-value of $2.87 \times 10^{-7}$ or
smaller.  }
\label{zn_lp_5}
\end{figure}

When using the $\zn$ recipe to analyze the on/off experiments
(Figs.~\ref{zn_lp_1.28} through \ref{zn_lp_5}), there is a large
region in which the method undercovers (by as much as two units of $Z$
at very low $\toffoverton$) with the extent of the region depending on
$\zclaim$.  This is in accord with Cranmer~\cite{cranmer}, who, using
the Monte Carlo method, finds for a specific case ($\mubkgnd=$100,
$\toffoverton=$1), that the $\zn$ recipe undercovers for $\zclaim=5$,
with a Type I error rate corresponding to $\ztrue = 4.2$.  Again,
there is overcoverage due to discreteness at small values of
$\mubkgnd$ and $\toffoverton$.

\begin{figure}[htbp]
\centering
\includegraphics*[width=2.7in]{\epsdir/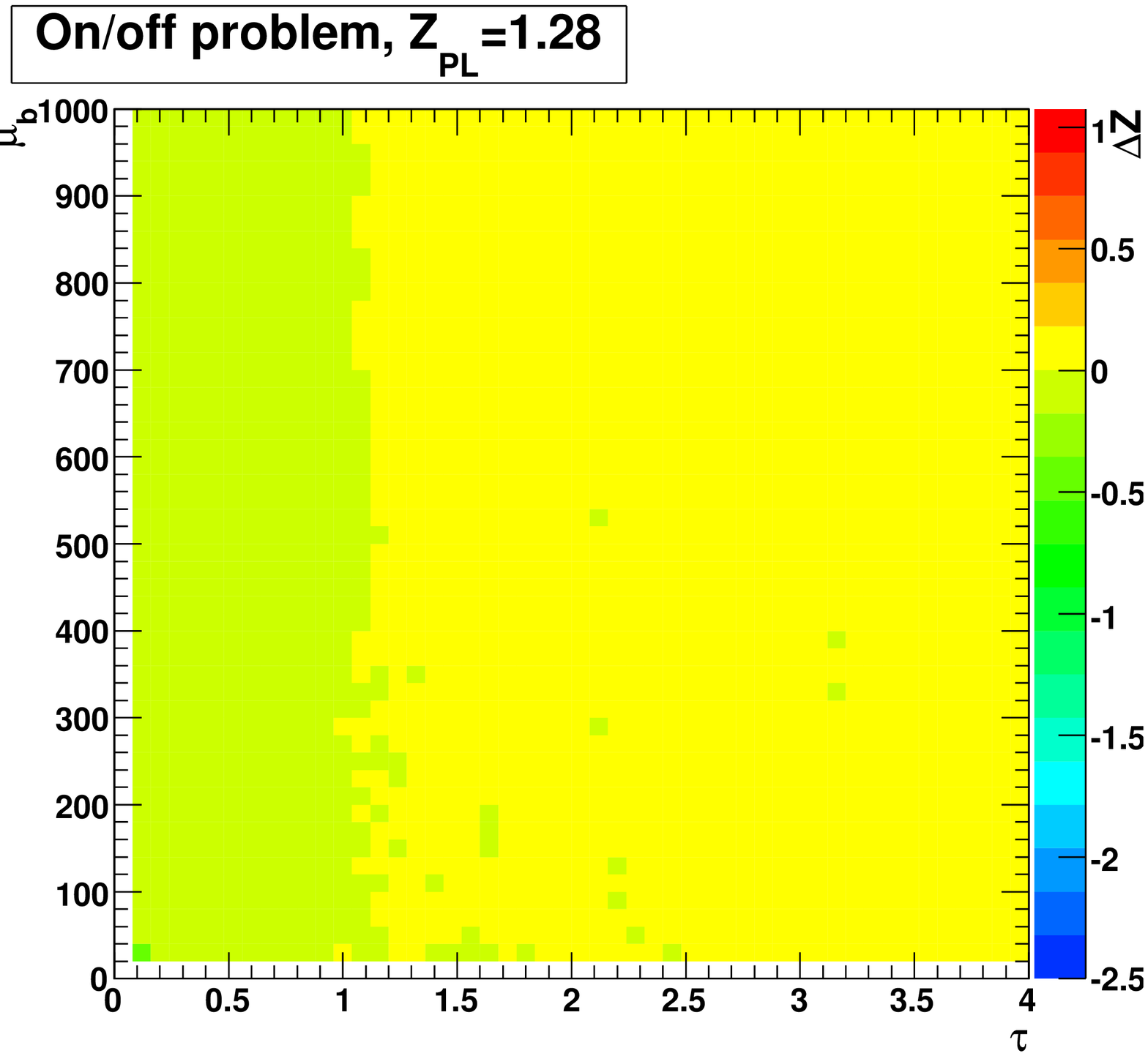}
\includegraphics*[width=2.7in]{\epsdir/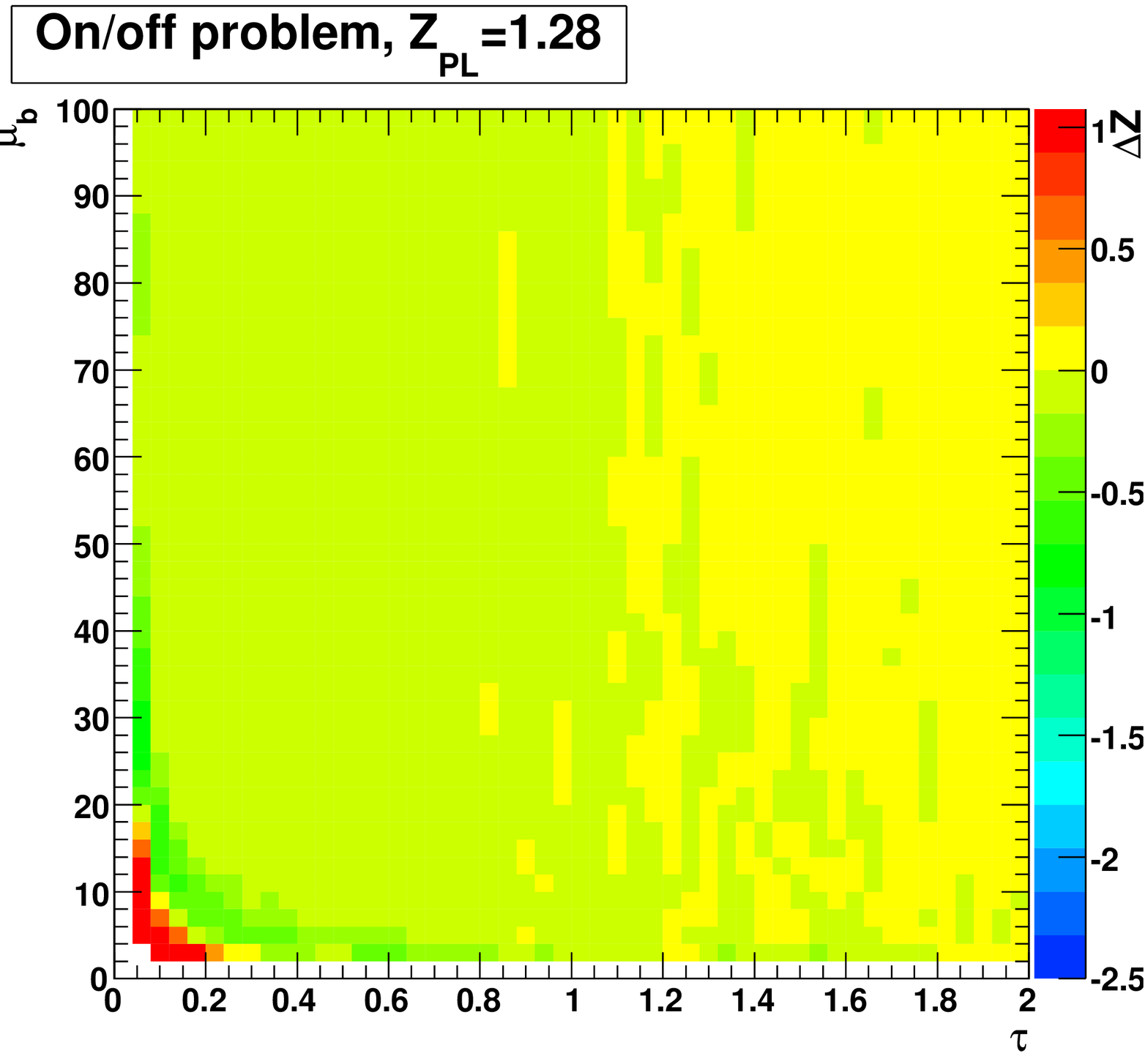}
\caption{ For the on/off problem analyzed using the profile likelihood
method, for each fixed value of $\toffoverton$ and $\mubkgnd$, the
plot indicates the calculated $\zdiff$ for the ensemble of experiments
quoting $\zclaim \ge 1.28$, i.e., a $p$-value of $0.1$ or smaller.  }
\label{proflik_lp_1.28}
\end{figure}

\begin{figure}[htbp]
\centering
\includegraphics*[width=2.7in]{\epsdir/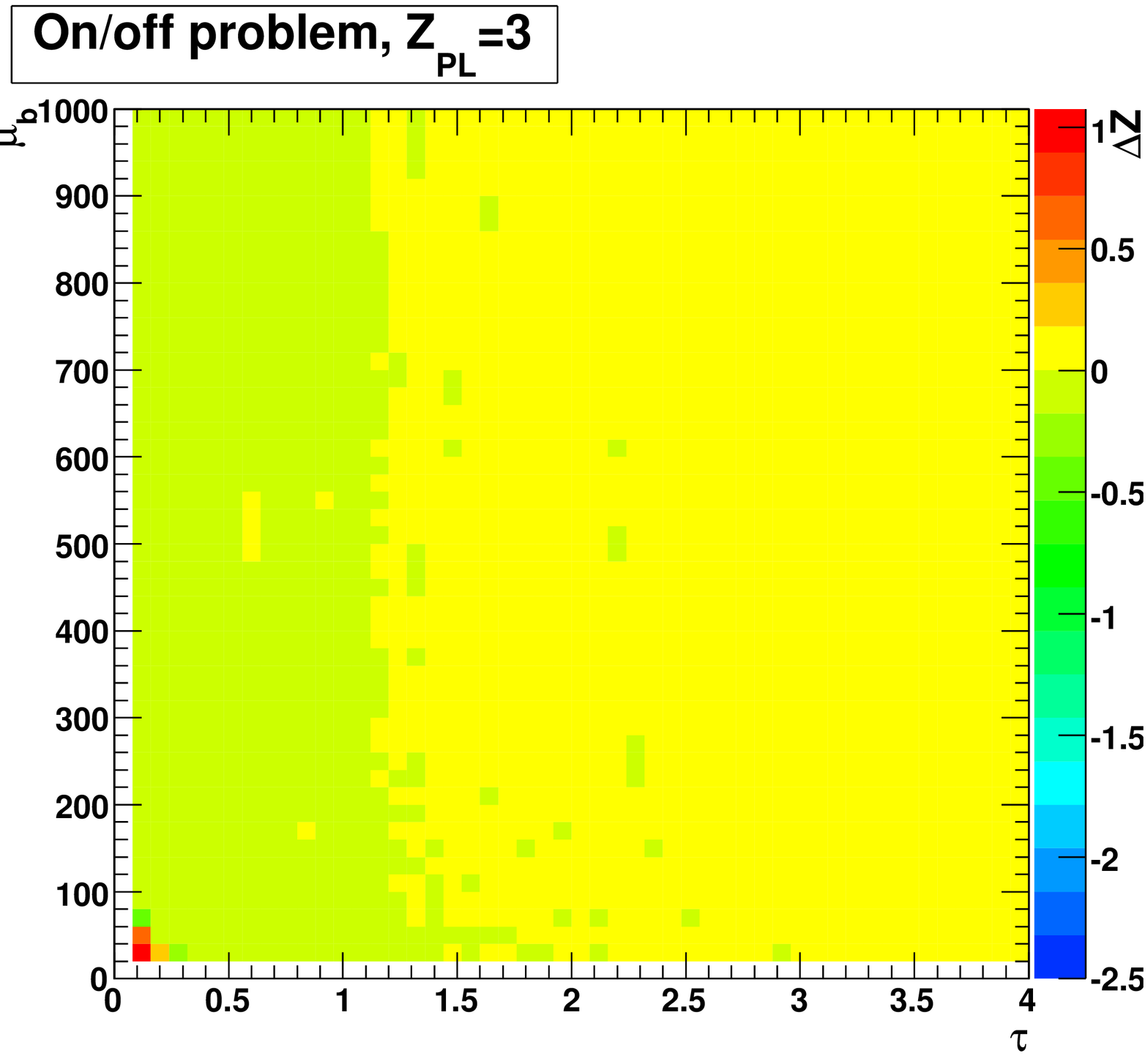}
\includegraphics*[width=2.7in]{\epsdir/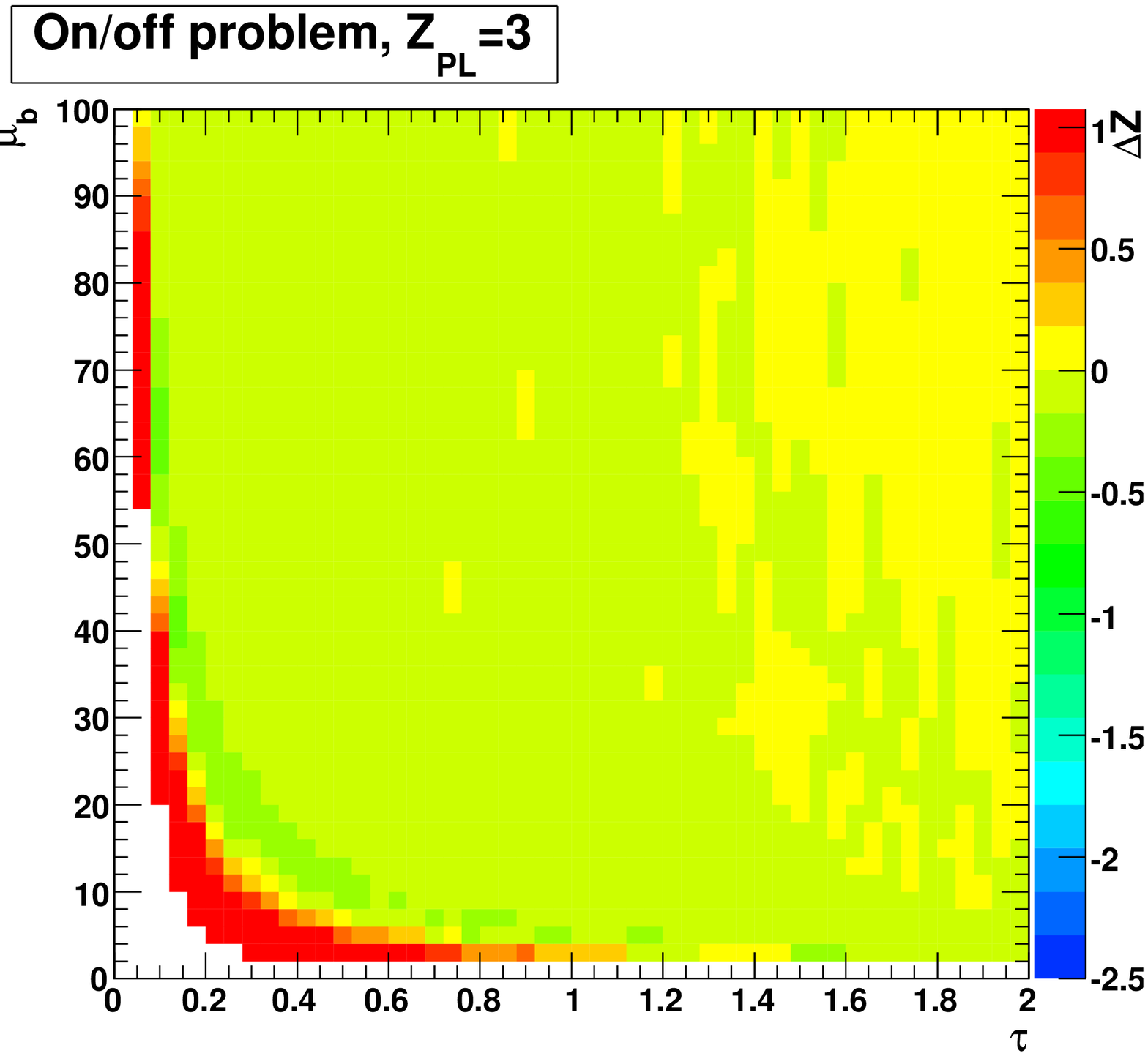}
\caption{ For the on/off problem analyzed using the profile likelihood
method, for each fixed value of $\toffoverton$ and $\mubkgnd$, the
plot indicates the calculated $\zdiff$ for the ensemble of experiments
quoting $\zclaim \ge 3$, i.e., a $p$-value of $1.35 \times 10^{-3}$ or
smaller.  }
\label{proflik_lp_3}
\end{figure}

\begin{figure}[htbp]
\centering
\includegraphics*[width=2.7in]{\epsdir/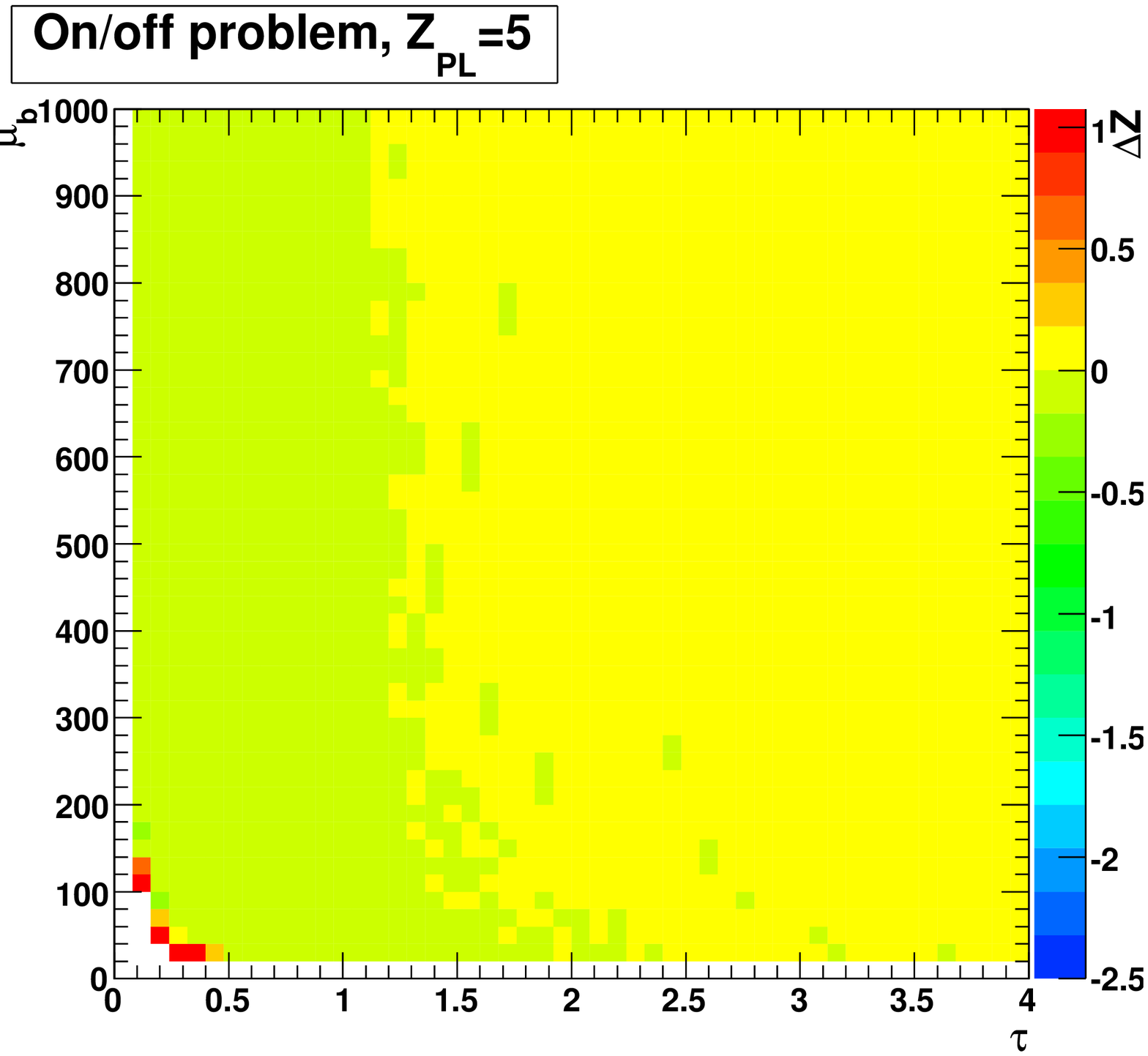}
\includegraphics*[width=2.7in]{\epsdir/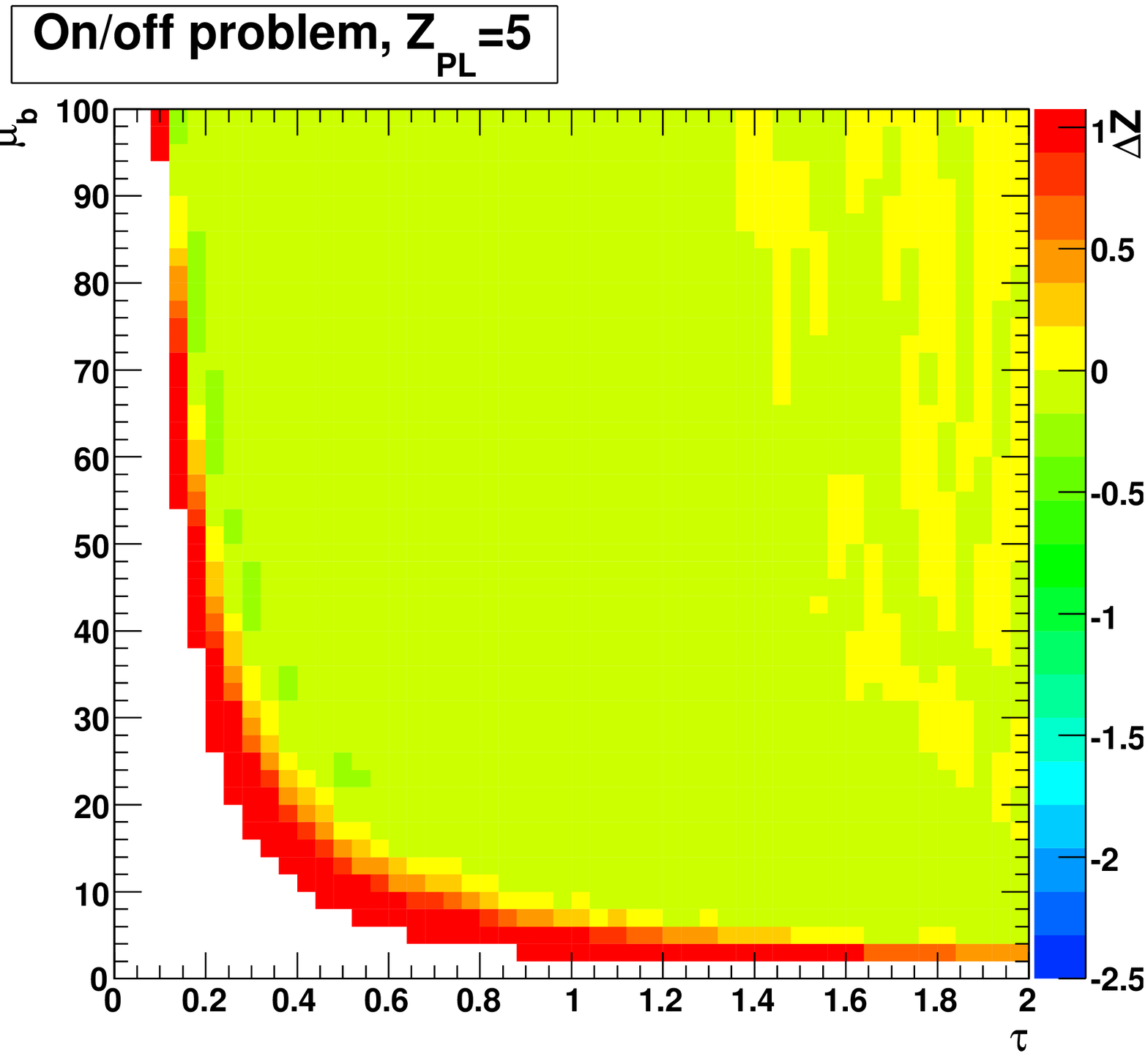}
\caption{ For the on/off problem analyzed using the profile likelihood
method, for each fixed value of $\toffoverton$ and $\mubkgnd$, the
plot indicates the calculated $\zdiff$ for the ensemble of experiments
quoting $\zclaim \ge 5$, i.e., a $p$-value of $2.87 \times 10^{-7}$ or
smaller.  }
\label{proflik_lp_5}
\end{figure}

The results of using the profile likelihood method to analyze the
on/off experiments are shown in Figs.~\ref{proflik_lp_1.28} through
\ref{proflik_lp_5}. There is slight undercoverage over much of the
parameter space, by at most half a unit of $Z$ or so.  As the true
parameters $\mubkgnd$ and $\toffoverton$ move away from the origin,
initially there is overcoverage caused by discreteness, giving way to
the region of largest undercoverage for $\zpl$, which then becomes
only slight undercoverage as near-asymptotic performance is reached.
At the point considered by Cranmer~\cite{cranmer} of $\mubkgnd=$100,
$\toffoverton=$1, we calculate $\ztrue=4.99$, in good agreement with
the result from his MC method of $\ztrue=5.0$. For small $\mubkgnd$
and large $\toffoverton$ (with the qualifiers small and large becoming
stricter for increasing $\zclaim$), the nominal coverage is achieved.

\begin{figure}[htbp]
\centering
\includegraphics*[width=2.7in]{\epsdir/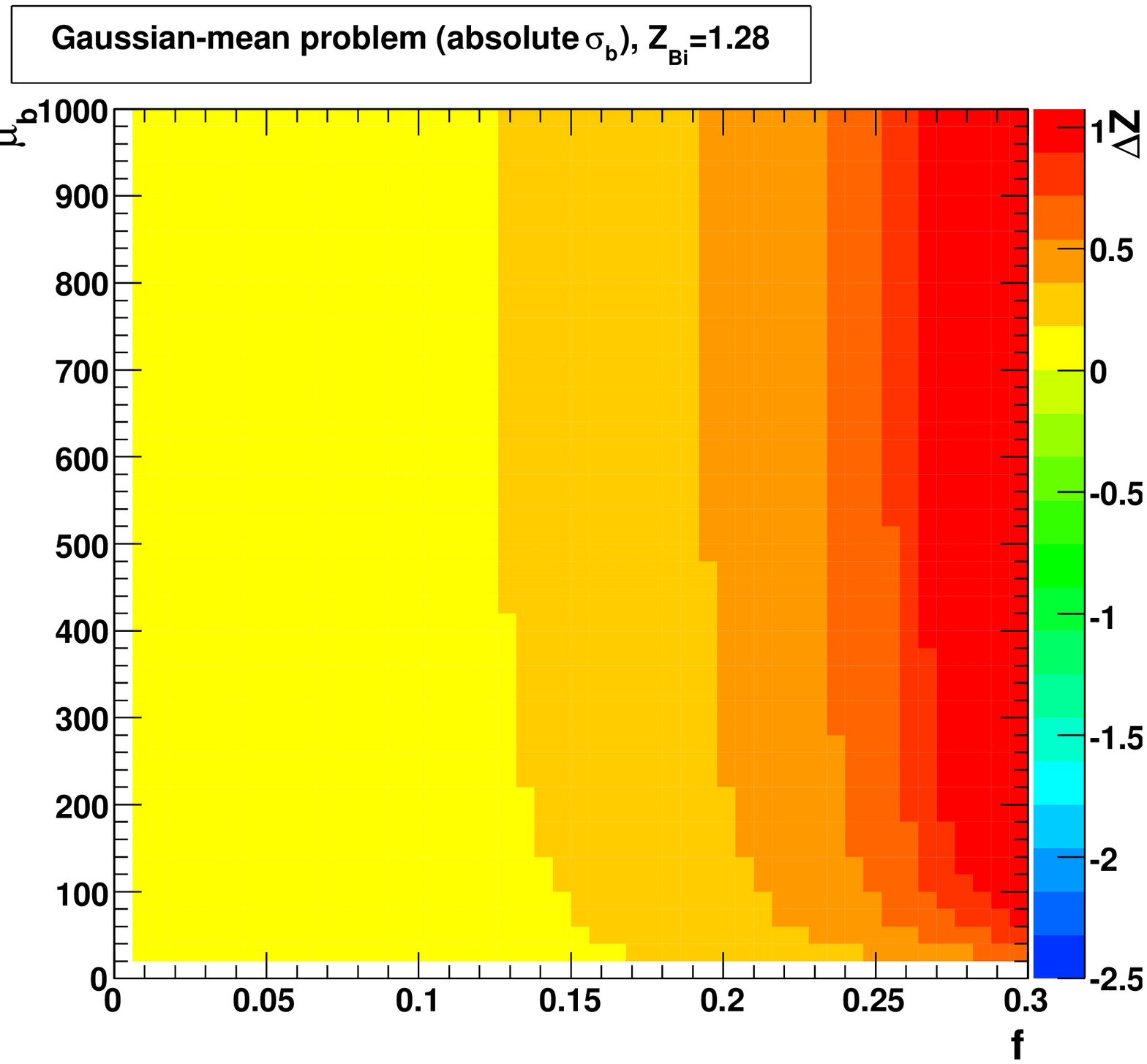}
\includegraphics*[width=2.7in]{\epsdir/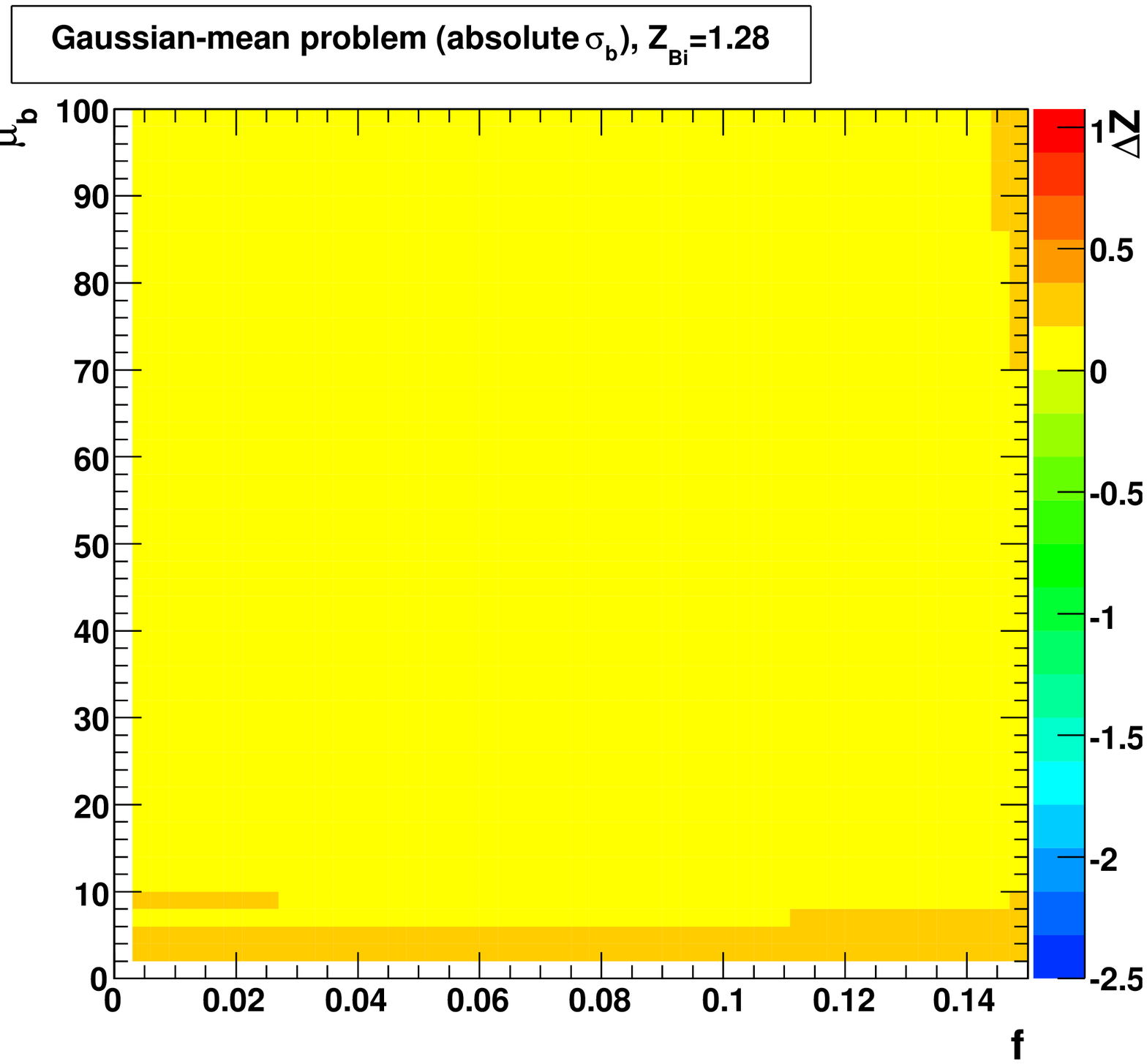}
\caption{ For the Gaussian-mean background problem with exactly known
$\sigmab$, analyzed using the $\zbi$ recipe, for each fixed value of
$f=\sigmab/\mubkgnd$ and $\mubkgnd$, the plot indicates the calculated
$\zdiff$ for the ensemble of experiments quoting $\zclaim \ge 1.28$,
i.e., a $p$-value of $0.1$ or smaller.  }
\label{zbi_lg_abs_1.28}
\end{figure}

\begin{figure}[htbp]
\centering
\includegraphics*[width=2.7in]{\epsdir/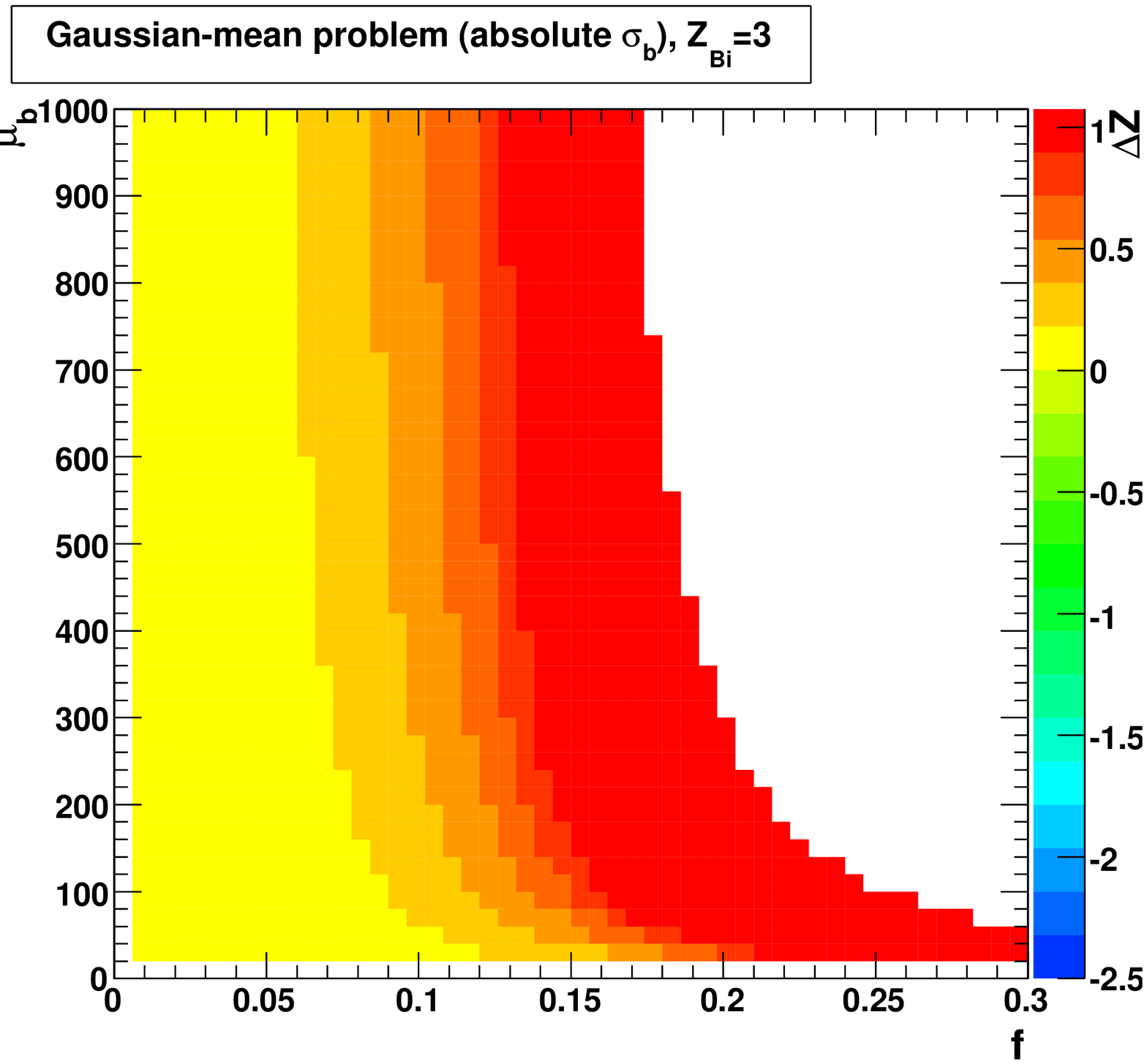}
\includegraphics*[width=2.7in]{\epsdir/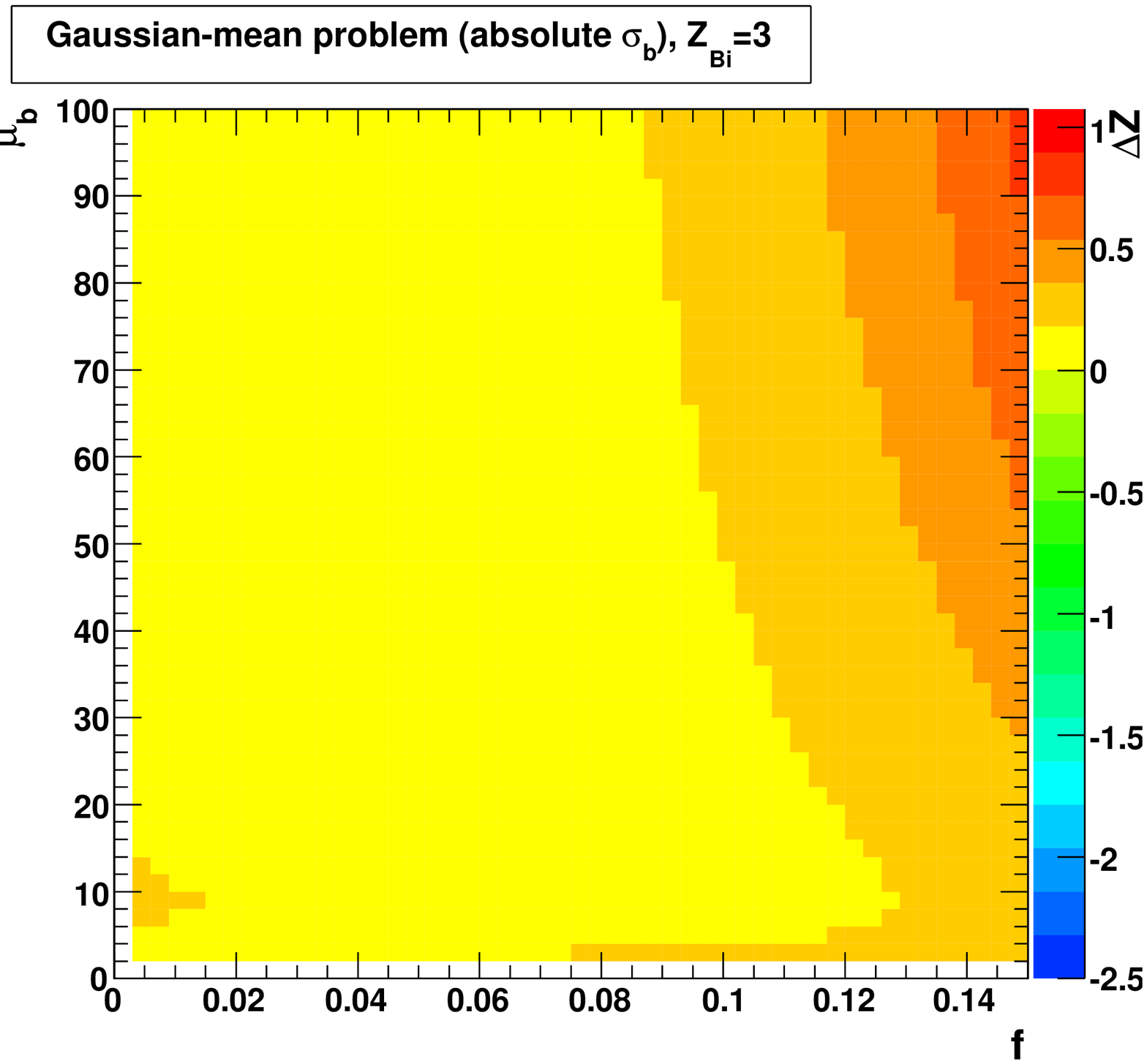}
\caption{ For the Gaussian-mean background problem with exactly known
$\sigmab$, analyzed using the $\zbi$ recipe, for each fixed value of
$f=\sigmab/\mubkgnd$ and $\mubkgnd$, the plot indicates the calculated
$\zdiff$ for the ensemble of experiments quoting $\zclaim \ge 3$,
i.e., a $p$-value of $1.35 \times 10^{-3}$ or smaller.  }
\label{zbi_lg_abs_3}
\end{figure}

\begin{figure}[htbp]
\centering
\includegraphics*[width=2.7in]{\epsdir/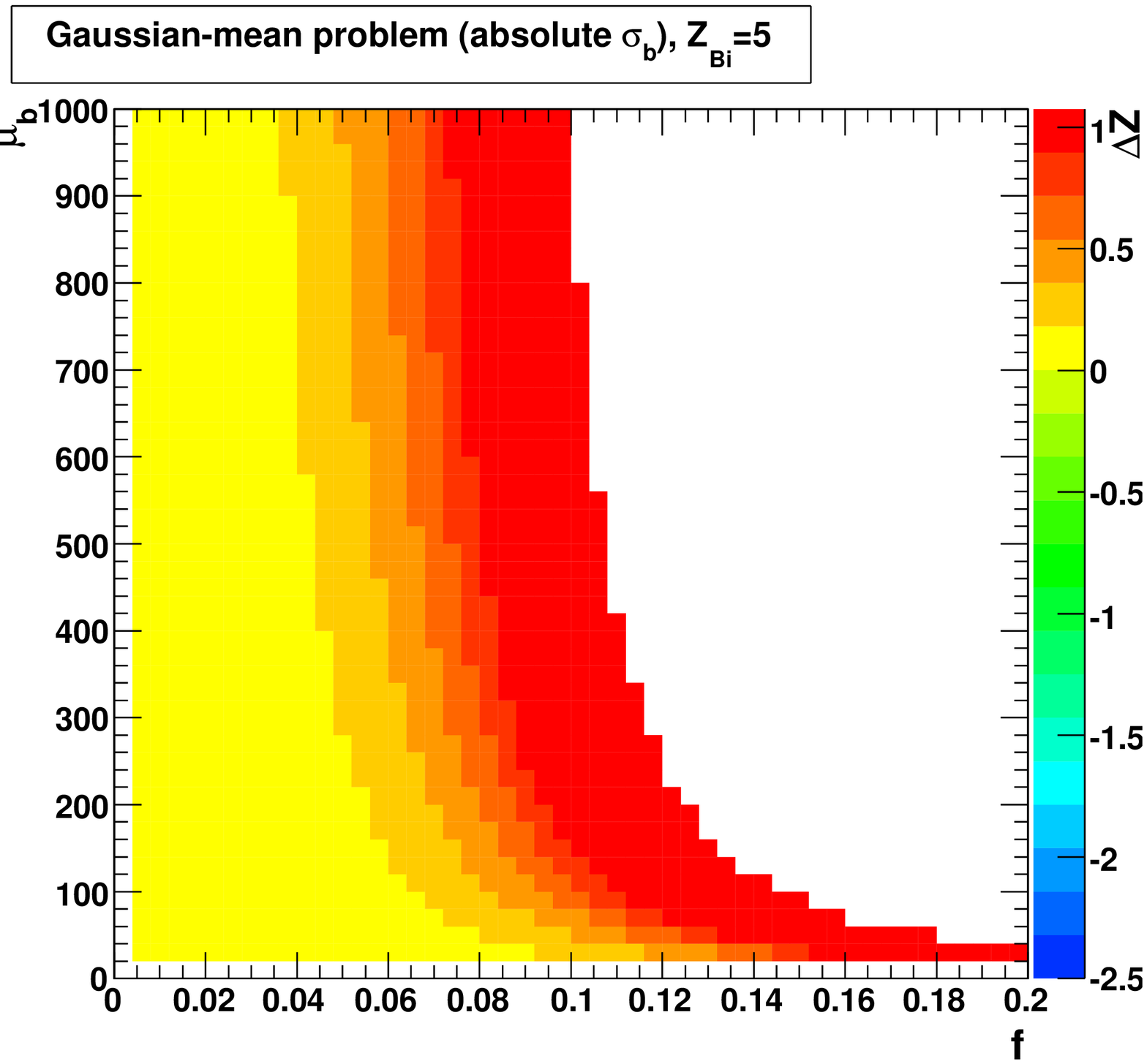}
\includegraphics*[width=2.7in]{\epsdir/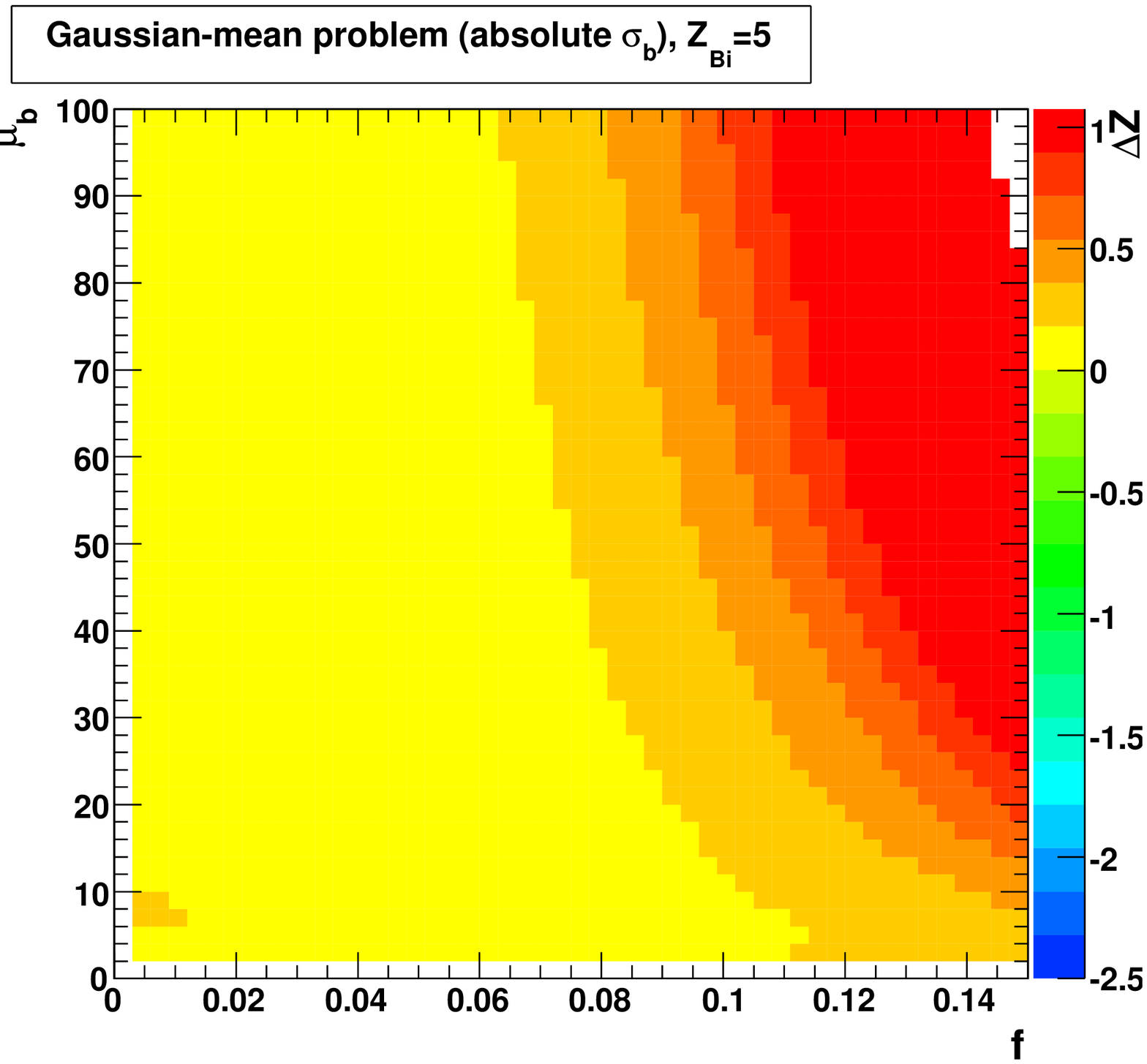}
\caption{ For the Gaussian-mean background problem with exactly known
$\sigmab$, analyzed using the $\zbi$ recipe, for each fixed value of
$f=\sigmab/\mubkgnd$ and $\mubkgnd$, the plot indicates the calculated
$\zdiff$ for the ensemble of experiments quoting $\zclaim \ge 5$,
i.e., a $p$-value of $2.87 \times 10^{-7}$ or smaller.  }
\label{zbi_lg_abs_5}
\end{figure}

\begin{figure}[htbp]
\centering
\includegraphics*[width=2.7in]{\epsdir/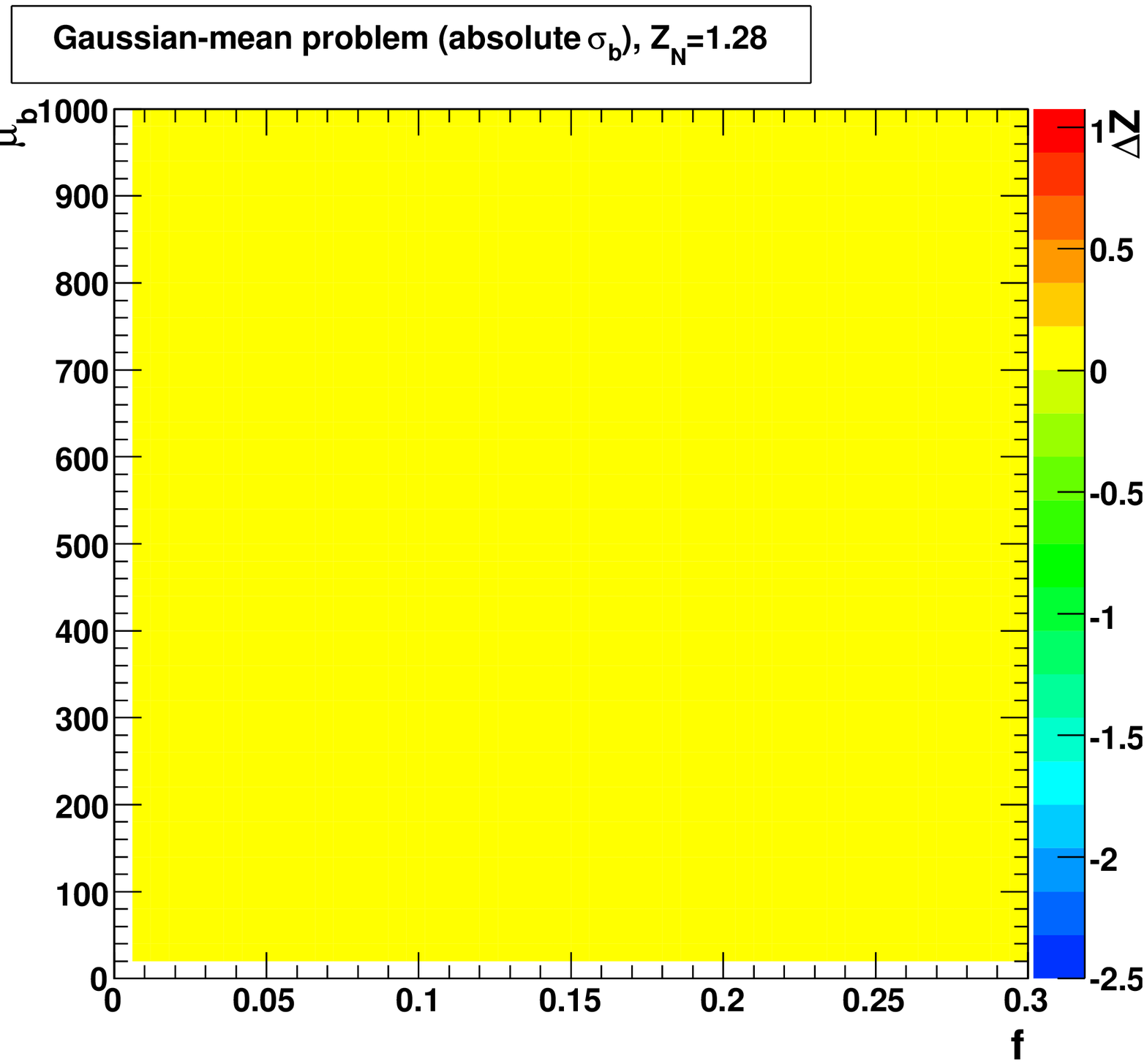}
\includegraphics*[width=2.7in]{\epsdir/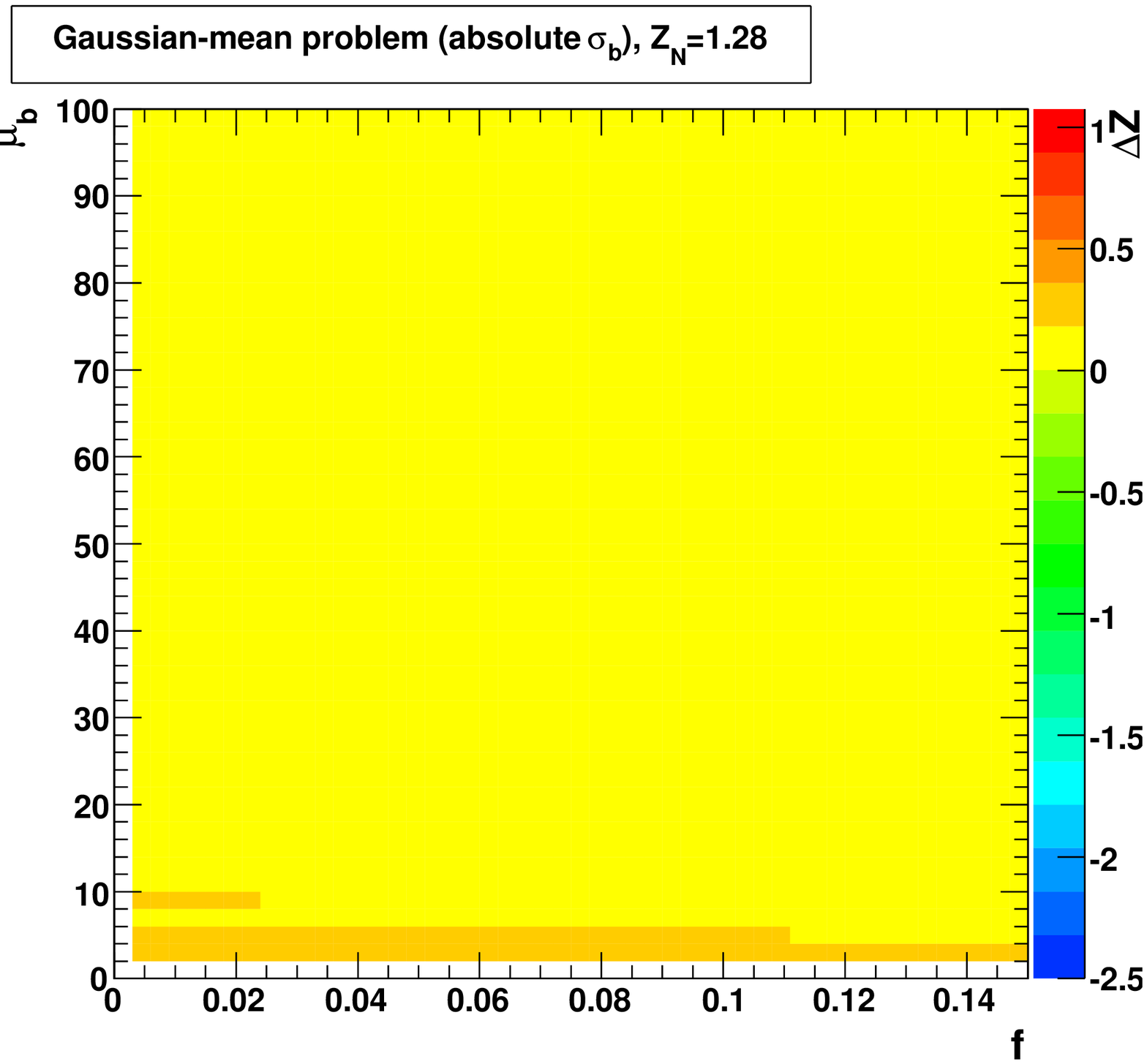}
\caption{ For the Gaussian-mean background problem with exactly known
$\sigmab$, analyzed using the $\zn$ recipe, for each fixed value of
$f=\sigmab/\mubkgnd$ and $\mubkgnd$, the plot indicates the calculated
$\zdiff$ for the ensemble of experiments quoting $\zclaim \ge 1.28$,
i.e., a $p$-value of $0.1$ or smaller.  }
\label{zn_lg_abs_1.28}
\end{figure}

\begin{figure}[htbp]
\centering
\includegraphics*[width=2.7in]{\epsdir/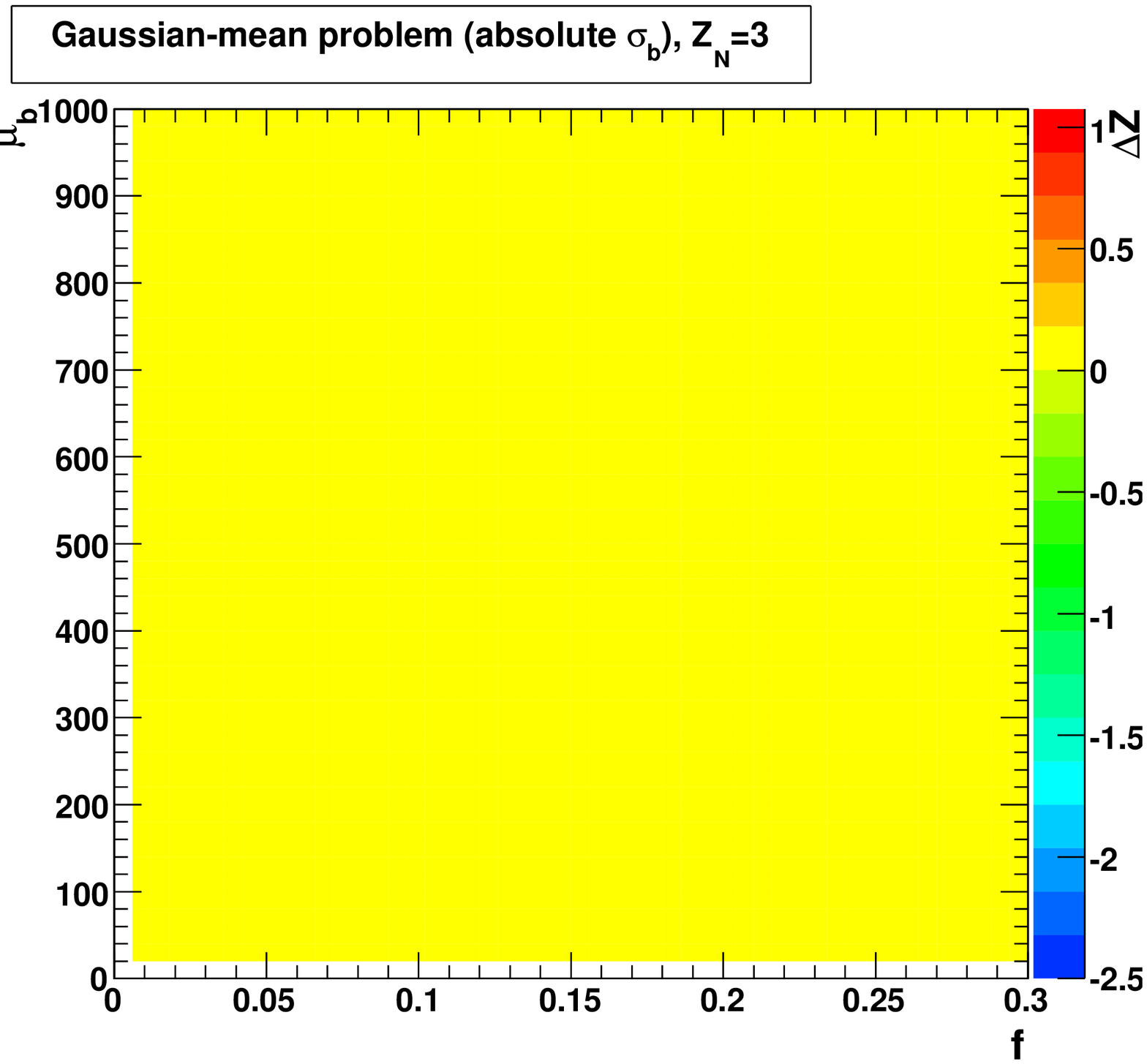}
\includegraphics*[width=2.7in]{\epsdir/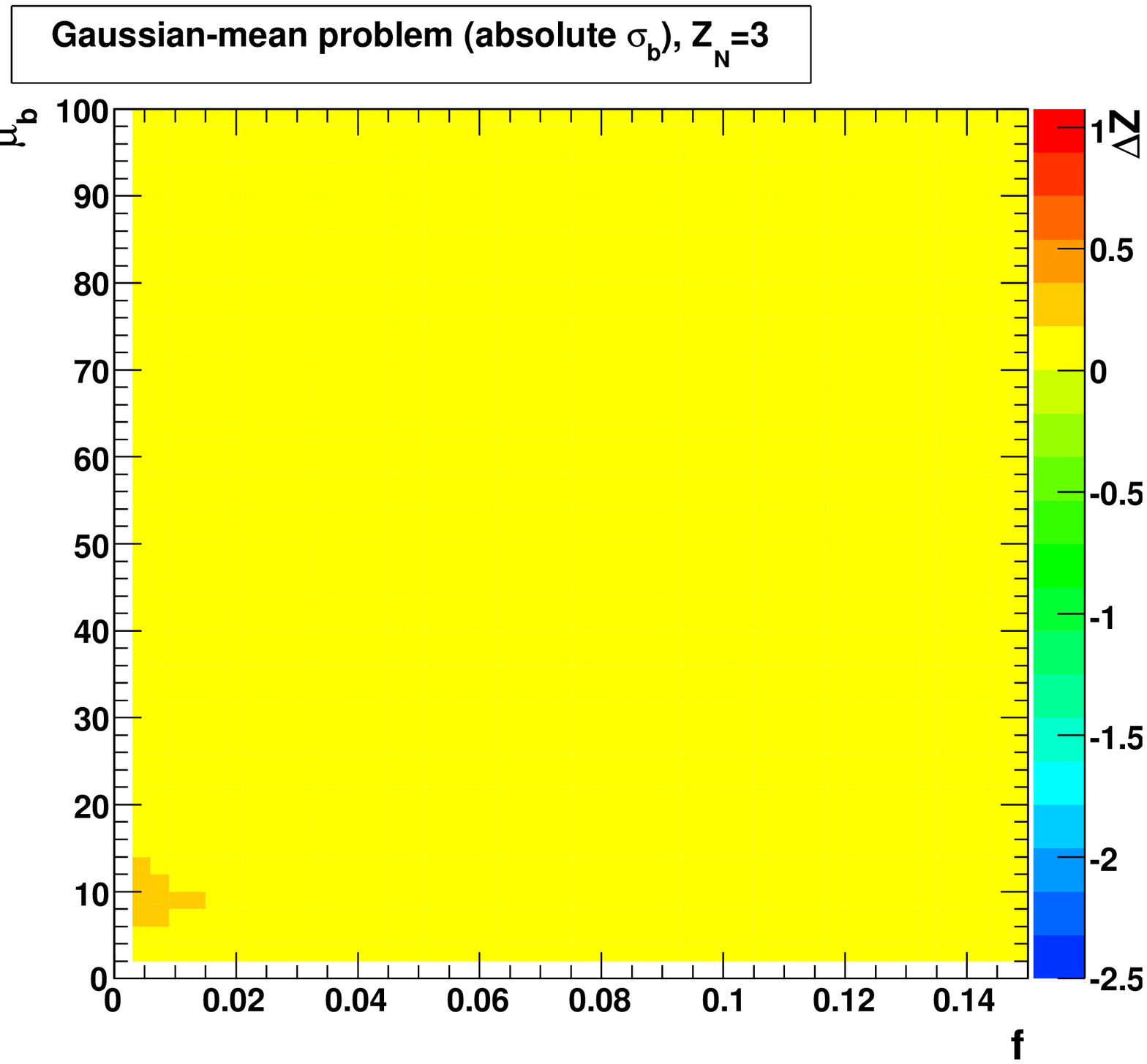}
\caption{ For the Gaussian-mean background problem with exactly known
$\sigmab$, analyzed using the $\zn$ recipe, for each fixed value of
$f=\sigmab/\mubkgnd$ and $\mubkgnd$, the plot indicates the calculated
$\zdiff$ for the ensemble of experiments quoting $\zclaim \ge 3$,
i.e., a $p$-value of $1.35 \times 10^{-3}$ or smaller.  }
\label{zn_lg_abs_3}
\end{figure}

\begin{figure}[htbp]
\centering
\includegraphics*[width=2.7in]{\epsdir/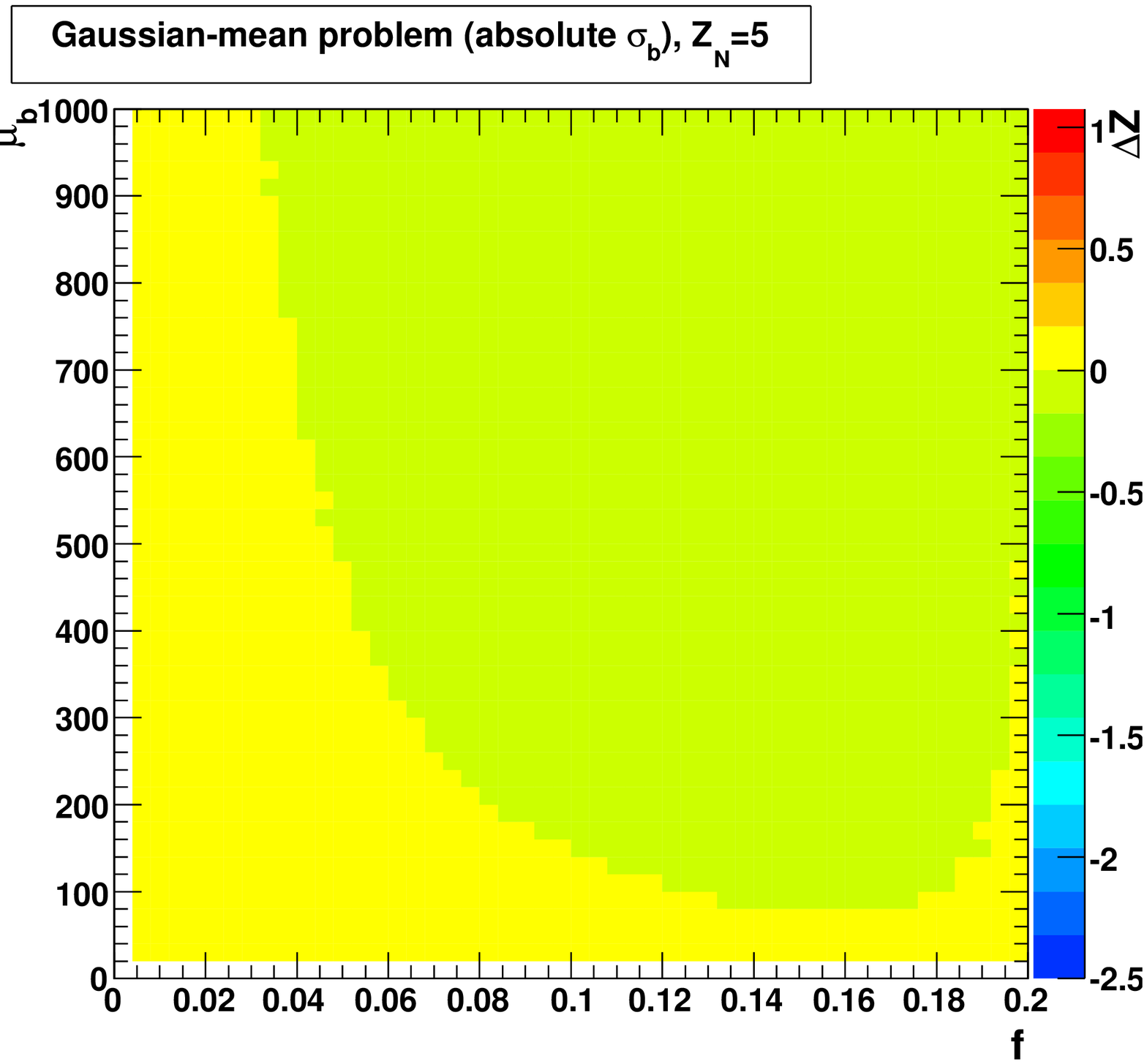}
\includegraphics*[width=2.7in]{\epsdir/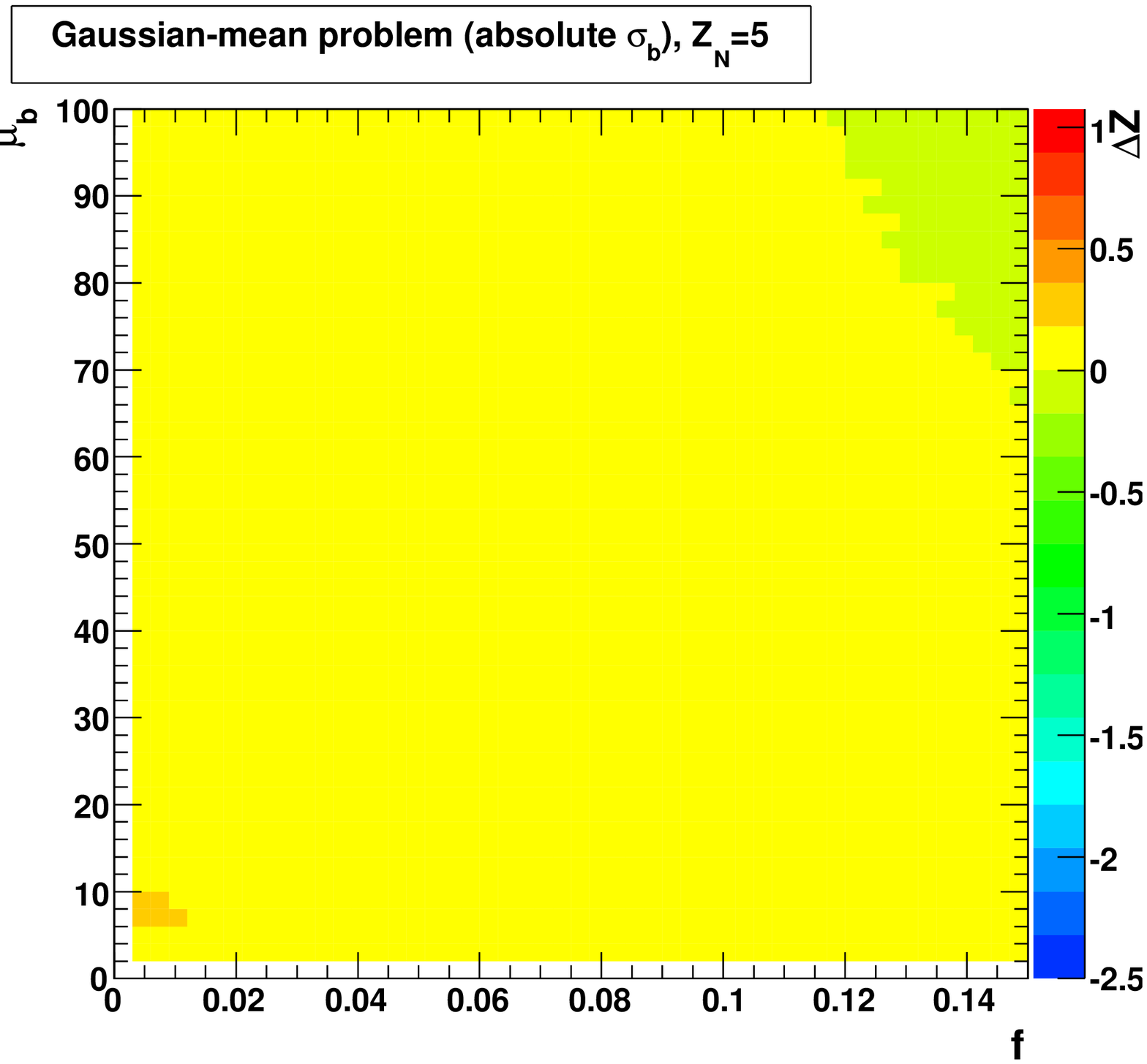}
\caption{ For the Gaussian-mean background problem with exactly known
$\sigmab$, analyzed using the $\zn$ recipe, for each fixed value of
$f=\sigmab/\mubkgnd$ and $\mubkgnd$, the plot indicates the calculated
$\zdiff$ for the ensemble of experiments quoting $\zclaim \ge 5$,
i.e., a $p$-value of $2.87 \times 10^{-7}$ or smaller.  }
\label{zn_lg_abs_5}
\end{figure}

\begin{figure}[htbp]
\centering
\includegraphics*[width=2.7in]{\epsdir/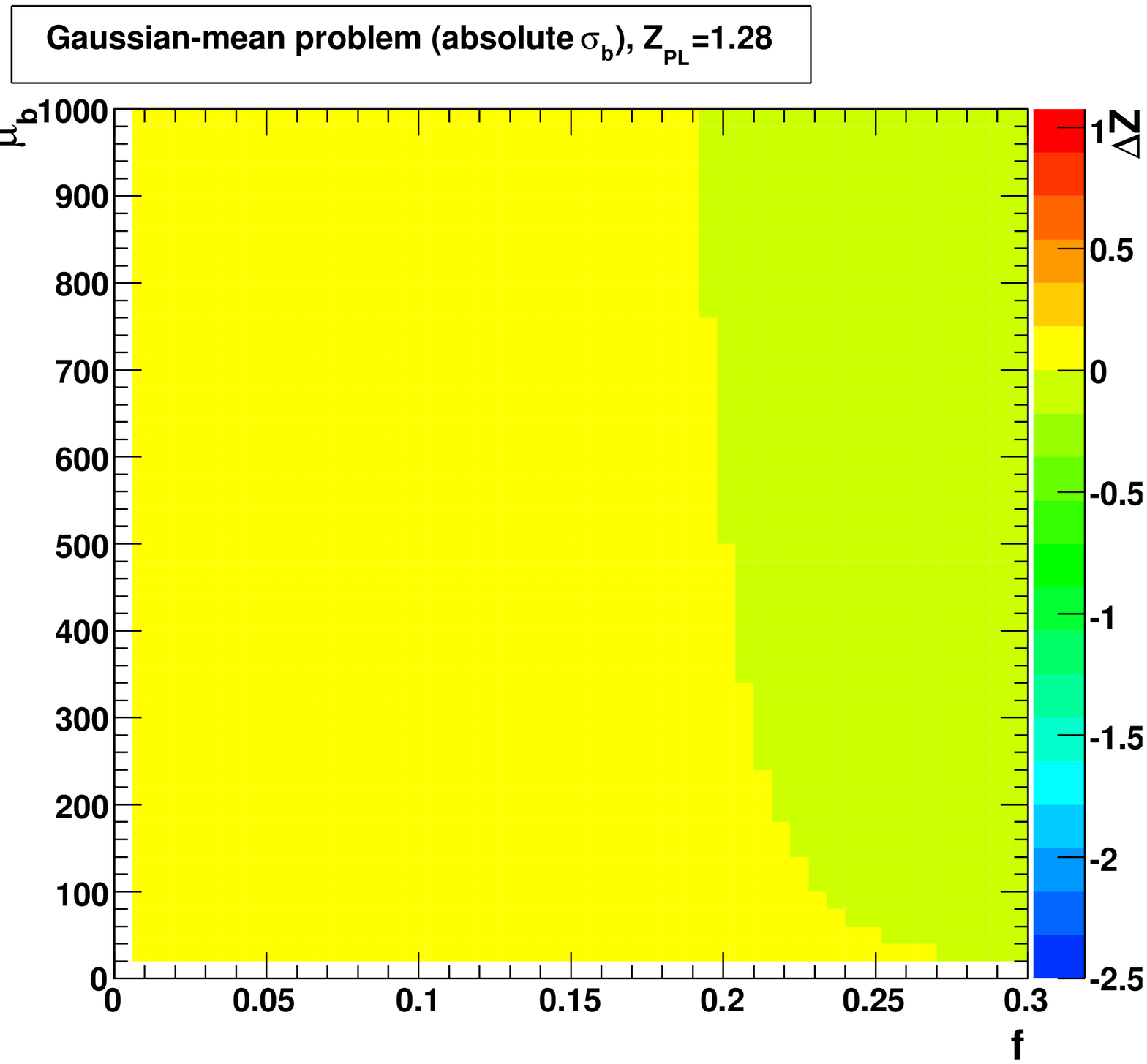}
\includegraphics*[width=2.7in]{\epsdir/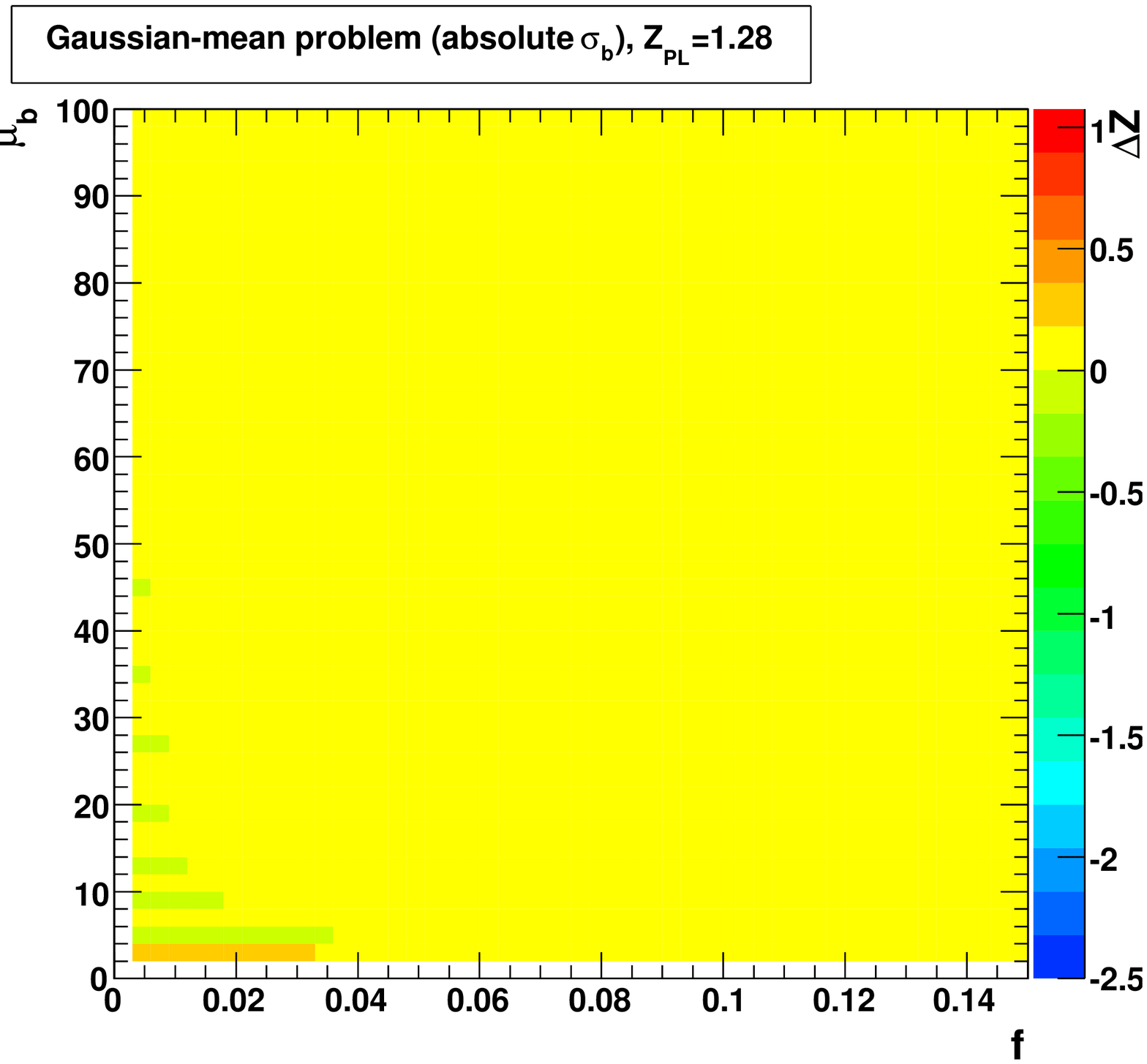}
\caption{ For the Gaussian-mean background problem with exactly known
$\sigmab$, analyzed using the profile likelihood method, for each
fixed value of $f=\sigmab/\mubkgnd$ and $\mubkgnd$, the plot indicates
the calculated $\zdiff$ for the ensemble of experiments quoting
$\zclaim \ge 1.28$, i.e., a $p$-value of $0.1$ or smaller.  }
\label{proflik_lg_abs_1.28}
\end{figure}

\begin{figure}[htbp]
\centering
\includegraphics*[width=2.7in]{\epsdir/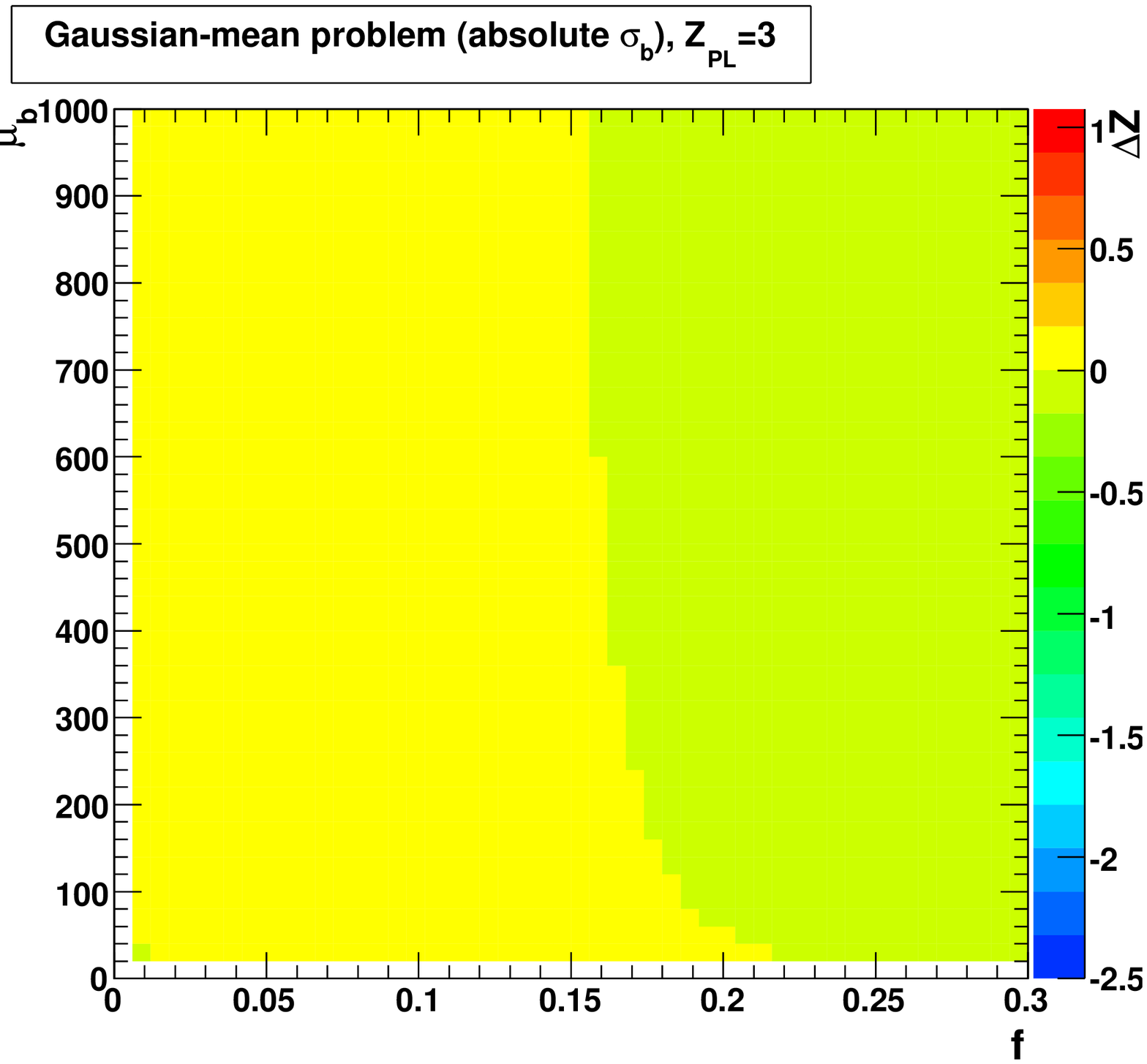}
\includegraphics*[width=2.7in]{\epsdir/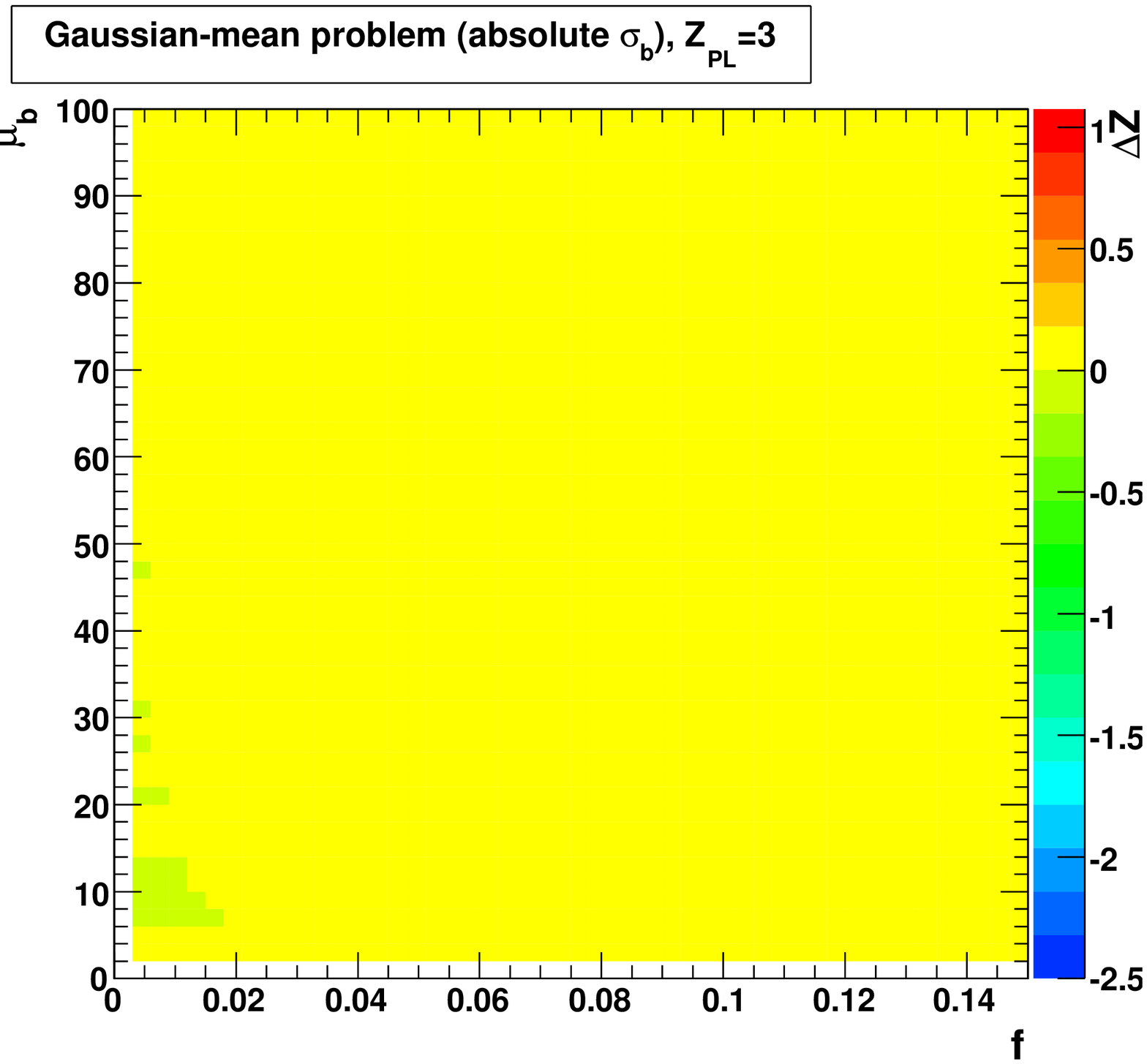}
\caption{ For the Gaussian-mean background problem with exactly known
$\sigmab$, analyzed using the profile likelihood method, for each
fixed value of $f=\sigmab/\mubkgnd$ and $\mubkgnd$, the plot indicates
the calculated $\zdiff$ for the ensemble of experiments quoting
$\zclaim \ge 3$, i.e., a $p$-value of $1.35 \times 10^{-3}$ or
smaller.  }
\label{proflik_lg_abs_3}
\end{figure}

\begin{figure}[htbp]
\centering
\includegraphics*[width=2.7in]{\epsdir/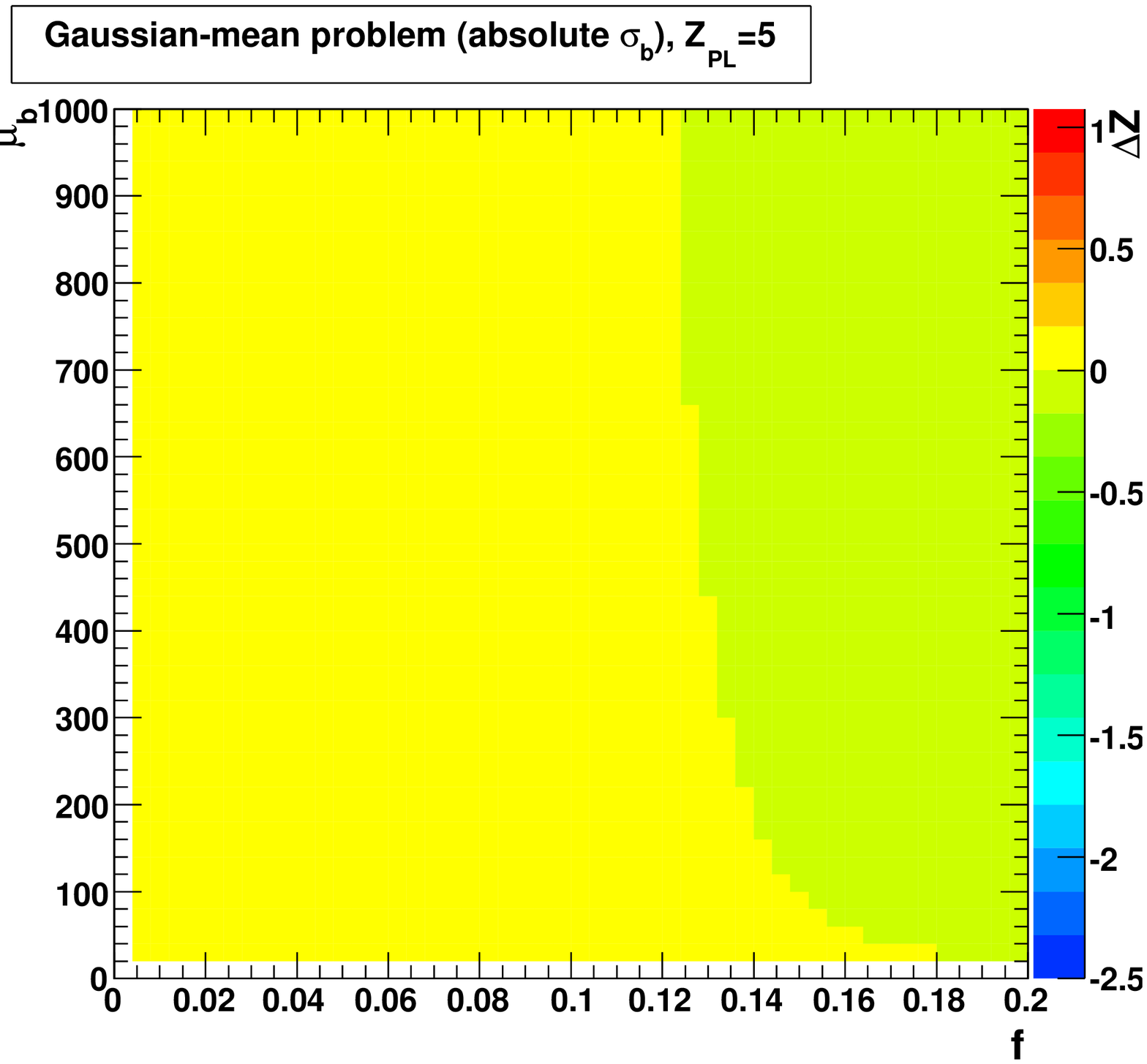}
\includegraphics*[width=2.7in]{\epsdir/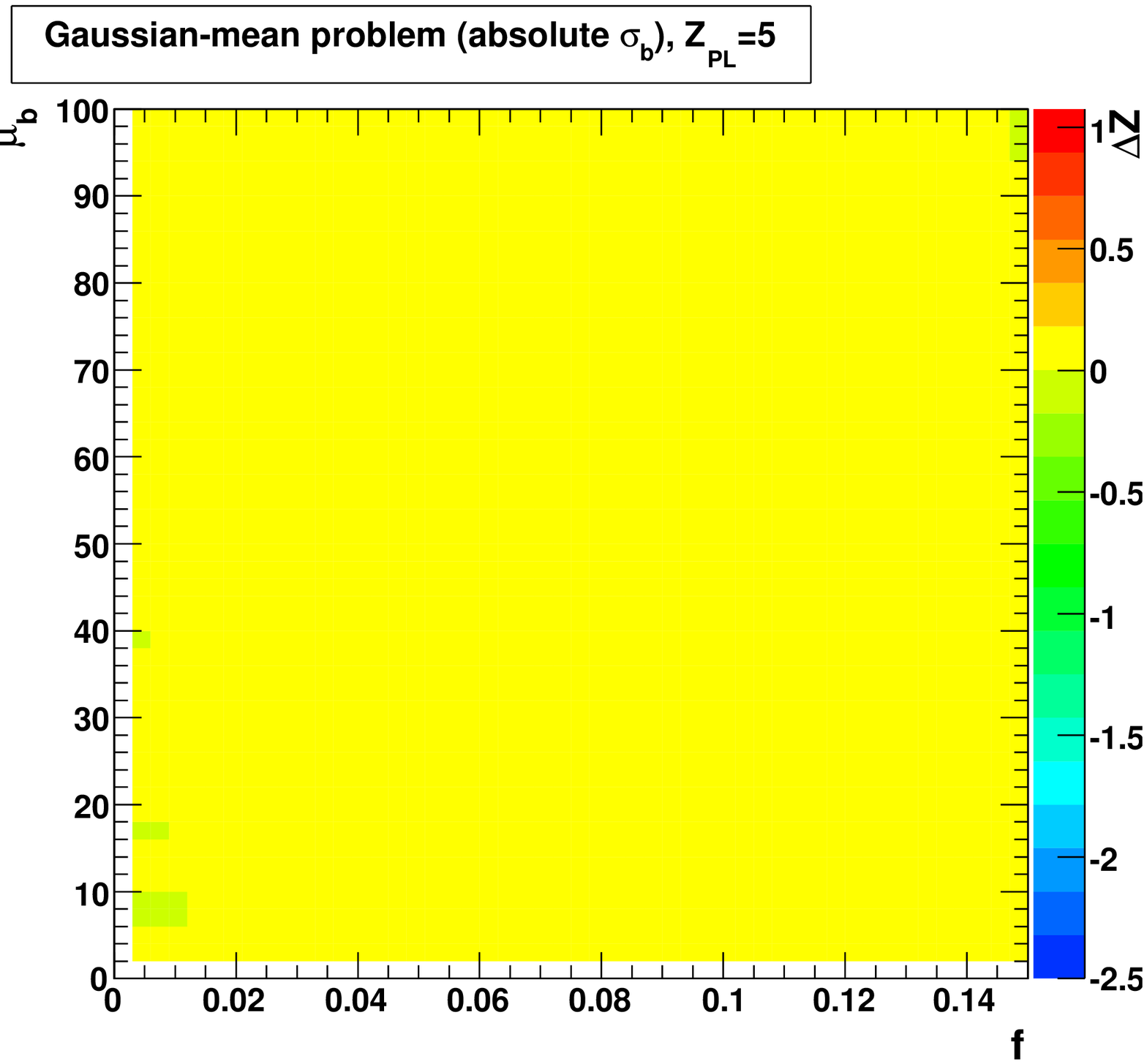}
\caption{ For the Gaussian-mean background problem with exactly known
$\sigmab$, analyzed using the profile likelihood method, for each
fixed value of $f=\sigmab/\mubkgnd$ and $\mubkgnd$, the plot indicates
the calculated $\zdiff$ for the ensemble of experiments quoting
$\zclaim \ge 5$, i.e., a $p$-value of $2.87 \times 10^{-7}$ or
smaller.  }
\label{proflik_lg_abs_5}
\end{figure}

For the Gaussian-mean background problem with exactly known $\sigmab$,
the results are in Figs.~\ref{zbi_lg_abs_1.28} through
\ref{zbi_lg_abs_5} when analyzed with $\zbi$; in
Figs.~\ref{zn_lg_abs_1.28} through \ref{zn_lg_abs_5} when analyzed
with $\zn$; and in Figs.~\ref{proflik_lg_abs_1.28} through
\ref{proflik_lg_abs_5} when analyzed with $\zpl$.  $\zbi$ overcovers
everywhere, quite severely for the larger values of $f$ considered;
this is an effect of small estimates of $\estb$ leading by the rough
correspondence of Eqn.~\ref{eqn:corr} to underestimates of the
shape-controlling parameter $\toffoverton = \estb/\sigmab^2$, and thus
to an overly broad and shifted gamma distribution which in turn leads
to estimated tail probabilities which are inappropriately large.
$\zn$ provides slight over-coverage and no undercoverage for
$\zclaim=1.28$ and $\zclaim=3$, but it undercovers for for $\zclaim=5$
at larger values of $f$ and $\mubkgnd$.  For the largest values of $f$
in Fig.~\ref{zn_lg_abs_5}, the reduction in undercoverage is an
artifact of using the truncated Gaussian model for the uncertainty in
the mean background, as the condition of Eqn.~\ref{zf} comes into
play.  $\zpl$ has good coverage over the entire parameter space shown,
with some effect of discreteness observable.

\begin{figure}[htbp]
\centering
\includegraphics*[width=2.7in]{\epsdir/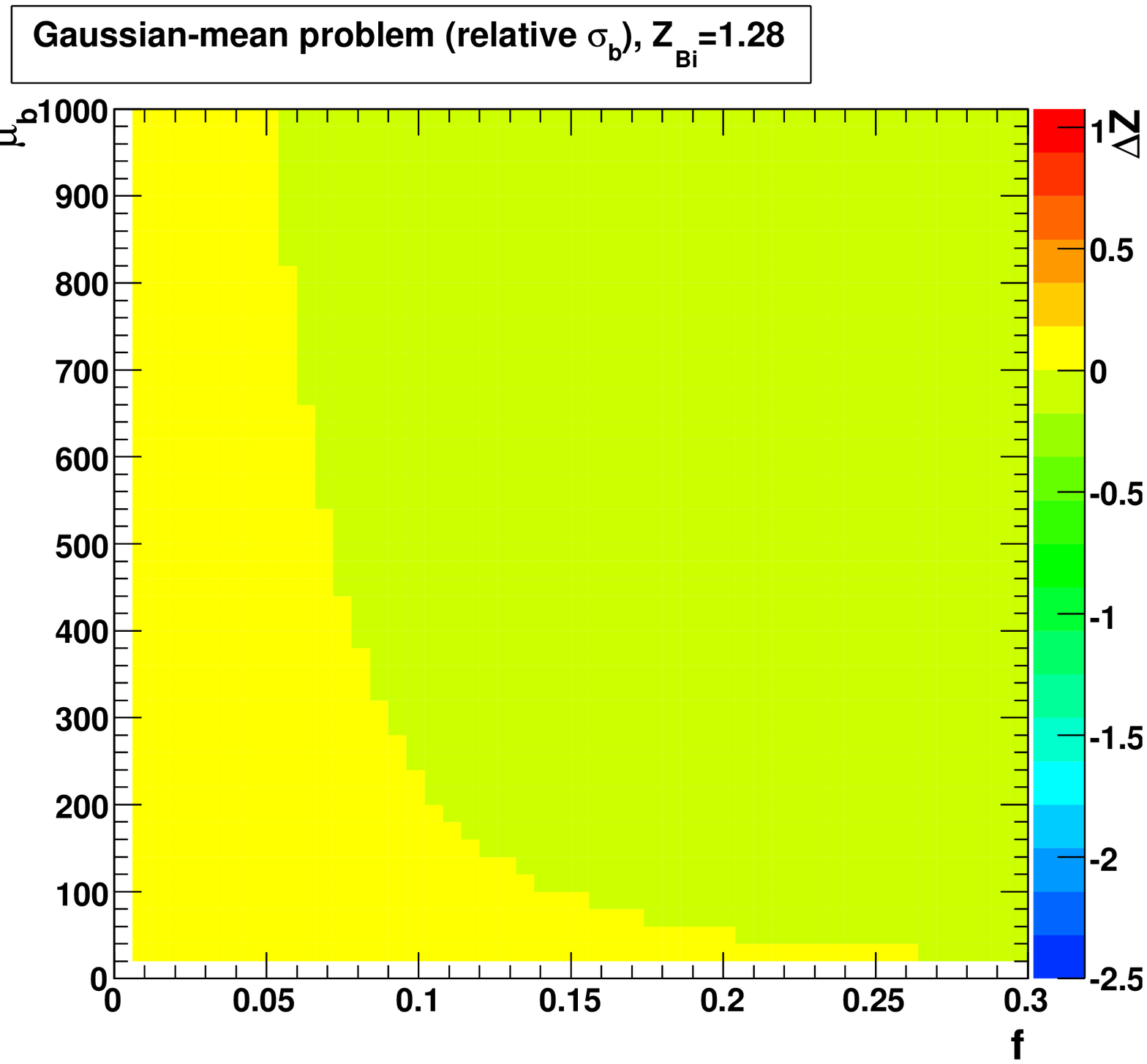}
\includegraphics*[width=2.7in]{\epsdir/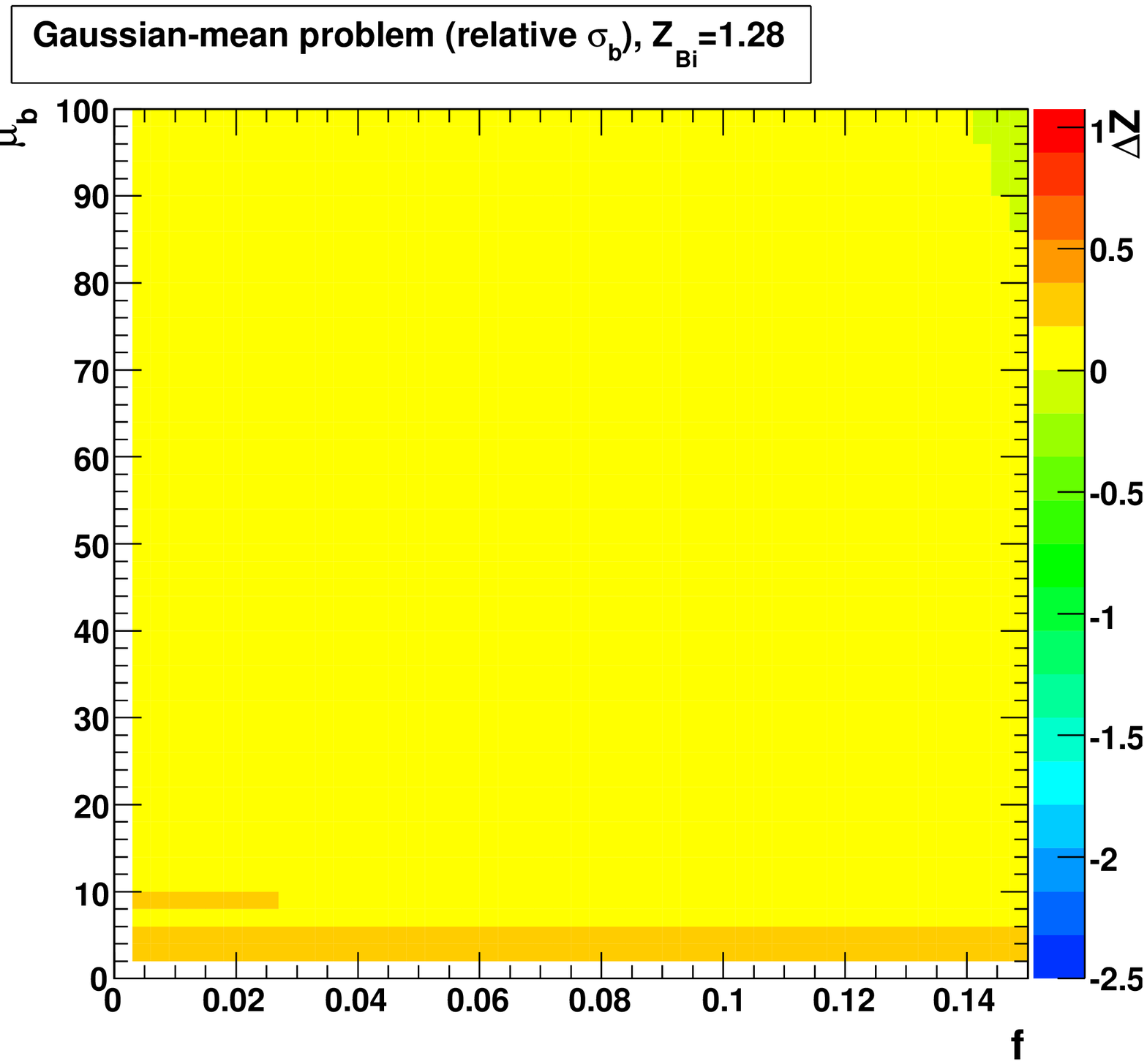}
\caption{ For the Gaussian-mean background problem with exactly known
relative uncertainty $f$, analyzed using the $\zbi$ recipe, for each
fixed value of $f$ and $\mubkgnd$, the plot indicates the calculated
$\zdiff$ for the ensemble of experiments quoting $\zclaim \ge 1.28$,
i.e., a $p$-value of $0.1$ or smaller.  }
\label{zbi_lg_rel_1.28}
\end{figure}

\begin{figure}[htbp]
\centering
\includegraphics*[width=2.7in]{\epsdir/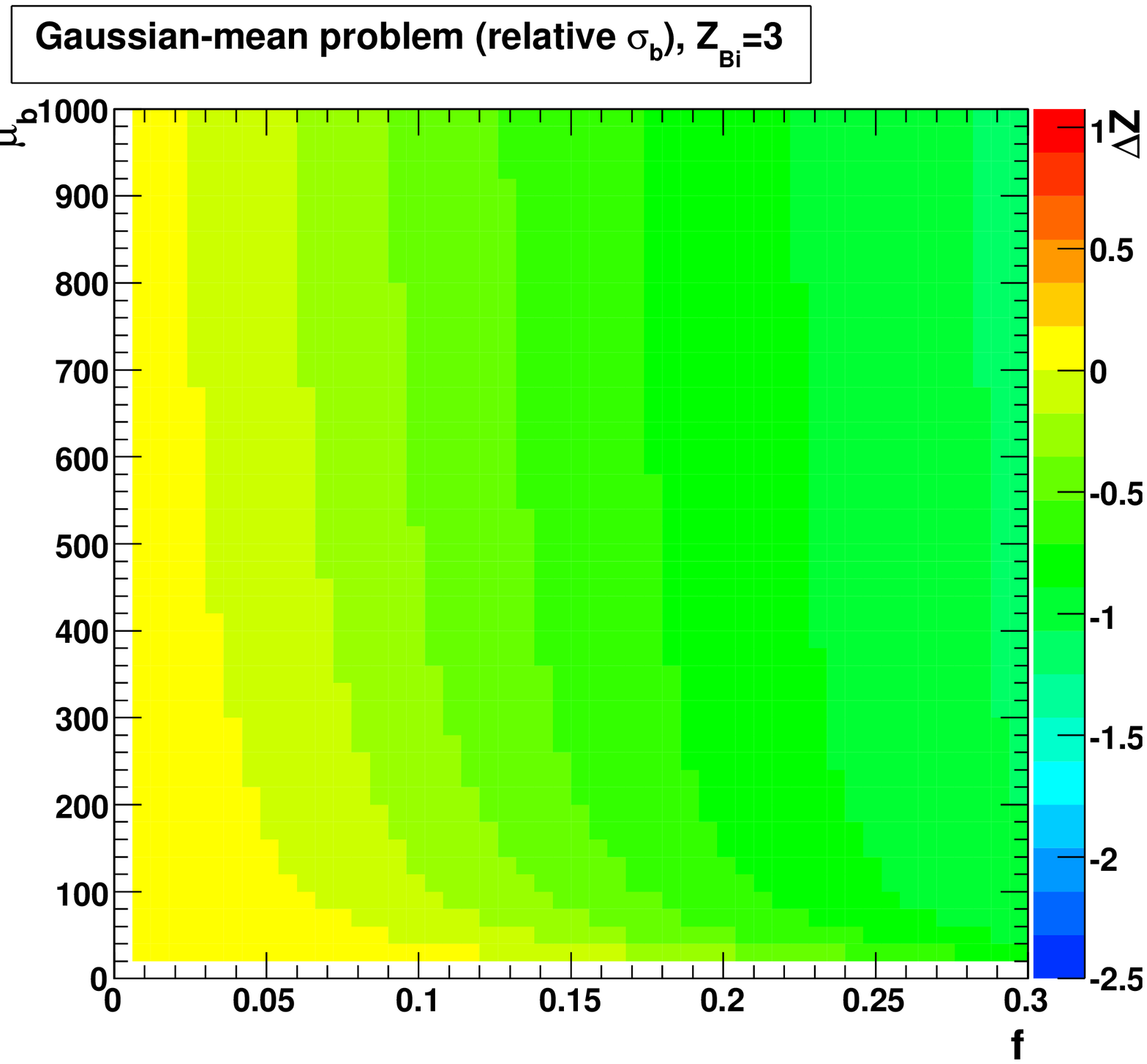}
\includegraphics*[width=2.7in]{\epsdir/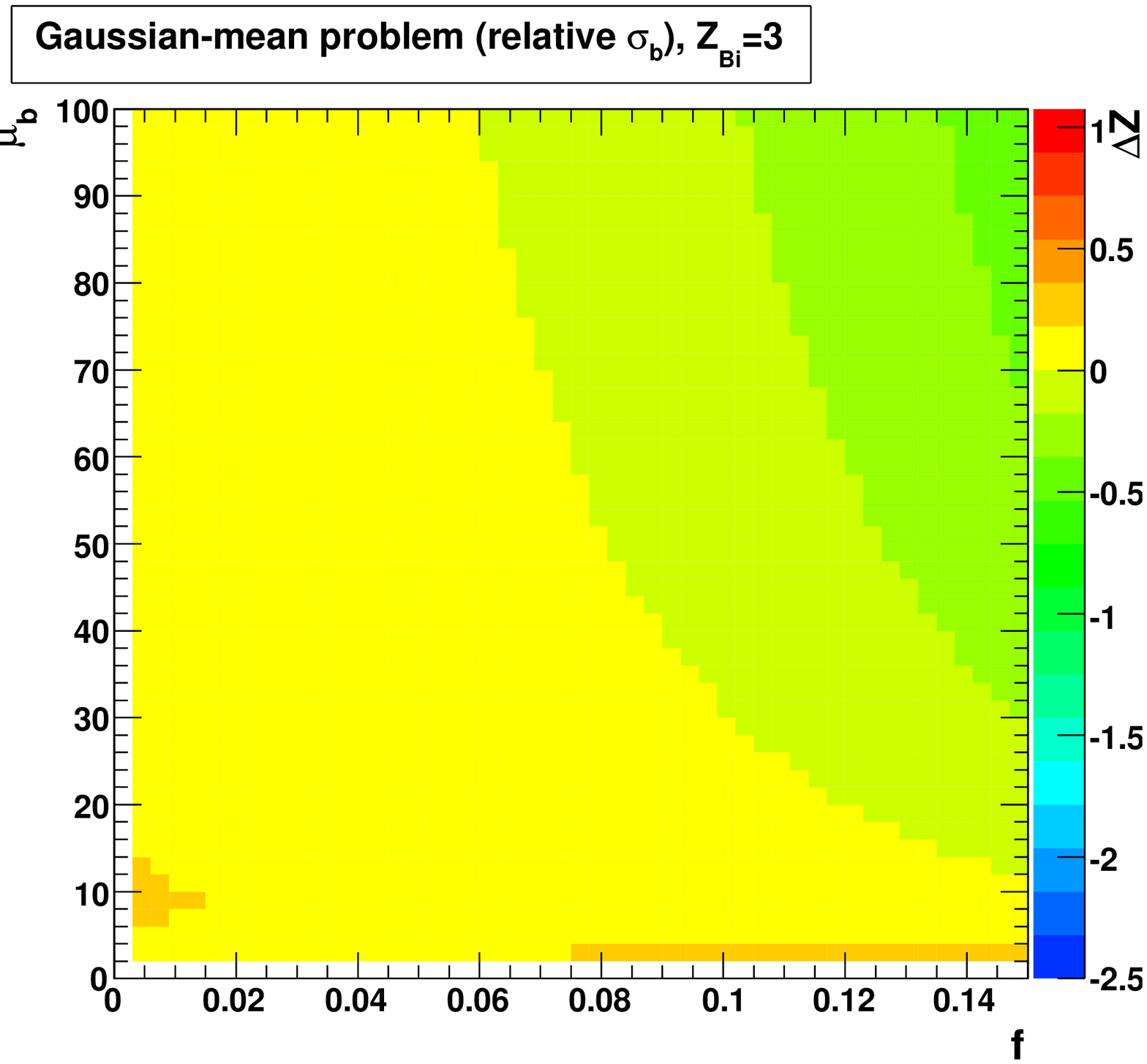}
\caption{ For the Gaussian-mean background problem with exactly known
relative uncertainty $f$, analyzed using the $\zbi$ recipe, for each
fixed value of $f$ and $\mubkgnd$, the plot indicates the calculated
$\zdiff$ for the ensemble of experiments quoting $\zclaim \ge 3$,
i.e., a $p$-value of $1.35 \times 10^{-3}$ or smaller.  }
\label{zbi_lg_rel_3}
\end{figure}

\begin{figure}[htbp]
\centering
\includegraphics*[width=2.7in]{\epsdir/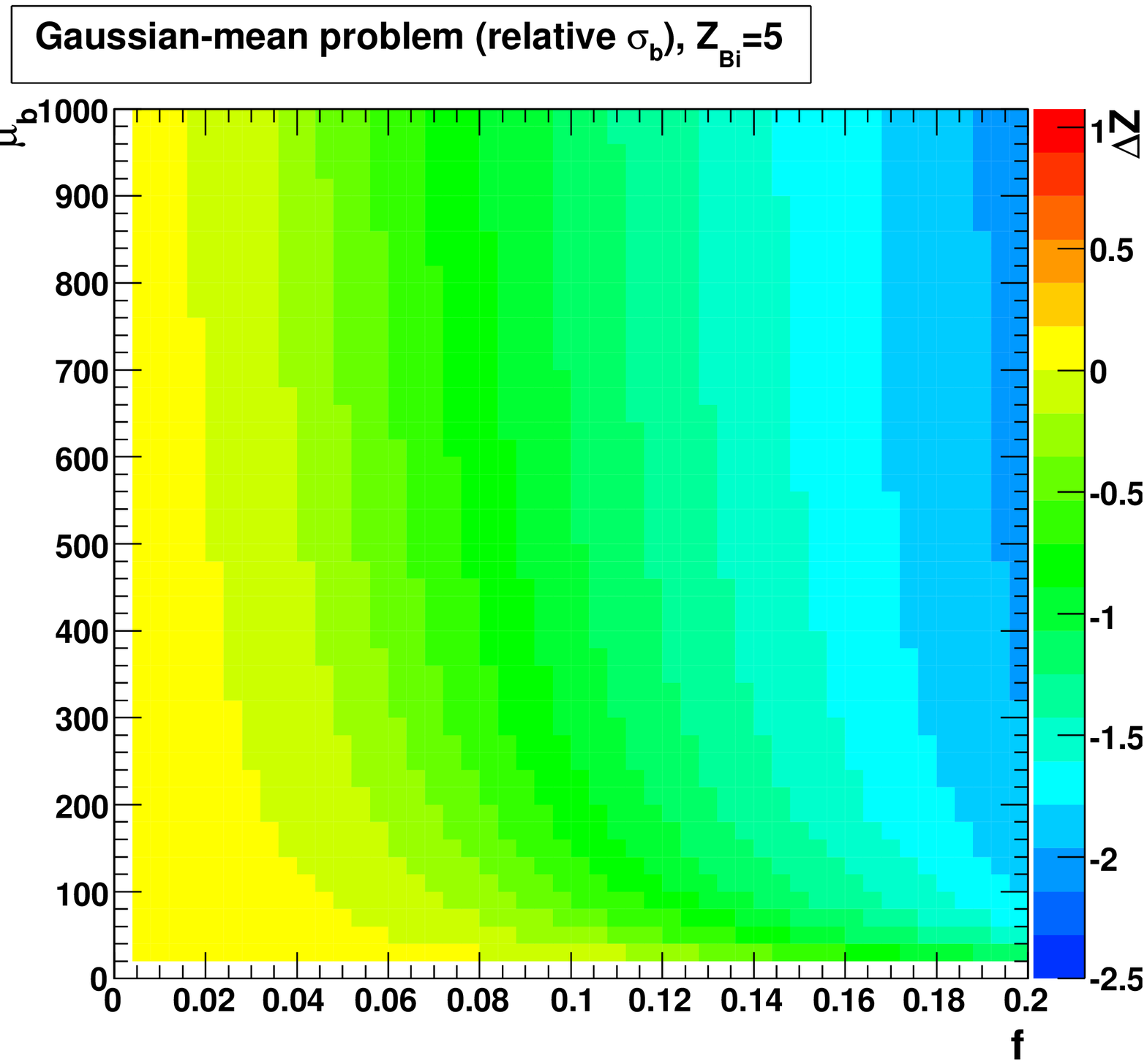}
\includegraphics*[width=2.7in]{\epsdir/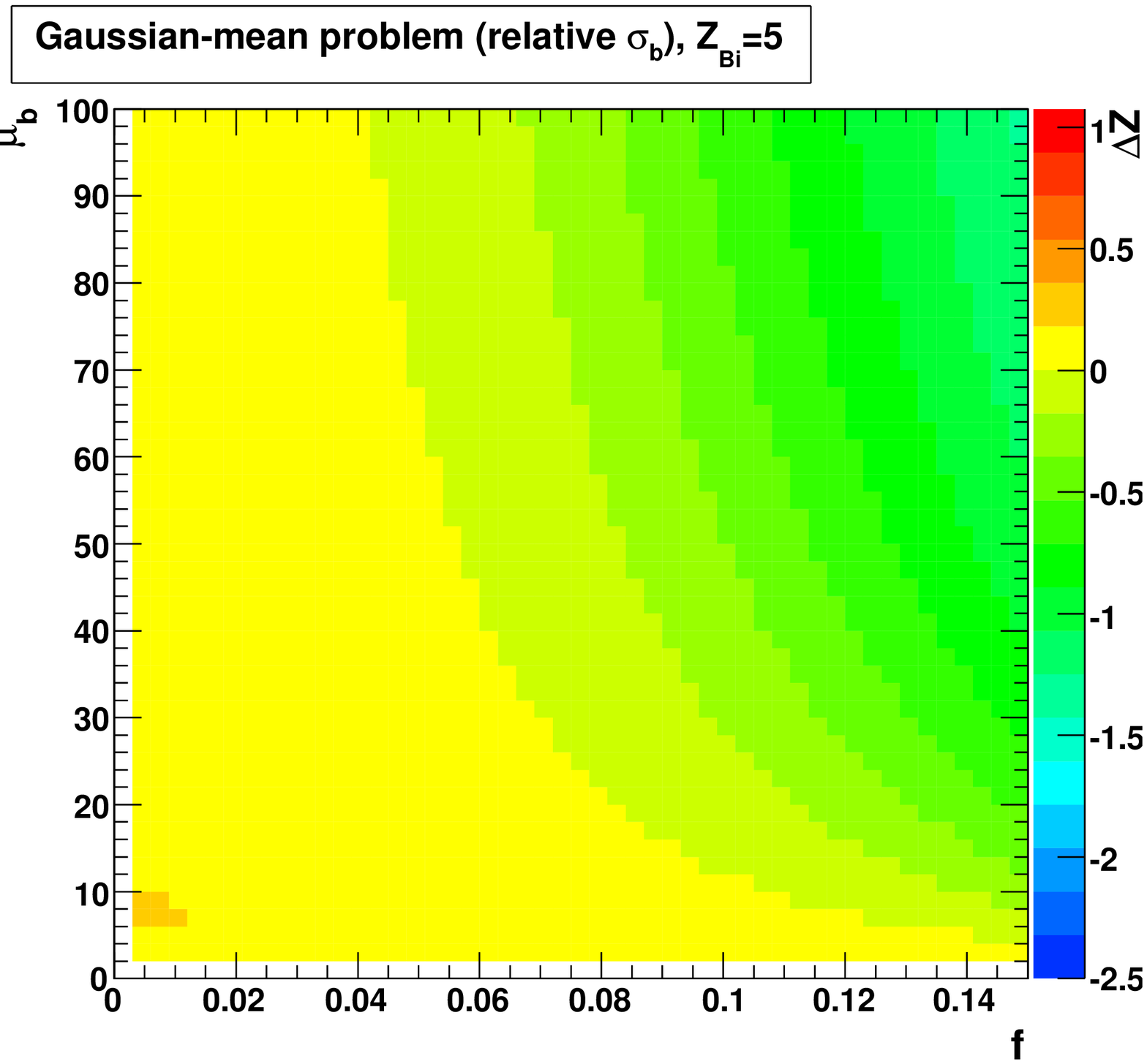}
\caption{ For the Gaussian-mean background problem with exactly known
relative uncertainty $f$, analyzed using the $\zbi$ recipe, for each
fixed value of $f$ and $\mubkgnd$, the plot indicates the calculated
$\zdiff$ for the ensemble of experiments quoting $\zclaim \ge 5$,
i.e., a $p$-value of $2.87 \times 10^{-7}$ or smaller.  }
\label{zbi_lg_rel_5}
\end{figure}

\begin{figure}[htbp]
\centering
\includegraphics*[width=2.7in]{\epsdir/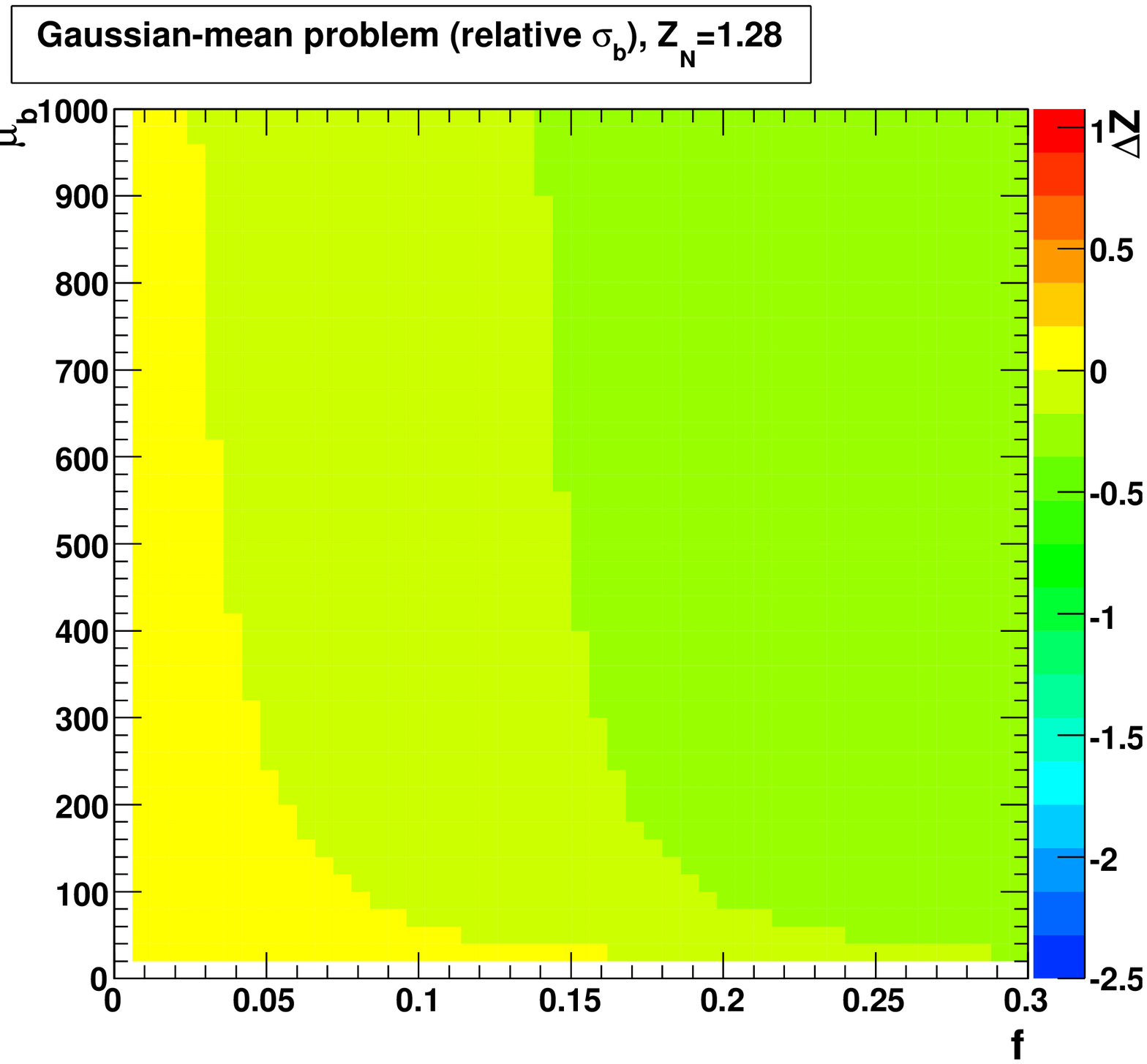}
\includegraphics*[width=2.7in]{\epsdir/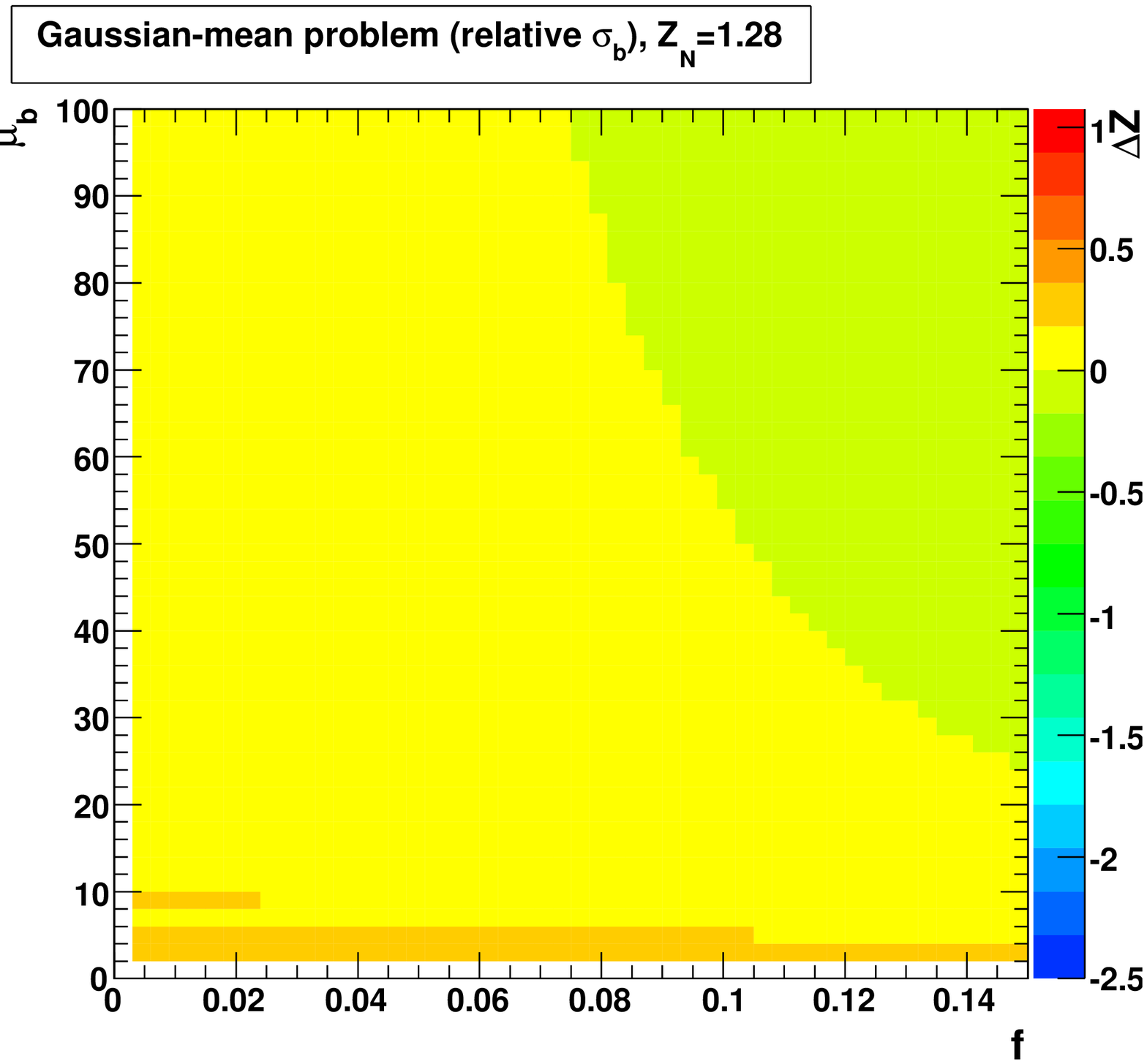}
\caption{ For the Gaussian-mean background problem with exactly known
relative uncertainty $f$, analyzed using the $\zn$ recipe, for each
fixed value of $f$ and $\mubkgnd$, the plot indicates the calculated
$\zdiff$ for the ensemble of experiments quoting $\zclaim \ge 1.28$,
i.e., a $p$-value of $0.1$ or smaller.  }
\label{zn_lg_rel_1.28}
\end{figure}

\begin{figure}[htbp]
\centering
\includegraphics*[width=2.7in]{\epsdir/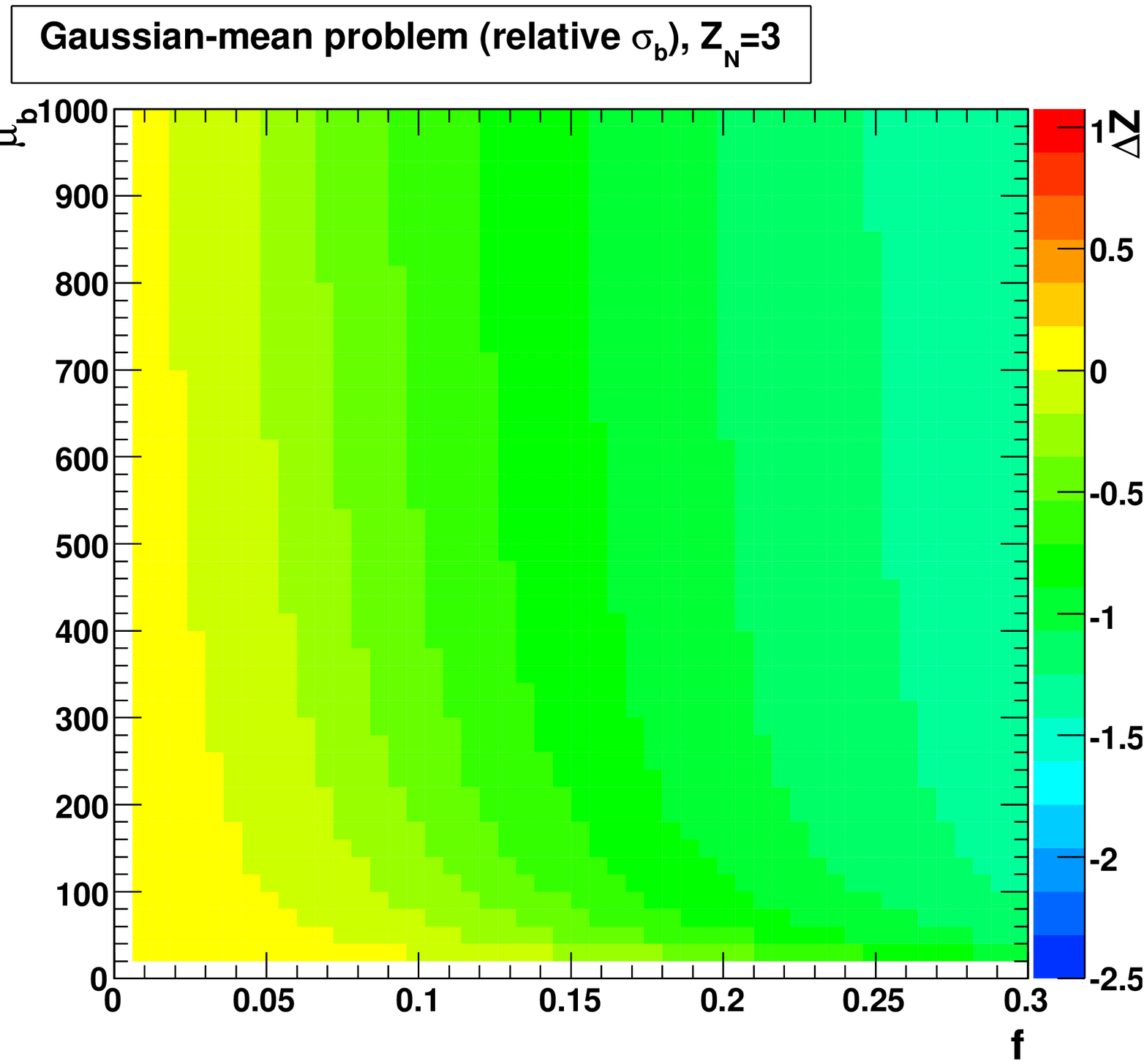}
\includegraphics*[width=2.7in]{\epsdir/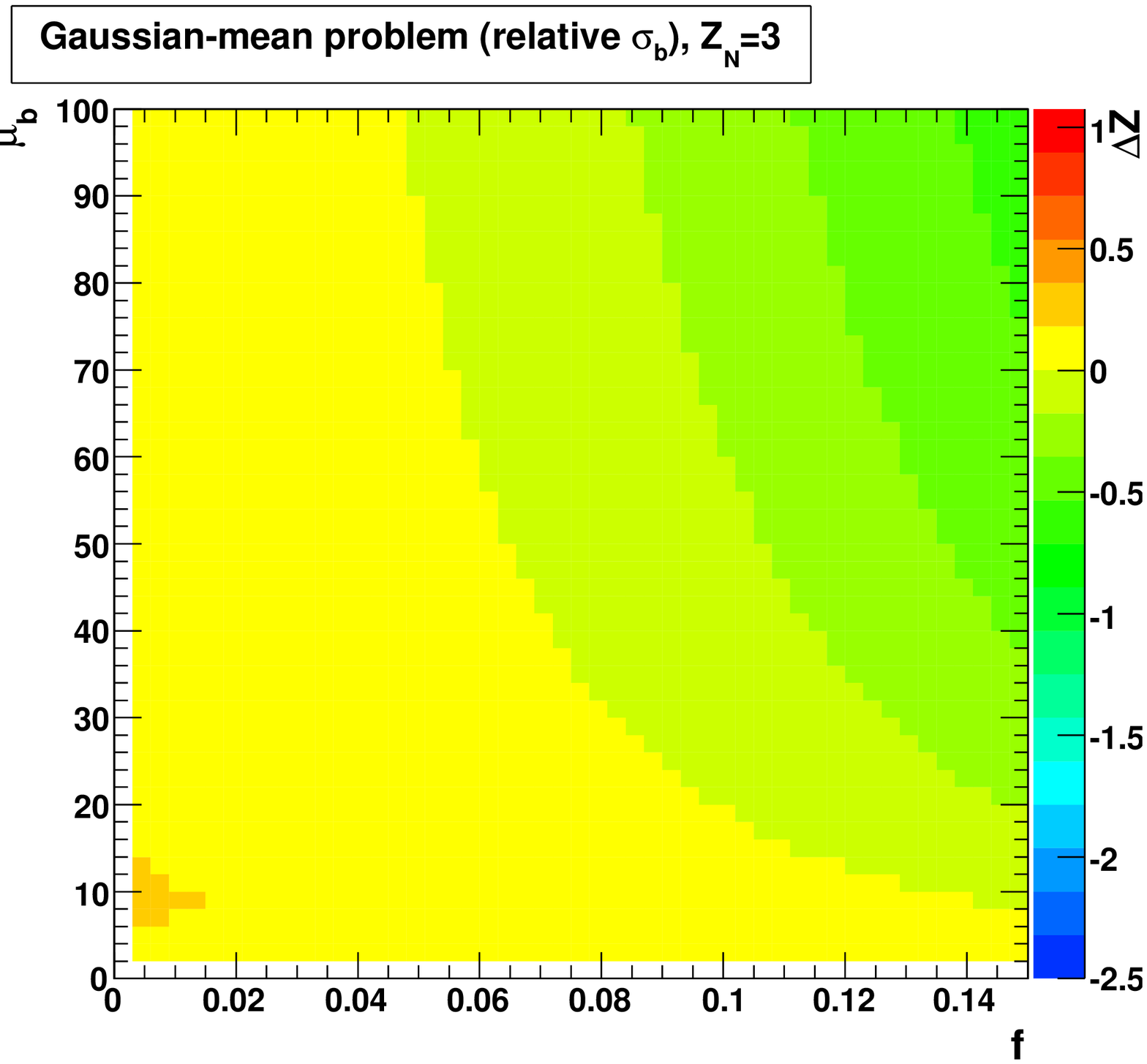}
\caption{ For the Gaussian-mean background problem with exactly known
relative uncertainty $f$, analyzed using the $\zn$ recipe, for each
fixed value of $f$ and $\mubkgnd$, the plot indicates the calculated
$\zdiff$ for the ensemble of experiments quoting $\zclaim \ge 3$,
i.e., a $p$-value of $1.35 \times 10^{-3}$ or smaller.  }
\label{zn_lg_rel_3}
\end{figure}

\begin{figure}[htbp]
\centering
\includegraphics*[width=2.7in]{\epsdir/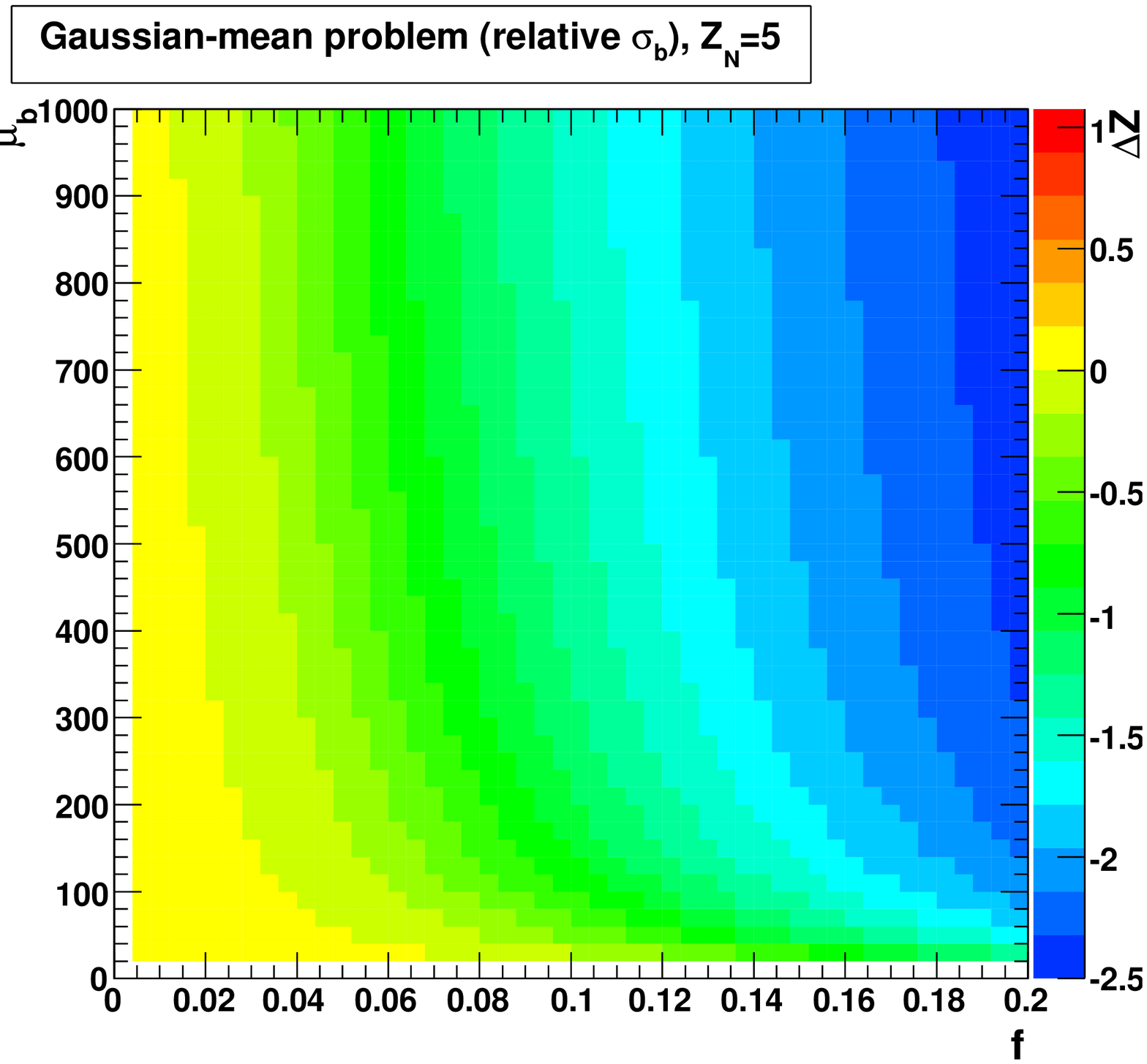}
\includegraphics*[width=2.7in]{\epsdir/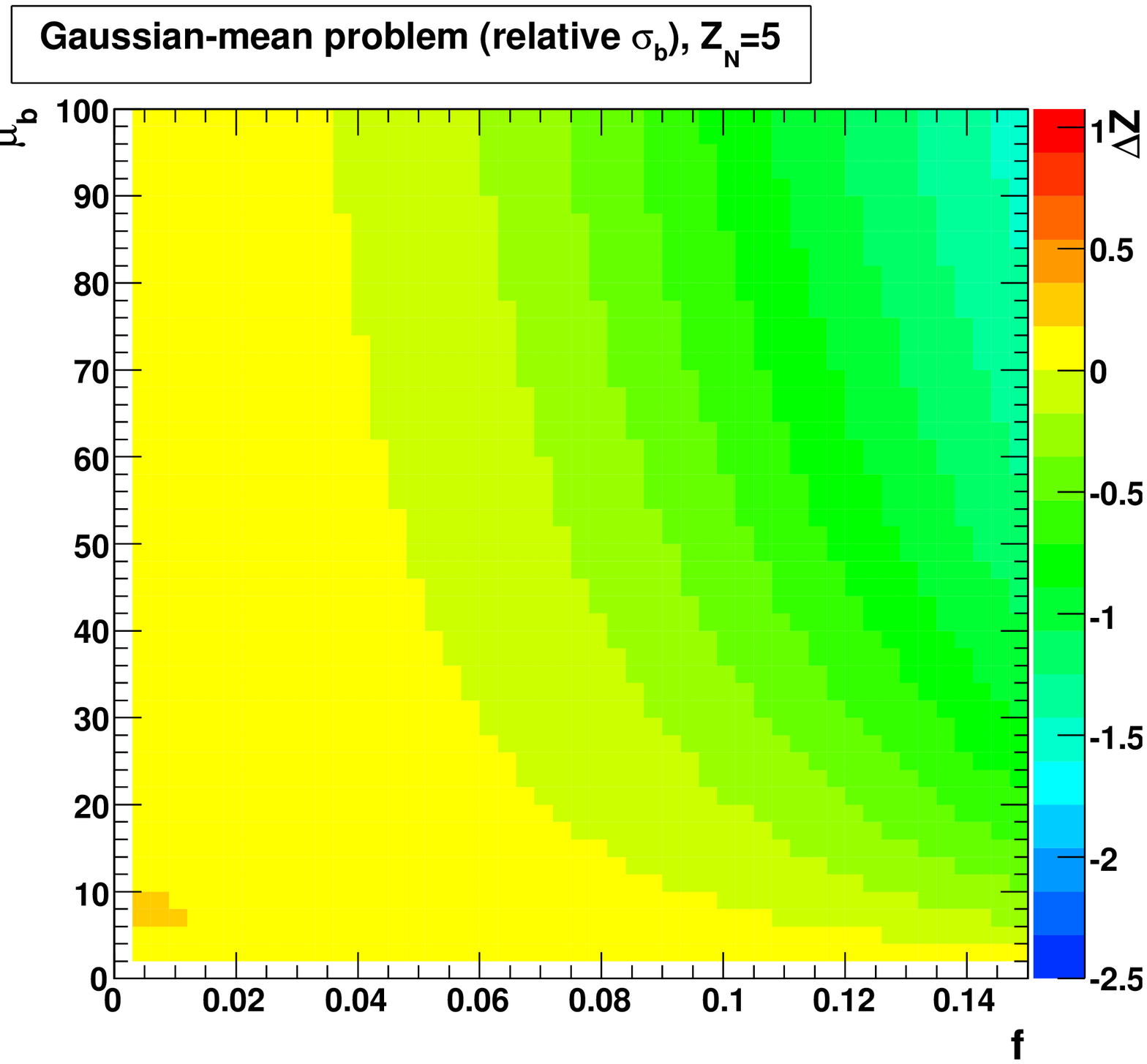}
\caption{ For the Gaussian-mean background problem with exactly known
relative uncertainty $f$, analyzed using the $\zn$ recipe, for each
fixed value of $f$ and $\mubkgnd$, the plot indicates the calculated
$\zdiff$ for the ensemble of experiments quoting $\zclaim \ge 5$,
i.e., a $p$-value of $2.87 \times 10^{-7}$ or smaller.  }
\label{zn_lg_rel_5}
\end{figure}

\begin{figure}[htbp]
\centering
\includegraphics*[width=2.7in]{\epsdir/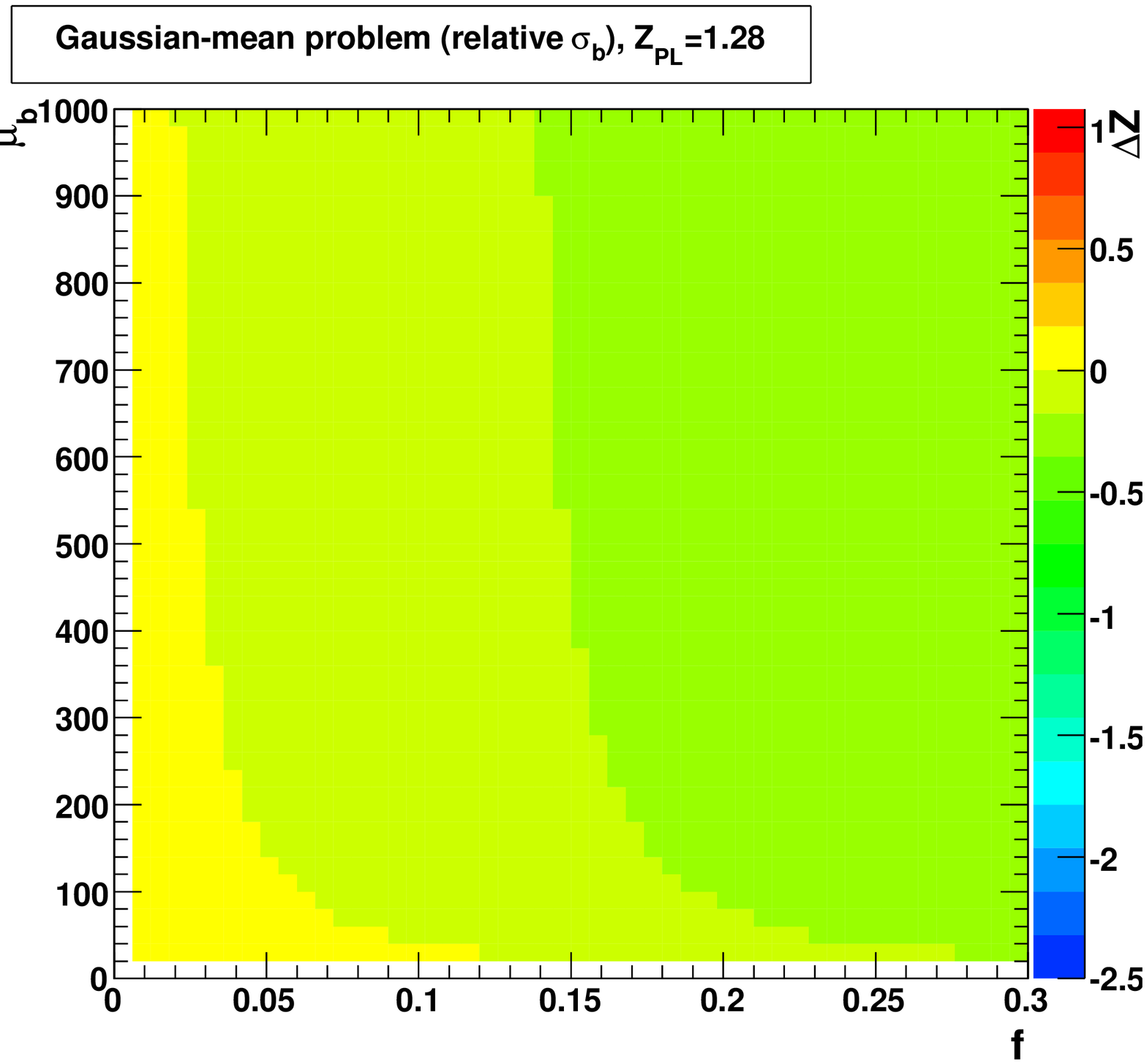}
\includegraphics*[width=2.7in]{\epsdir/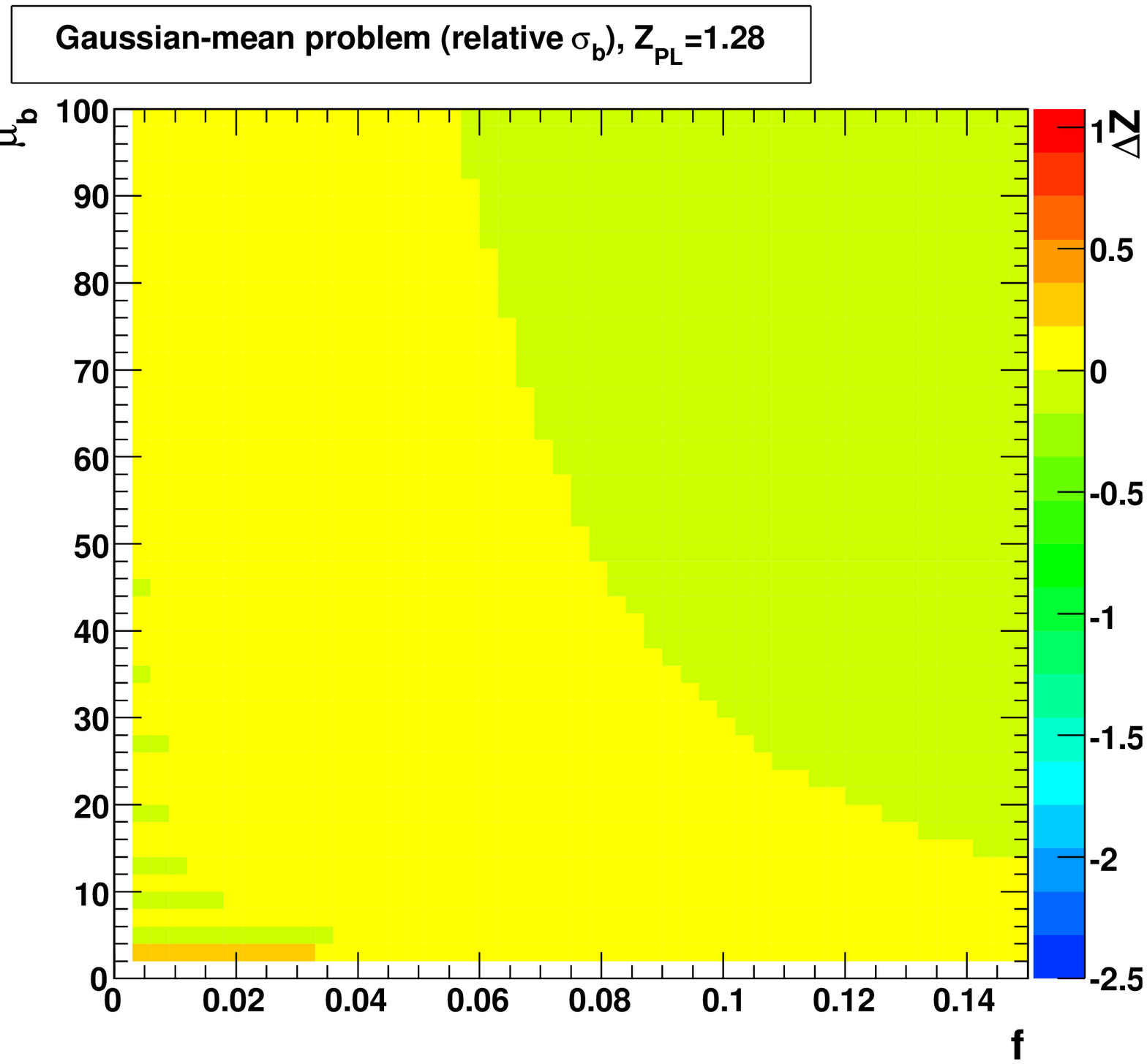}
\caption{ For the Gaussian-mean background problem with exactly known
relative uncertainty $f$, analyzed using the profile likelihood
method, for each fixed value of $f$ and $\mubkgnd$, the plot indicates
the calculated $\zdiff$ for the ensemble of experiments quoting
$\zclaim \ge 1.28$, i.e., a $p$-value of $0.1$ or smaller.  }
\label{proflik_lg_rel_1.28}
\end{figure}

\begin{figure}[htbp]
\centering
\includegraphics*[width=2.7in]{\epsdir/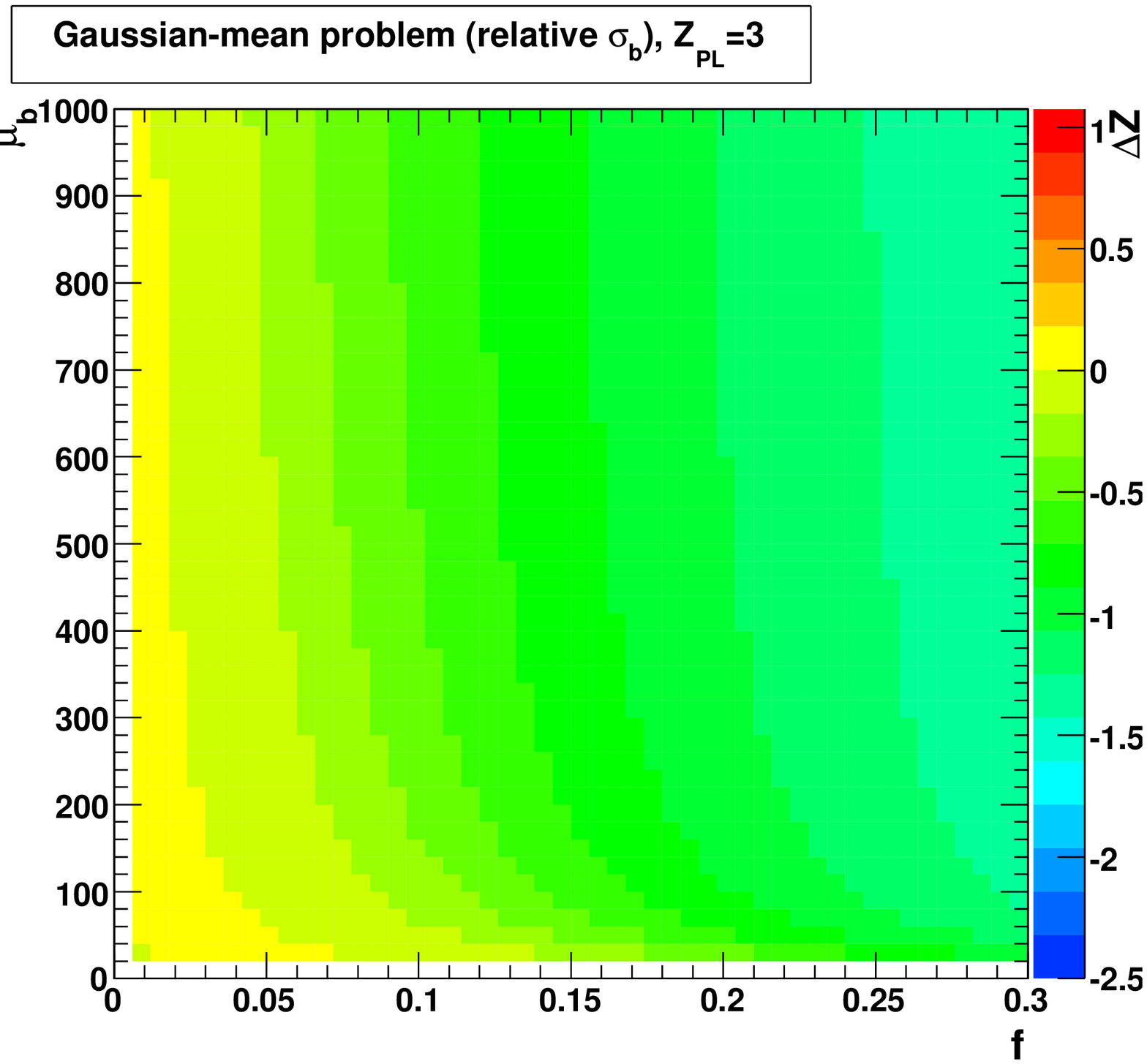}
\includegraphics*[width=2.7in]{\epsdir/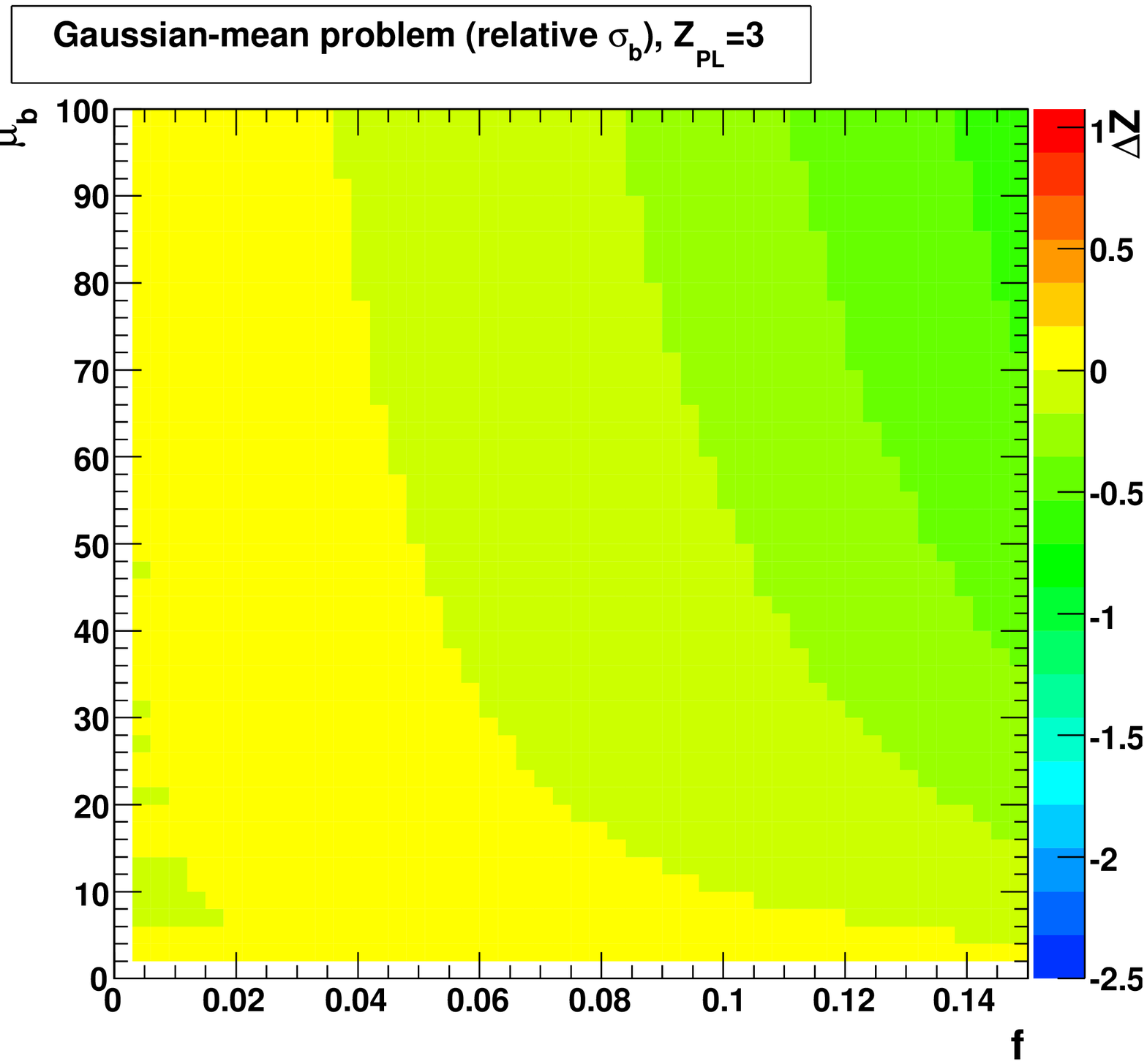}
\caption{ For the Gaussian-mean background problem with exactly known
relative uncertainty $f$, analyzed using the profile likelihood
method, for each fixed value of $f$ and $\mubkgnd$, the plot indicates
the calculated $\zdiff$ for the ensemble of experiments quoting
$\zclaim \ge 3$, i.e., a $p$-value of $1.35 \times 10^{-3}$ or
smaller.  }
\label{proflik_lg_rel_3}
\end{figure}

\begin{figure}[htbp]
\centering
\includegraphics*[width=2.7in]{\epsdir/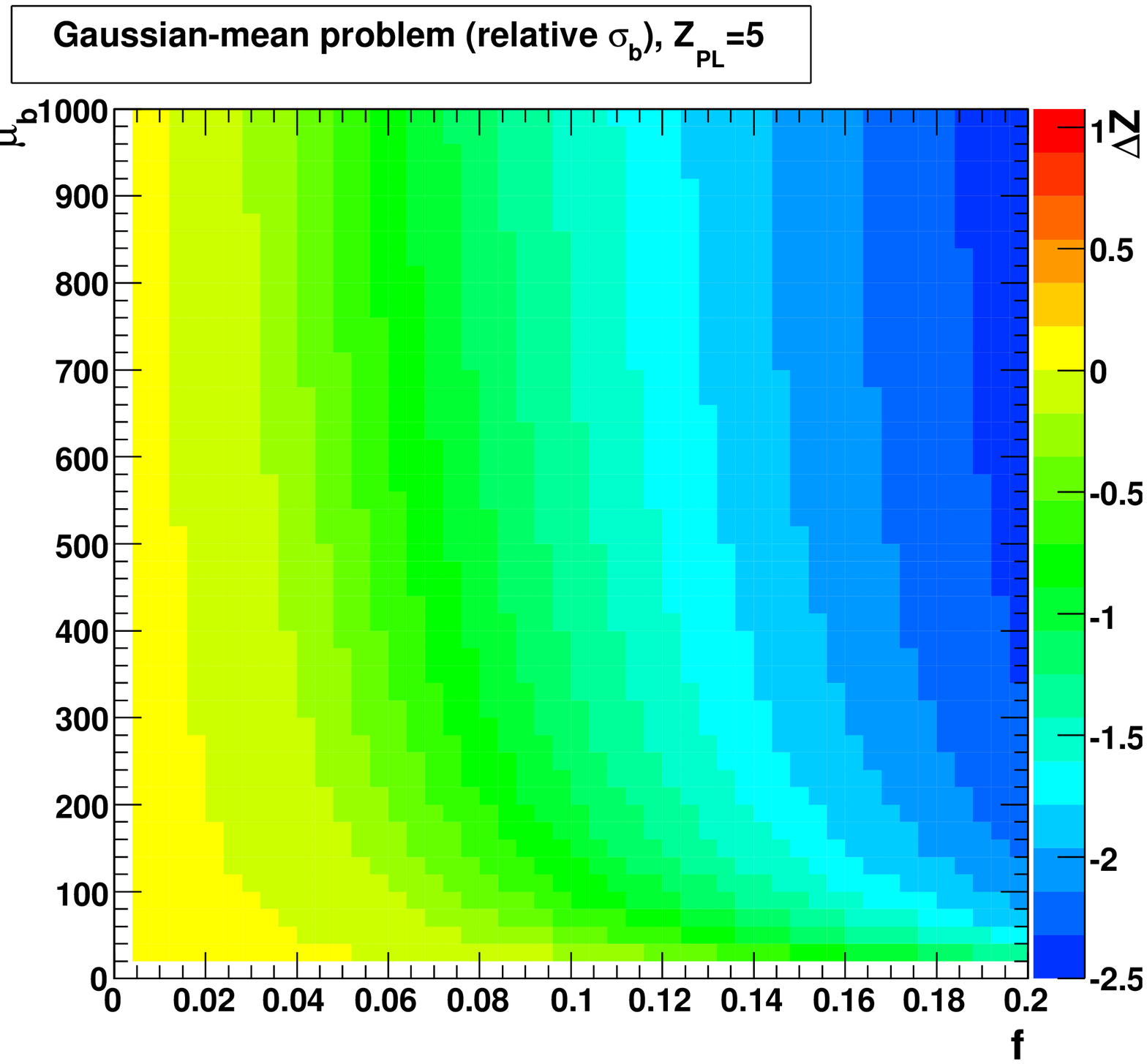}
\includegraphics*[width=2.7in]{\epsdir/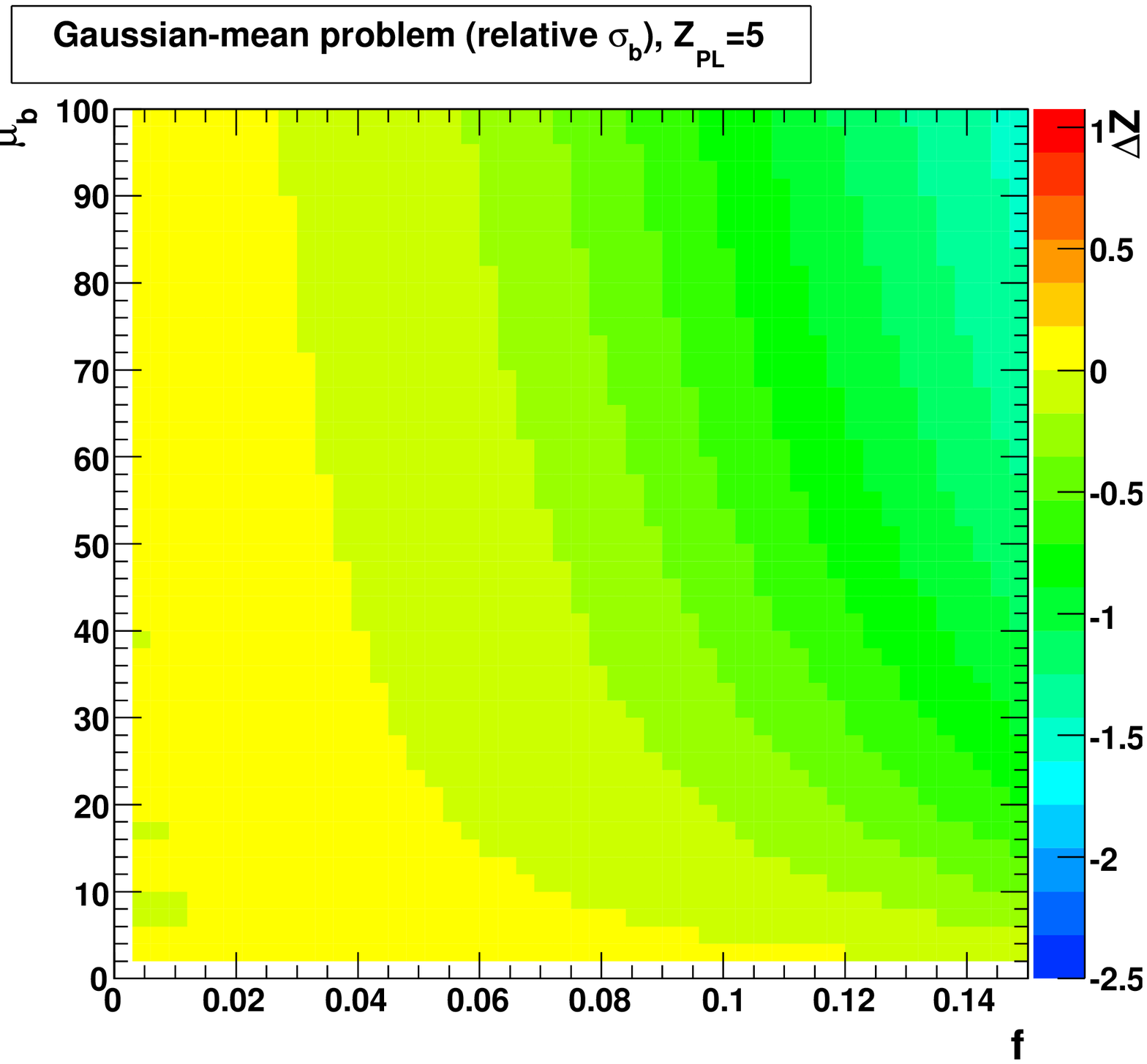}
\caption{ For the Gaussian-mean background problem with exactly known
relative uncertainty $f$, analyzed using the profile likelihood
method, for each fixed value of $f$ and $\mubkgnd$, the plot indicates
the calculated $\zdiff$ for the ensemble of experiments quoting
$\zclaim \ge 5$, i.e., a $p$-value of $2.87 \times 10^{-7}$ or
smaller.  }
\label{proflik_lg_rel_5}
\end{figure}

For the Gaussian-mean background problem with exactly known relative
uncertainty $f$, the results are in Figs.~\ref{zbi_lg_rel_1.28}
through \ref{zbi_lg_rel_5} when analyzed with $\zbi$; in
Figs.~\ref{zn_lg_rel_1.28} through \ref{zn_lg_rel_5} when analyzed
with $\zn$; and in Figs.~\ref{proflik_lg_rel_1.28} through
\ref{proflik_lg_rel_5} when analyzed with $\zpl$.  Both $\zbi$ and
$\zn$ give good coverage for small values of $f$ and small $\mubkgnd$,
but both undercover for large regions of the parameter space, with
$\zbi$ performing slightly better in some regions.  
The undercoverage of $\zbi$ is an effect of small estimates of $\estb$
leading by the rough correspondence of Eqn.~\ref{eqn:corr} to
overestimates of the shape-controlling $\toffoverton = 1/(f^2 \estb)$,
and thus to an overly narrow gamma distribution, which in turn leads
to estimated tail probabilities which are inappropriately small.  
For $\zpl$, the
region of good coverage is smaller in $\mubkgnd$ and $f$ than for
either of $\zbi$ or $\zn$, but like the latter two, the profile
likelihood method also undercovers for a large part of the parameter
space for this problem.

In all of the results shown for $\zpl$ for the Gaussian-mean
background problem (Figs.~\ref{proflik_lg_abs_1.28} through
\ref{proflik_lg_abs_5} and Figs.~\ref{proflik_lg_rel_1.28} through
\ref{proflik_lg_rel_5}), we assume (as we believe to be common
practice) that the experimenter is truncating the Gaussian pdf for
$\estb$ at zero, i.e., set $P(\estb|\mubkgnd)=0$ for $\estb\le 0$ and
renormalized. This results in a denominator for $\lhoodg$ which
depends on $\mubkgnd$, and the determination of $\zpl$ is performed
numerically.  As with the discussion of Gaussian truncation above for
$p(\mubkgnd)$, if this ad hoc procedure makes any material difference,
one should explore other functional forms.  As a check, we removed the
truncation, i.e., used Eqn.~\ref{eqn:LG} as it is written.  The only
perceptible difference is in Figs.~\ref{proflik_lg_abs_1.28} through
\ref{proflik_lg_abs_5}, where the slight undercoverage at $f>0.1$
disappears.

\section{Conclusion}
\label{conclusion}

As seen in these simple prototype problems, naive use of a recipe for
including systematic errors can lead to significant departure from the
claimed $Z$.  For a true on/off problem (sideband estimate of
background in a binned analysis), $\zbi=\zgamma$ avoids undercoverage
by construction, but can be quite conservative for small numbers of
events, at least when the standard intervals for ratio of Poisson
means are used.  Since undercoverage is usually considered to be worse
than overcoverage, we recommend $\zbi$ be considered for general use
in this problem; for a range of values, it is conveniently implemented
in ROOT, as illustrated in Appendix~\ref{app:root}.  However, one
should be aware of the overcoverage with small numbers of events, and
perhaps consider use of alternative intervals for the binomial
parameter or the ratio of Poisson means.  Consistent with long
experience in HEP and GRA and as noted by Rolke et
al. \cite{rolkelopez,rolkeconrad}, the profile likelihood-derived
$\zpl$ provides a strikingly good approximation in most of the
parameter space, with at most modest under-coverage; thus $\zpl$
should also be routinely calculated, especially given the easy use of
the formula of Li and Ma, Eqn.~\ref{eqn:lima}.

For the Gaussian-mean background problem, $\zbi$ works as well as or
better than $\zn$ in much of the space; for extremely small
uncertainties on a large mean background, the implementation in ROOT
can be supplemented using Ref.~\cite{newbetainc}.  The profile
likelihood method performs extremely well for exactly known $\sigmab$
For the case of exactly known relative $f$, all three methods have
severe under-coverage for high values of $\zclaim$ and $f>0.1$.  Since
$\zbi$ and $\zn$ are not well-founded for the Gaussian-mean background
problem, and since the profile likelihood is based on asymptotic
theory, checks of coverage in the region of application are essential.

This paper explores only three recipes for two simple problems; of
course, it is of interest to extend the studies to other recipes and
more complex problems.  For example, if the background in the signal
region has several components, each estimated in a separate subsidiary
experiment, one can attempt to summarize this information
approximately and apply single-components methods.  (One could try
both an approximate $\zn$ and a scaled $\zgamma$ where the scaling
reflects the ratio $f =\sigmab/\mubkgnd$.)  $\zpl$ can be extended to
likelihood functions describing all components.  As problems become
more complex, exact coverage by construction is not likely to be
achieved, since even when a full-blown Neyman construction is feasible
(guaranteeing no undercoverage), it typically leads to overcoverage.
When approximations such as combining background components are made,
one should check the coverage with a full simulation reflecting the
individual components.

As Monte Carlo simulation or numerical integration is often used even
for the simplest problems, the fact the $\zpl$ has the simple
expression in Eqn.~\ref{eqn:lima} is extremely useful both for
checking the results of a simulation, or for providing a speedy
evaluation (e.g. in GRA when data from many segments of the sky must
be monitored in real time).  While for some parameters evaluation of
$\zbi$ encounters numerical problems, its expression in terms of the
incomplete beta function is also quite convenient.

All of these issues become even more severe as $Z$ values as high as 5
or even higher are sought or quoted, as is common in high energy
physics.  The implied tail probability of $2.87\times10^{-7}$ should
be used with caution, as it can be extremely sensitive to underlying
assumptions.  While this paper explores the coverage assuming that the
model is correct, for high $Z$ values one is of course also
susceptible to modeling errors, for example non-Gaussian tails in the
uncertainties.

\ack{We thank Kyle Cranmer and Luc Demortier for numerous enlightening
discussions, pointers to references, and for insightful comments on
earlier versions of this work.  J.L. wishes to thank LANL for
hospitality and financial support during his sabbatical; Tom Loredo
for Ref.~\cite{zhang}; and James Berger for hospitality at the
SAMSI 2006 Institute, and acknowledges useful conversations there with
professors John Hartigan and Joel Heinrich, which helped him toward
the proof that $\zbi=\zgamma$.}  This work was partially supported by
the U.S. Department of Energy and the National Science Foundation.

\clearpage

~\vspace{0.5cm}

\vfill

\vfill

\clearpage

~\vspace{0.5cm}

\vfill

\vfill

\clearpage

\appendix
\section{Notation}
\label{app:notation}
Table~\ref{tab:notation} defines the variables used in this paper.

\begin{table}[h]
\caption{}
\label{tab:notation}
\begin{tabular}{lll} \hline
Symbol & definition \\ \hline
$\noff        $ 
&  total observed in ``off'' (background) region      \\ 
$\non         $ 
&  total observed in ``on'' (signal) region  \\
$\ntot        $ 
&  $\non + \noff$		  \\
$\musignal    $ 
&   true signal mean in ``on'' (signal) region		  \\
$\mubkgnd     $ 
&   true background mean in ``on'' (signal) region	  \\
$\estb        $ 
&   estimate of background mean in ``on'' (signal) region	  \\
$\sigmab      $ 
&   uncertainty on estimate $\estb$ in ``on'' region	  \\
$s$ 
&   estimate of signal events in the ``on'' region = $\non - \estb$ \\
$f            $ 
&   relative uncertainty on $\estb$; $\sigmab/\mu_b$ \\
$\muon        $ 
&   true total mean in signal region = $\mu_s + \mu_b$ \\
$\muoff       $ 
&   true background mean in ``off'' (background) region \\
$\mutot       $ 
&   true total mean in ``on'' plus ``off'' regions = $\muon+\muoff$ \\
$\toffoverton $ 
&   ratio of background means in ``off'' and ``on'' regions:
                                       $\muoff/\mubkgnd$  \\
$\ratmean     $ 
&   ratio of Poisson means $\muoff/\muon$           \\
$\binparam    $ 
&   binomial parameter $\muon/\mutot$	  
\end{tabular}
\end{table}

\section{Derivation of approximate tail area of normal distribution}
\label{app:zapprox}
With $\Phi(Z) = 1 - p$ defined in Eqns.~\ref{zdef}-\ref{Phidef}, we
derive Eqn.~\ref{eqn:zapprox} by starting with the large-$Z$ expansion
of $1-\Phi(Z)$, the cumulative distribution of the normal density
$\phi(Z)$, given as asymptotic expansion 26.2.12 in Ref.~\cite{abro}:
\begin{equation}
p = 1- \Phi(Z) \approx \frac{\phi(Z)}{Z}\ (1 - \frac{1}{Z^2} + \dots)
\end{equation}
Then we follow Ref.~\cite{Linnemann}
by neglecting the higher order terms,
\begin{equation}
p = \frac{1}{\sqrt{2\pi}} \frac{\exp(-Z^2/2)}{Z},
\end{equation}
\begin{equation}
\ln ( p \sqrt{2\pi}) = -Z^2 / 2 -  \ln Z.
\end{equation}
Defining $u$ by further manipulation of the left side,
\begin{equation}
u = -2 \ln ( p \sqrt {2\pi}) = Z^2 + \ln Z^2,
\end{equation}
and substituting $Z^2 \approx u$ into $\ln Z^2$ by initially
neglecting this term, we obtain
\begin{equation}
Z^2 = u -  \ln Z^2 \approx u - \ln u.
\end{equation}
Thus 
\begin{equation}
 Z \approx \sqrt { u - \ln u}.
\end{equation} 

\section{\boldmath Proof of the identity of $\zbi$ and $\zgamma$}
\label{app:proof}
The essence of the proof is to tie together two established
identities. The first is a ``parameter mixing''
\cite{Kendall_vol1,rose} identity that relates the negative binomial
distribution to a mix of Poisson distributions with mean drawn from a
Gamma density (as found in $\pgamma$); the second connects binomial
tail probabilities (as found in $\pbi$) to negative binomial tail
probabilities.

We start by combining Eqns.~\ref{pp}, \ref{eqn:bayesavg}, and
\ref{eqn:lhood}, arriving at the expression (equivalent to Eqn.~6 of
Ref.~\cite{alexandreas}, also in Sec.~III of Ref.~\cite{Linnemann}),
using $\alpha = 1/\toffoverton$,
\begin{equation}
\label{eqn:pgexplicit}
\pgamma = \sum_{j=\non}^{\infty}
\frac{\alpha^j (1/(1+\alpha))^{1+j+\noff} (j+\noff)!}
    {j!\, \noff!}.
\end{equation}
Substituting for $\alpha$ in terms of 
$\binparam = 1/(1+\toffoverton) = \alpha/(1+\alpha)$:
\begin{equation}
\pgamma = \sum_{j=\non}^{\infty}
\frac{\rho^j (1 - \rho)^{1+\noff}(j+\noff)! } {j!\, \noff!}
\end{equation}
\begin{equation}
 = \sum_{j=\non}^{\infty} \nbi(\nfail=j;\, \nsuc=\noff+1,\,
{\rm prob_{\rm success} = 1-\binparam}).
\end{equation}
As indicated, the term summed can be identified as the {\em negative
binomial} \cite{Kendall_vol1,rose} probability NBi for observing $j$
counts on-source (confusingly corresponding to number of ``failures''
in the usual exposition of NBi) in less time than it takes to observe
exactly $\noff+1$ counts off-source (number of ``successes'' $\nsuc$),
where as above, $\binparam = \muon/\mutot$ is the ratio of the mean
numbers of counts on-source to the total mean on and off source (and
hence $1-\binparam$ corresponds to the usual probability for
``success'' in NBi).  Thus, in more compact notation,
\begin{equation}
\label{pnbi}
\pgamma = \nbi( \nfail >  \non-1 | \nsuc=\noff+1),
\end{equation}
thus completing the first main identity.

Now the probability for {\em more} than $k$ on-source counts while
waiting for $m$ off-source counts is precisely equal
\cite{Kendall_vol1} to the probability of finding {\em fewer} than $m$
off-source counts in {\em exactly} $k+m$ total counts for the same
ratio of (on/total) means $\binparam$.  This relates a negative
binomial tail probability to the (ordinary) binomial tail probability:
\begin{equation}
\nbi(\nfail > k | \nsuc=m ) = \bi( \nsuc < m | k+m ).
\end{equation}
The left hand side of this identify matches the right hand side of
Eqn.~\ref{pnbi} for $k=\non-1$ and $m=\noff+1$, so Eqn.~\ref{pnbi}
becomes
\begin{equation}
\pgamma =   \bi( \nsuc < \noff+1 | \non  + \noff )
=\bi(\nsuc < \noff +1 | \ntot ).
\end{equation}
Since the sum of counts is constrained in the Binomial probability,
the latter expression can be re-written in terms of complementary
outcomes:
\begin{equation}
\pgamma  = \bi(\nfail \geq  \non  | \ntot ).
\end{equation}
Comparing with Eqn.~\ref{pbisum} confirms that $\pbi = \pgamma$ and
hence $\zbi = \zgamma$.  This relation was first proved by other
methods in 2003 \cite{kim}, but the present proof seems to be more
illuminating.

\section{\boldmath Details of the Calculations of $\ztrue$}
\label{app:ztrue}
This Appendix provides more details of the calculation of $\ztrue$ in
Sec.~\ref{performance}.
\subsection{Details of calculation of $\ztrue$ for the on/off problem}
For each point in $(\mubkgnd, \toffoverton)$ space for which one
calculates $\ztrue$, one has a plane of discrete points $(\noff,
\non)$, with each point having the joint probability $P(\non |
\mubkgnd) \cdot P(\noff | \toffoverton \mubkgnd)$, where $P$ is the
Poisson probability.  The joint probabilities of all the points
$(\noff, \non)$ for which the recipe studied returns $Z\ge\zclaim$ are
summed to obtain the Type I error rate for a test with the implied
significance level.  Navigating in the plane of $(\noff, \non)$ is
facilitated making use of Eqn.~\ref{eqn-jointProb} and thus
considering lines of constant $\ntot$, along which binomial
probabilities are calculated to obtain efficiently the contour
bounding the region with $Z\ge\zclaim$.

\subsubsection{The $\zbi$ recipe applied to the on/off problem}
\label{detailonoffzbi}
In this simplest case, $\toffoverton$ is fixed and given, so for each
$(\noff, \non)$ point, $\pbi$ and $\zbi$ are calculated from
Eqns. \ref{betaincomplete} and \ref{eqn:z}, and compared to $\zclaim$.

\subsubsection{The $\zn$ recipe applied to the on/off problem}
Starting with $\non$, $\noff$, and $\toffoverton$, one obtains $\estb$
from Eqn.~\ref{bnofftau}), $\sigmab$ from Eqn.~\ref{sigmab_corr}, and
proceeds as usual.  ($f$ is thereby equal to $1/\sqrt{\noff}$.)

\subsubsection{The profile likelihood method applied to the on/off 
problem}
We proceed exactly as in Sec.~\ref{detailonoffzbi}, but instead of
calculating a $p$-value and then $Z$ at each point $(\noff, \non)$,
$Z$ is calculated directly from 
Eqn.~\ref{eqn:lratioCIequiv}.

\subsection{Details of calculation of $\ztrue$ for the 
Gaussian-mean background problem} 
For each point in $(f,\mubkgnd)$ space for which one calculates
$\ztrue$ corresponding to a particular $\zclaim$, one considers all
values of $\non$, and for each value of $\non$ one finds (via a binary
search) the critical value of $\estb$ such that $Z=\zclaim$.  Then the
Type I error rate is the sum of the products of the probability of
obtaining each $\non$ and the Gaussian tail probability for $\estb$
such that $Z\ge\zclaim$ for that $\non$.  The tail probability is
obtained using the error function and true values of $\mubkgnd$ and
$\sigmab=f\mubkgnd$.

\subsubsection{The $\zn$ recipe applied to the 
Gaussian-mean background problem} 
In the case where $\sigmab$ is assumed known, $\zn$ is directly
computed; in the case where $f$ is known, $\sigmab$ is first estimated
by $f\estb$.

\subsubsection{The $\zbi$ recipe applied to the 
Gaussian-mean background problem}
\label{detailgaus}
This again uses the rough correspondence of Eqn.~\ref{eqn:corr}.  In
the case where $\sigmab$ is known exactly, then for each $\non$, one
searches for $\estb$ such that when $\estb$ is used in
Eqns.~\ref{eqn:corr} and \ref{bnofftau} to obtain $\toffoverton$ and
$\noff$, the resulting $\zbi$ from Eqns.~\ref{betaincomplete} and
\ref{eqn:z} is equal to $\zclaim$.  In the case where $f$ is known
exactly, as usual one first estimates $\sigmab$ by $f\estb$ and then
in the same way finds the critical value of $\estb$.  (I.e., one
computes $\toffoverton=\estb/(f\estb)^2$ and
$\noff=\estb\toffoverton$, from which one obtains $\zbi$.)

\subsubsection{The profile likelihood method applied to the
Gaussian-mean background problem} 
Once more using the rough correspondence of Eqn.~\ref{eqn:corr}, we
search for the $\estb$ such that when calculating $\toffoverton$ and
$\noff$ as in Sec.~\ref{detailgaus}, the resulting $Z$ using
Eqn.~\ref{eqn:lratioCIequiv} is equal to $\zclaim$.

\section{\boldmath Implementation of $\zbi$ in ROOT}
\label{app:root}
As noted in Sec.~\ref{binom}, the ratio in Eqn.~\ref{betaincomplete}
is implemented in ROOT \cite{root} following the algorithm in {\em
Numerical Recipes} \cite{numrep}; therefore one simply calls
BetaIncomplete to obtain the $p$-value, and then ErfInverse to convert
it to $Z$ according to Eqn.~\ref{eqn:z}.

For the simple on/off problem with $\non=140$, $\noff=100$, and
$\toffoverton=1.2$, the ROOT commands are:
\begin{verbatim}
double n_on = 140.
double n_off = 100.
double tau = 1.2 
double P_Bi = TMath::BetaIncomplete(1./(1.+tau),n_on,n_off+1)
double Z_Bi = sqrt(2)*TMath::ErfInverse(1 - 2*P_Bi)     
\end{verbatim}
yielding $\pbi = 4.19 \times 10^{-5}$ and $\zbi = 3.93$.

In order to apply $\zbi$ to the Gaussian-mean background problem,
consider for example the observations $\non=140$ and $\estb = 83.3 \pm
8.33$.  Using the correspondence in Eqn.~\ref{eqn:corr} to obtain
$\toffoverton$, and then Eqn.~\ref{bnofftau} to obtain $\noff =
\estb\,\toffoverton$, the ROOT commands are similarly
\begin{verbatim}
double n_on = 140.
double mu_b_hat = 83.33
double sigma_b = 8.333
double tau = mu_b_hat/(sigma_b*sigma_b)
double n_off = tau*mu_b_hat   
double P_Bi = TMath::BetaIncomplete(1./(1.+tau),n_on,n_off+1)
double Z_Bi = sqrt(2)*TMath::ErfInverse(1 - 2*P_Bi)     
\end{verbatim}
The result in this example is then identical to the on/off example
within round-off error, since the chosen $\estb$ and $\sigmab$ were
chosen to reproduce the same $\toffoverton$ and $\noff$.

As $\sigmab$ becomes small, $\toffoverton$ and $\noff$ become large,
so ironically this implementation encounters numerical trouble for
{\em small} uncertainty on the background (and in particular
background known exactly).  For such small errors on background,
neglecting them using Eqn.~\ref{pp} seems reasonable but should be
studied further.  The implementation of the incomplete beta function
by Majumder and Bhattacharjee~\cite{newbetainc} used for the coverage
calculations in Sec.~\ref{binom} provides some expanded capability.
Beyond that, one may consider using the asymptotic formulas in
Eqn.~\ref{eqn:zapprox}.

\end{document}